%% file: wjets_nloew_v2.tex
\documentclass[a4paper]{article}
\pdfoutput=1
\usepackage[T1]{fontenc}
\usepackage{graphicx}
\usepackage{amsmath}
\usepackage{xspace}
\usepackage[numbers,sort&compress]{natbib}
\usepackage{subfig}
\usepackage[shortlabels]{enumitem}

\usepackage{pstricks}
\usepackage{graphicx}
\usepackage{xspace}

\usepackage{jheppub}
\hypersetup{
  pdfauthor={S. Kallweit, J. M. Lindert, P. Maierh\"ofer, S. Pozzorini, M. Sch\"onherr},
  pdftitle={NLO electroweak automation and precise predictions for W + multijet production at the LHC}
}

\newcommand{\Mt}{m_{\Pt}}

\newcommand{\GF}{{G_\mu}}

\newcommand{\LO}{\mathrm{LO}\xspace}
\newcommand{\NLO}{\mathrm{NLO}\xspace}
\newcommand{\QCD}{\mathrm{QCD}}
\newcommand{\EW}{\mathrm{EW}}
\newcommand{\QCDpEW}{\mathrm{QCD+EW}}
\newcommand{\QCDtEW}{\mathrm{QCD\times EW}}

\newcommand{\Gt}{\Gamma_{\Pt}}

\newcommand{\Comix}{{\rmfamily\scshape Comix}\xspace}
\newcommand{\Amegic}{{\rmfamily\scshape Amegic++}\xspace}
\newcommand{\Collier}{{\rmfamily\scshape Collier}\xspace}
\newcommand{\Munich}{{\rmfamily \scshape Munich}\xspace}
\newcommand{\Sherpa}{{\rmfamily\scshape Sherpa}\xspace}
\newcommand{\OpenLoops}{{\rmfamily\scshape OpenLoops}\xspace}
\newcommand{\MunichOpenLoops}{{\rmfamily \scshape Munich+OpenLoops}\xspace}

\newcommand{\calN}{\mathcal{N}}
\newcommand{\calI}{\mathcal{I}}
\newcommand{\X}{X}
\newcommand{\Y}{Y}
\newcommand{\Z}{Z}

\def\wjet#1{\mbox{$W+{#1}\jet$}}
\def\mathswitchr#1{\relax\ifmmode{\mathrm{#1}}\else$\mathrm{#1}$\fi}
\newcommand{\Pt}{\mathswitchr t}

\newcommand{\PWp}{W^+}

\newcommand{\jet}{j}

\newcommand{\ri}{\mathrm{i}}
\newcommand{\rF}{\mathrm{F}}
\newcommand{\rR}{\mathrm{R}}

\newcommand{\rT}{\mathrm{T}}
\newcommand{\rd}{\mathrm{d}}

\newcommand{\rS}{\mathrm{S}}

\newcommand{\MW}{M_\mathrm{W}}
\newcommand{\MZ}{M_\mathrm{Z}}
\newcommand{\MH}{M_\mathrm{H}}

\newcommand{\GeV}{\mathrm{GeV}}
\newcommand{\TeV}{\mathrm{TeV}}

\newcommand{\alphaS}{\alpha_{\rS}}
\newcommand{\ord}{\mathcal{O}}
\newcommand{\gmu}{G_{\mu}}

\newcommand{\HTtot}{H_{\mathrm{T}}^{\mathrm{tot}}}
\newcommand{\HTtotcut}{H_{\mathrm{T,cut}}^{\mathrm{tot}}}
\newcommand{\pT}{p_{\mathrm{T}}}

\newcommand{\pTjone}{p_{\mathrm{T},\jet_1}}

\newcommand{\pTwp}{p_{\mathrm{T,W^+}}}
\newcommand{\minvjj}{m_{\jet_1\jet_2}}
\newcommand{\deltajj}{\Delta\phi_{\jet_1\jet_2}}

\newcommand{\beqar}{\begin{eqnarray}}
\newcommand{\eeqar}{\end{eqnarray}}
\newcommand{\beq}{\begin{equation}}
\newcommand{\eeq}{\end{equation}}
\newcommand{\bit}{\begin{itemize}}
\newcommand{\eit}{\end{itemize}}
\newcommand{\xs}[2]{#1}

\def\refeq#1{\mbox{(\ref{#1})}}

\def\reffi#1{\mbox{Fig.~\ref{#1}}}
\def\reffis#1{\mbox{Figures~\ref{#1}}}
\def\refta#1{\mbox{Table~\ref{#1}}}

\def\refse#1{\mbox{Section~\ref{#1}}}
\def\refses#1{\mbox{Sections~\ref{#1}}}

\def\citeres#1{\mbox{Refs.~\cite{#1}}}

\preprint{
\begin{flushright}
LPN14-127 \\ IPPP/14/107 \\ DCPT/14/214 \\ MCNET-14-26 \\ ZU-TH 42/14 \\ MITP/14-102
\end{flushright}
}

\author[a,c]{S.~Kallweit,}
\author[a]{J.~M.~Lindert,}
\author[a,b]{P.~Maierh\"ofer,}
\author[a]{S.~Pozzorini,}
\author[a,b]{and M.~Sch{\"o}nherr}

\affiliation[a]{Physik-Institut, Universit\"at Z\"urich,
Winterthurerstrasse 190, 
	CH-8057 Z\"urich,
	Switzerland }
\affiliation[b]{Institute for Particle Physics Phenomenology, 
           Durham University, 
           Durham DH1 3LE, 
           UK}
\affiliation[c]{Institut f\"ur Physik \& PRISMA Cluster of Excellence, 
Johannes Gutenberg Universit\"at, 55099 Mainz, Germany}

\emailAdd{kallweit@physik.uzh.ch}
\emailAdd{lindert@physik.uzh.ch}
\emailAdd{philipp.maierhoefer@durham.ac.uk}
\emailAdd{pozzorin@physik.uzh.ch}
\emailAdd{marek.schoenherr@physik.uzh.ch}

\title{
NLO electroweak automation and 
precise predictions for 
$W+$\,multijet production at the LHC}

\abstract{
We present a fully automated implementation of next-to-leading order
electroweak (NLO EW) corrections in the \OpenLoops 
matrix-element generator combined with the 
\Sherpa and \Munich Monte Carlo frameworks. The 
process-independent character of the implemented algorithms opens the door to NLO
QCD+EW simulations for a vast range of Standard
Model processes, up to high particle multiplicity, at current
and future colliders. As a first application, we present 
NLO QCD+EW predictions for 
the production of positively charged on-shell $W$ bosons
in association with up to three jets at the Large
Hadron Collider. At the TeV energy scale, due to the presence of
large Sudakov logarithms, EW corrections reach the
20--40\% level and play an important role for searches of
physics beyond the Standard Model. The dependence of NLO EW
effects on the jet multiplicity is investigated in detail, and
we find that $W+$\,multijet final states feature genuinely
different EW effects as compared to the case of $W+1$\,jet.


}

\begin{document}
\maketitle
\flushbottom

%

\section{Introduction}

The production of a $W$ boson in association with jets represents one of the
most prominent classes of processes at the Large Hadron Collider (LHC).
Thanks to the high cross section and clean experimental signature,
$W+$\,jet production can be probed with high accuracy over a wide range of
jet multiplicities and energy scales~\cite{Aad:2010ab,Aad:2012en,Aad:2014qxa,
Chatrchyan:2011ne,Chatrchyan:2012jra,Khachatryan:2014uva}.
Such measurements provide a powerful testing ground for the Standard Model
as well as for perturbative QCD methods and tools that build the
fundament of all theoretical simulations of high-energy collisions at hadron
colliders.
The process $pp\to W+$\,jets represents also an important background to
various benchmark Standard Model reactions, such as $t\bar t$, single-top,
diboson and Higgs-boson production.  Moreover $W+$\,multijet production is
the dominant background in several searches of physics beyond the Standard
Model (BSM) that are based on signatures with leptons, missing energy, and
jets.
In this context, precise theoretical predictions and reliable uncertainty
estimates for the $W+$\,multijet background can play a critical role for the
precision of the measurements and the sensitivity to new phenomena.
In particular, the accuracy of theoretical simulations 
of $W+$\,multijet production at large transverse momentum and 
high jet multiplicity is very
important for BSM searches at the TeV scale.

Predictions for $W+1\jet$ and \wjet{2}
production at next-to-leading order (NLO) in QCD 
have been known for many 
years~\cite*{Arnold:1988dp,Arnold:1989ub,Ellis:1998fv,Campbell:2002tg,
FebresCordero:2006sj,Campbell:2006cu,Campbell:2008hh,
Cordero:2009kv,Badger:2010mg,Frederix:2011qg,Caola:2011pz}.
More recently,
the advent of on-shell methods~\cite{Ellis:2011cr,Ita:2011hi}
lead to the completion of 
NLO QCD calculations for $W+$\,multijet 
production with three~\cite{Berger:2009zg,Ellis:2009zw,KeithEllis:2009bu,Berger:2009ep},
four~\cite{Berger:2010zx}, and even five~\cite{Bern:2013gka} associated jets.
The inclusion of NLO QCD corrections strongly reduces the
renormalisation and factorisation scale 
dependence  of $W+\,$multijet predictions,
especially for high-multiplicity final states.

At NLO QCD, scale uncertainties for $W+$\,multijet production are typically below 10\%
and can be regarded as a realistic estimate of the error
due to missing NNLO QCD corrections.  However, QCD scale variations do not
reflect the uncertainty due to missing electroweak (EW) corrections.
This is particularly relevant at high transverse momenta, where
EW corrections are strongly enhanced by logarithmic contributions of Sudakov
type~\cite{Fadin:1999bq,Kuhn:1999nn,Denner:2000jv,Denner:2001gw,
Ciafaloni:2000df,Baur:2006sn,Mishra:2013una},
which can reach several tens of percent at the TeV scale.  
Electroweak NLO effects are thus the dominant source of theoretical
uncertainty in NLO QCD simulations of $W+$multijet production at high
transverse momenta, and their inclusion can significantly improve the
sensitivity to BSM searches at the energy frontier.

Electroweak NLO predictions for $W$-boson production in association with a
single jet have been presented in~\cite{Kuhn:2007qc,Kuhn:2007cv} for the
case of stable $W$ bosons, and in~\cite{Denner:2009gj} for the
related process $pp\to \ell\nu \jet$, which includes resonant and non resonant 
contributions to $W\to\ell\nu$ decays.  
At high transverse momenta the EW corrections to $pp\to W+1\jet$ are
negative and very large.  They reach about $-40\%$ at
2~TeV~\cite{Kuhn:2007qc,Kuhn:2007cv}.  The impact of NLO EW corrections on
vector-boson plus multijet processes is expected to be similarly sizable.
However, due to their higher technical complexity, 
NLO EW calculations for multijet final states are almost completely
unexplored to date.  The importance of EW Sudakov logarithms for the $Z+$\,multijet
background to Supersymmetry searches has been estimated
in~\cite{Chiesa:2013yma}, using the next-to-leading logarithmic  
approximation of~\cite{Denner:2000jv}.
Very recently, using the automated one-loop 
generator {\sc Recola}~\cite{Actis:2012qn,Actis:2013dfa}, 
Denner et al.~have presented NLO EW predictions for 
$pp\to\ell^+\ell^-\jet\jet$~\cite{Denner:2013fca,Denner:2014ina}, 
which represents the first  NLO EW calculation for 
vector-boson production in association with more than one jet.
Important steps towards the automation of NLO EW corrections have been undertaken also within the
{\sc Madgraph5}\_{aMC@NLO} 
framework~\cite{Frixione:2014qaa,Alwall:2014hca}
and by the {\sc GoSam}~\cite{Cullen:2014yla} collaboration.

In this paper we present a fully automated implementation of NLO EW corrections 
based on the \OpenLoops one-loop generator~\cite{Cascioli:2011va} 
in combination with
the \Munich~\cite{munich} and 
\Sherpa~\cite{sherpaqedbrems,Gleisberg:2007md,Gleisberg:2008ta} 
Monte Carlo programs. 
The implemented algorithms are highly efficient and fully general. They
support NLO QCD and EW
simulations of high-energy collisions for any Standard Model process 
up to high particle multiplicity.
As an application we consider
$W+$multijet production and, for the first time, 
we present NLO QCD+EW predictions for $pp\to W+2\jet$ and $pp\to W+3\jet$
at the LHC.
Given that, at least for the case of $W+1j$ production,
the EW corrections feature a neglible dependence on the $W$-boson 
charge~\cite{Kuhn:2007cv}, in this paper we restrict ourselves 
to the case of positively charged $W$ bosons.

Virtual EW corrections are automated within the
\OpenLoops framework, which is based on a fast numerical recursion 
for the generation of one-loop scattering amplitudes in the Standard
Model~\cite{Cascioli:2011va}.  The \OpenLoops program has already been
applied to various nontrivial NLO QCD~\cite{Cascioli:2013gfa,
Cascioli:2013era, Cascioli:2013wga, Maierhofer:2013sha, Hoeche:2014qda,
Hoeche:2014rya} and NNLO QCD~\cite{Abelof:2014fza,Grazzini:2013bna,
Cascioli:2014yka, Gehrmann:2014fva} simulations,\footnote{In the context of the NNLO
calculations of~\cite{Abelof:2014fza,Grazzini:2013bna,
Cascioli:2014yka, Gehrmann:2014fva}
\OpenLoops was used for the evaluation of all relevant real--virtual and real--real amplitudes.} 
and its first public version was released very recently~\cite{hepforge}.  
As compared to QCD corrections,
in the EW sector virtual corrections are significantly more involved 
as they receive contributions from a wider set of particles ($\gamma$, $Z$, $W$, $H$), 
which are characterised by a nontrivial mass spectrum.  Moreover, while NLO QCD 
corrections are usually dominated by real-emission effects, in the case of NLO EW
corrections the most prominent role is typically played by the one-loop virtual contributions.
In particular, the exchange of 
virtual EW gauge bosons can give rise to large Sudakov logarithms.

Within our computational framework virtual EW corrections 
are complemented by two independent and fully automated 
implementations of NLO QED bremsstrahlung.
The first one is based on \Munich~\cite{munich},
a fully generic and very fast parton-level Monte Carlo integrator that 
has already been applied to various nontrivial
multi-particle NLO calculations~\cite{Denner:2010jp,
Denner:2012yc,
Denner:2012dz,
Cascioli:2013gfa,
Cascioli:2013wga}
and also to NNLO calculations~\cite{Grazzini:2013bna,
Cascioli:2014yka,
Gehrmann:2014fva}
based on $q_\rT$-subtraction~\cite{Catani:2007vq}.
The second implementation of QED bremsstrahlung is based on the
\Sherpa Monte Carlo generator~\cite{Gleisberg:2007md,Gleisberg:2008ta}, 
which was used in the pioneering 
NLO QCD calculations of vector-boson plus multijet 
production~\cite{Berger:2009zg,Ellis:2009zw,KeithEllis:2009bu,
Berger:2009ep,Berger:2010zx,Bern:2013gka}, as well as for 
their matching to the parton shower~\cite{Hoeche:2012ft}
and the merging of multijet final states at NLO~\cite{Hoeche:2012yf}.
Both Monte Carlo tools, \Munich and \Sherpa, employ the dipole subtraction
scheme~\cite{Catani:1996vz,Catani:2002hc} for the cancellation of infrared
singularities.  The relevant one-loop and (in the case of \Munich) tree matrix 
elements are obtained from \OpenLoops through generic
built-in interfaces, and the full chain of
operations that are relevant for NLO EW and QCD simulations---from process definition 
to the calculation of
fully differential collider observables---is supported in
a completely automated way.
These tools have the potential to address NLO QCD+EW simulations for a very
wide range of processes.  As reflected in the 2013 Les Houches wish
list~\cite{Butterworth:2014efa}, this represents one of the key priorities
for the accurate theoretical interpretation of the data that will be
collected during Run2 of the LHC.

The paper is organised as follows.  
Sect.~\ref{se:generalities} is devoted
to general features of NLO EW corrections, including the interplay of NLO
EW and QCD contributions, the treatment of initial- and final-state photons,
and the real emission of weak gauge bosons.
The automation of NLO QCD+EW simulations 
is presented in Sect.~\ref{se:automation}, with emphasis on
genuinely new aspects that go beyond a mere extension of NLO automation from
the $SU(3)$ to the $SU(2)\times U(1)$ sector of the Standard Model.
The building blocks of the NLO QCD+EW calculation of $pp\to W+1,2,3$\,jets 
are introduced in
Sect.~\ref{se:Wjets}, where technical subtleties related to the on-shell
treatment of $W$ bosons are discussed in detail.
The setup of the simulation and numerical predictions for $W^+$
production in association with up to three jets at the 13\;TeV LHC are presented in
Sects.~\ref{se:setup} and~\ref{se:results}, respectively.  The dependence of
NLO EW effects on the jet multiplicity and new features that emerge in
multijet final states are studied in detail.  Our conclusions can be found
in Sect.~\ref{se:conclusions}.

\section{General aspects of NLO electroweak corrections}
\label{se:generalities}

In this section we discuss general aspects of NLO EW calculations that
play a nontrivial role in the definition of physical observables as well as
for the extension of automated NLO algorithms from the QCD to the EW sector
of the Standard Model.

\subsection[Power counting in $\alpha$ and $\alphaS$]{Power counting in $\mathbf{\alpha}$ and $\mathbf{\alphaS}$}
\label{se:powercounting}

In the case of simple scattering processes, where the Born cross section can
be associated with a unique perturbative order $\alphaS^n\alpha^m$ with
fixed powers $m$ and $n$, the NLO QCD and EW corrections can be
unambiguously identified as, respectively, the
$\ord(\alphaS^{n+1}\alpha^m)$ and $\ord(\alphaS^n\alpha^{m+1})$
contributions to the cross section.  However, in general, scattering
processes can receive various Born contributions of
$\ord(\alphaS^{n}\alpha^{m})$ with $n+m$ fixed, and $0\le n,m \le n+m$.  In this case,
which applies to processes that involve more than one external quark--antiquark pair,
the naive separation of NLO QCD and NLO EW effects is not possible, and
infrared singularities of QCD and EW type start ``overlapping''.
This feature is schematically depicted in Figs.~\ref{fig:4qEW}--\ref{fig:4qMIX}
for the case of $q\bar q \to q'\bar q'$ scattering, which is the
simplest process with a nontrivial EW--QCD interplay.
In general, at Born level it receives contributions\footnote{Mixed interference terms of 
$\ord({\alphaS\alpha})$
contribute only in case of equal quark flavours, 
$q'=q$.} of  order $\alphaS^2$, $\alphaS\alpha$ and $\alpha^2$.
The representative diagrams in Fig.~\ref{fig:4qEW} illustrate 
what might be {\it naively} regarded as the NLO EW correction to the 
$\ord(\alphaS^2)$ Born contribution, namely
terms of $\ord(\alphaS^2\alpha)$ that result from  
order $\alphaS\times\alphaS$ tree interferences 
(Fig.~\ref{fig:4qtreeQCD}) via insertions 
of real photons (Fig.~\ref{fig:4qrealEW})
or virtual EW particles (Fig.~\ref{fig:4qvirtEW}).
However, as illustrated in Fig.~\ref{fig:4qMIX},
contributions of the same order 
$\alphaS^2\alpha$ can be obtained
also from $\alphaS\times\alpha$ tree interferences
(Fig.~\ref{fig:4qtreeMIX}) via insertions 
of real (Fig.~\ref{fig:4qrealMIX})
or virtual QCD partons (Fig.~\ref{fig:4qvirtMIX}). The latter
can be {\it naively} regarded as the NLO QCD corrections to 
the $\ord(\alphaS\alpha)$ Born contribution.
However, a consistent separation of $\ord(\alphaS^2\alpha)$ corrections into 
NLO EW and NLO QCD terms, as suggested through 
Figs.~\ref{fig:4qEW}--\ref{fig:4qMIX}, is not possible.
First of all, the two categories overlap since 
diagrams like the one-loop topology in Fig.~\ref{fig:4qvirtEW}
can be regarded both as an EW or QCD correction 
to a gluon- or $\gamma/Z$-exchange tree amplitude, respectively.
Moreover, this type of diagrams involves infrared (IR)
singularities of EW {{\it and} QCD type, 
whose cancellation requires photon and gluon emission terms
of type~\ref{fig:4qrealEW} and~\ref{fig:4qrealMIX}, respectively.
It is thus clear that the full set of 
contributions of $\ord(\alphaS^2\alpha)$ must be taken into account.
These considerations can be extended to processes involving
additional external gluons, quarks and EW particles, and 
in general only the full set of 
contributions with a fixed order in $\alphaS$ and $\alpha$
can be considered as a well defined perturbative prediction.
As far as the terminology is concerned, 
the most transparent approach is to label each contribution 
with the respective order in $\alphaS$ and $\alpha$. 
However, depending on the context, it might be convenient to denote
the full set of $\ord(\alphaS^{n}\alpha^{m+1})$ 
terms as NLO EW correction with respect to 
$\ord(\alphaS^{n}\alpha^{m})$ or, alternatively,
as NLO QCD correction with respect to $\ord(\alphaS^{n-1}\alpha^{m+1})$.


\begin{figure*}[t]
\centering
\subfloat[][Leading QCD Born]{   
   \includegraphics[width=0.27\textwidth]{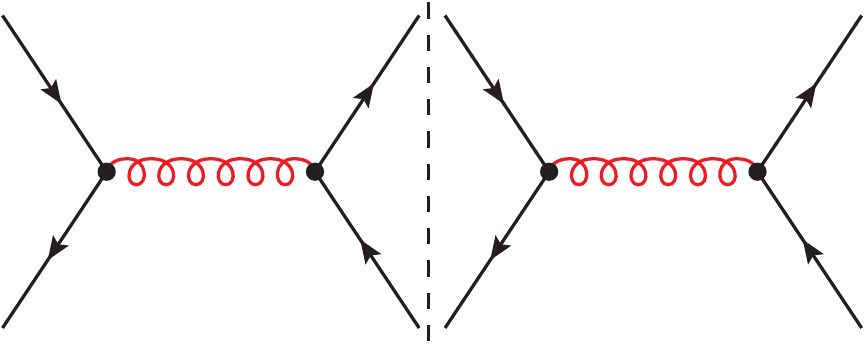}
   \label{fig:4qtreeQCD}
}\qquad
\subfloat[][Real $\ord(\alphaS^2\alpha)$ correction]{   
   \includegraphics[width=0.27\textwidth]{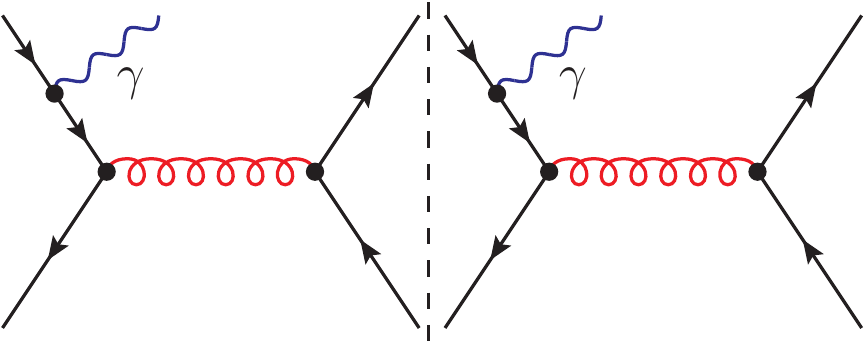}
   \label{fig:4qrealEW}
}\qquad
\subfloat[][Virtual  $\ord(\alphaS^2\alpha)$ correction]{   
   \includegraphics[width=0.27\textwidth]{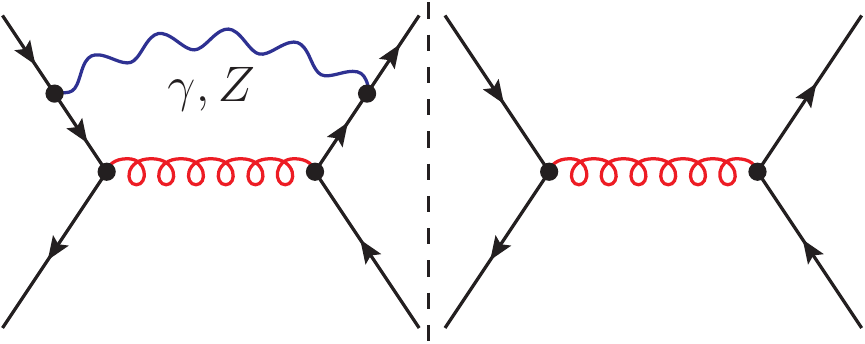}
   \label{fig:4qvirtEW}
}
\caption{Corrections of $\ord(\alphaS^2\alpha)$ that are generated by 
dressing $\ord(\alphaS^2)$ Born terms with 
real or virtual EW partons.}
\label{fig:4qEW}
\end{figure*}


\begin{figure*}[t]
\centering
\subfloat[][QCD-EW Born interference]{   
   \includegraphics[width=0.27\textwidth]{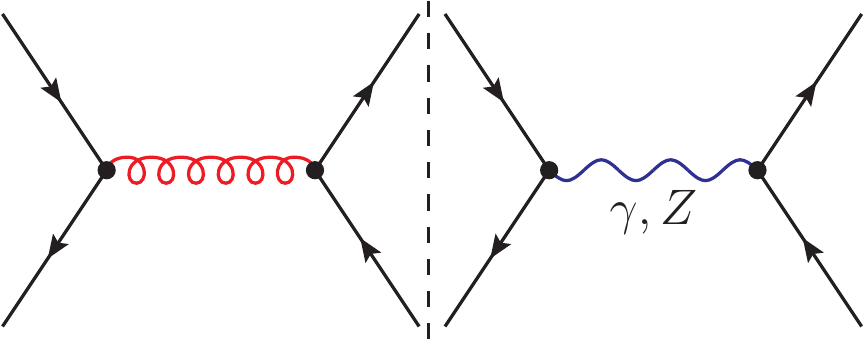}
   \label{fig:4qtreeMIX}
}\qquad
\subfloat[][Real  $\ord(\alphaS^2\alpha)$ correction]{   
   \includegraphics[width=0.27\textwidth]{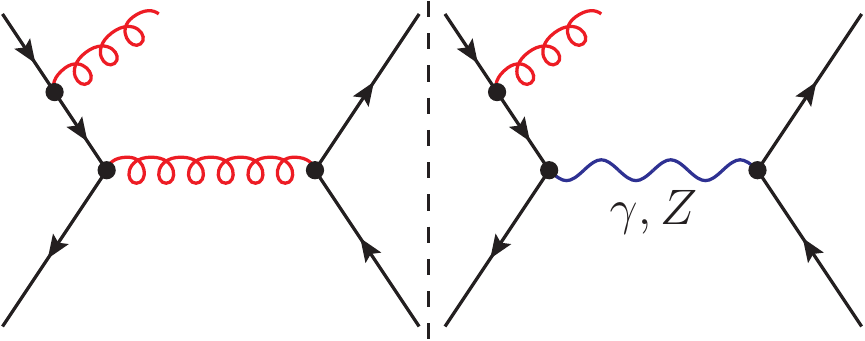}
   \label{fig:4qrealMIX}
}\qquad
\subfloat[][Virtual  $\ord(\alphaS^2\alpha)$ correction]{   
   \includegraphics[width=0.27\textwidth]{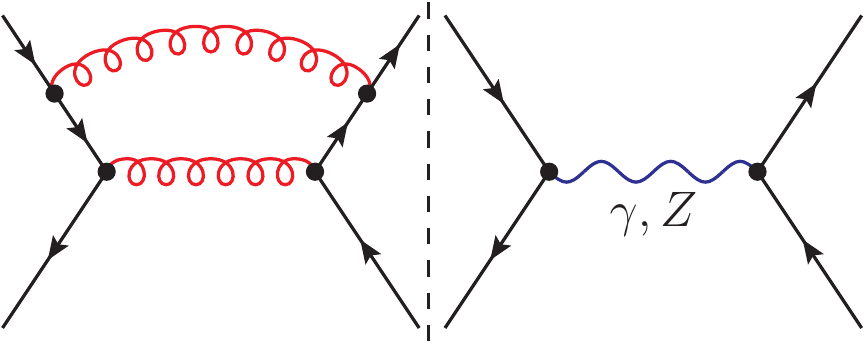}
   \label{fig:4qvirtMIX}
}
\caption{Corrections of $\ord(\alphaS^2\alpha)$ that are generated by 
dressing $\ord(\alphaS\alpha)$ Born terms with 
real or virtual QCD partons.
The
interference of the tree diagrams in Fig.~\ref{fig:4qtreeMIX}
vanishes as a result of their particular colour flow,
but this picture should be understood as a schematic illustration of 
non-vanishing EW--QCD interferences that arise 
between $s$-channel and $t$-channel contributions to same-flavour 
$q\bar q\to q\bar q$ scattering, or in processes with additional 
external gluons.
}
\label{fig:4qMIX}
\end{figure*}

\subsection{Virtual and real electroweak corrections}
\label{se:virtrealewcorr}

The infrared safe definition of physical observables 
requires the combination of virtual and real 
corrections at the same perturbative order.
As discussed above, the cancellation of all virtual IR singularities 
at a certain order $\alphaS^{n}\alpha^{m+1}$
can require various bremsstrahlung processes that involve
additional photons, QED charged particles (quarks and leptons) and also QCD 
partons (gluons and quarks).  The inclusion of such bremsstrahlung contributions 
is mandatory,
since the emission of {\it massless} partons cannot be resolved as a
separate process in the soft and collinear limits.  
As for the emission
of heavy particles, i.e.~$W,Z$ and Higgs bosons or top quarks, 
the situation is different. For instance, 
in QCD, from the viewpoint of $\alphaS$ power counting, 
top quark emissions can be included in the
definition of NLO bremsstrahlung on the same 
footing as light-quark emissions.
However, top-quark emissions are not indispensable for the cancellation
of IR singularities, and since they lead to 
completely different experimental signatures, 
final states with additional top quarks are 
most conveniently handled as separate processes. For example,
it is preferable to exclude $pp\to t\bar t W$ 
from the NLO QCD corrections to $pp\to W+1\jet$, and
to treat it as a separate $2\to 3$ process.

Similarly, at NLO EW, while the emission of heavy particles can be formally
treated as NLO bremsstrahlung together with photon emission, we advocate a
process bookkeeping approach where massive emissions are handled as separate
processes, and only massless (or light) emissions are included in the
definition of NLO EW corrections.
For instance, $pp\to WZ$ 
should not be included
in the NLO EW corrections to single $W$ production, and
should be kept as a separate diboson production process.  
Of course, certain observables
receive contributions both from $WZ$ and single $W$ final states, but the
different physics dynamics of the two processes, which are individually IR
finite, provides a strong motivation for a systematic separation of
theoretical predictions for $pp\to W$ and $pp\to WZ$.
Moreover, we point out that a systematic inclusion of 
massive EW bremsstrahlung at NLO can lead
to quite unpleasant ambiguities and double counting issues.  In particular,
besides the overlap between processes with different vector boson
multiplicity, such as $W$ and $WZ$ production, also 
processes involving different kinds of vector bosons would start overlapping.  
For example, $WZ$ production would contribute 
to the NLO EW corrections to both single $W$ and single $Z$ production.

Thus, in order to avoid overlap and double-counting issues, at the 
{\it technical} level it is preferable to adopt a process bookkeeping approach 
that keeps massive real emissions 
apart from the NLO EW
corrections to the respective ``no emission'' processes.
On the other hand, at the level of {\it physical} observables, one has to
keep in mind that these two contributions enter at the same perturbative
order and are related to each other in a subtle way. In particular, at the
TeV scale both contributions involve large Sudakov logarithms, whose effects
can partially cancel against each other in a way that bears some analogies
with the cancellation of IR singularities in QCD.
More precisely, at the TeV scale
one-loop EW amplitudes involve large negative logarithms, which 
originate from the exchange of virtual $Z/W$ bosons
in the soft and collinear regions 
and tend to be compensated by the real emission of 
soft and collinear $Z/W$
bosons~\cite{Ciafaloni:2000df,Baur:2006sn,Mishra:2013una}.
However, for realistic collider processes this
kind of cancellation is always
incomplete and often
rather modest.  Firstly, Sudakov logarithms of soft origin do not cancel
completely since initial- and final-state particles carry $SU(2)\times U(1)$ charges and
thus do not fulfill the conditions of the Bloch--Nordsieck theorem~\cite{Ciafaloni:2000df}.
Secondly, Sudakov logarithms from initial-state collinear weak-boson 
emission do not cancel at all, since they are not factorised into 
standard PDFs.  Thirdly, the
suppression of parton luminosities at high centre-of-mass energy and other
kinematic effects tend to reduce the quantitative impact of the emission of
extra heavy particles in a significant way.  Finally, as far as differential
observables and experimental cuts are concerned, one should keep in mind
that the contributions from virtual and real $Z/W$ bosons behave in a
completely different way.  

In summary, in presence of large EW Sudakov effects the interplay between
virtual EW corrections and massive EW bremsstrahlung deserves detailed
quantitative studies, but these different contributions can and should be
simulated as independent processes.

\subsection{Photon-induced processes}
\label{sec:photons}

Electroweak NLO corrections involve various types of massless real-emission
contributions that arise from $q\to q\gamma$, $\bar q\to \bar q\gamma$, and
$\gamma\to q\bar q$ splitting processes, as well as from analogous leptonic
and usual QCD splittings.
In the case of hadronic collisions, initial-state emissions of photons and
quarks give rise to $\ord(\alpha)$ collinear singularities that need to be
factorised into the PDFs.  This requires the introduction of a photon
distribution function and the inclusion of QED effects in the DGLAP
evolution of the (anti)quark and photon
densities~\cite{Martin:2004dh,Ball:2013hta}.  Consequently, hadronic cross
sections receive photon-induced contributions with photon--hadron and
photon--photon initial states.

For what concerns the power counting in $\alpha$, one option is 
to treat the photon density as $\ord(1)$ contribution, similarly as
for the quark and gluon PDFs.  In this case, for EW-induced
processes such as dilepton and $W^+W^-$ hadro-production, the
$\gamma\gamma$ channel can contribute already at LO, 
and the corresponding NLO
EW corrections involve $q\gamma$ and $\bar q\gamma$-induced bremsstrahlung
contributions with an additional final state (anti)quark.
In QCD-induced hadronic collisions where (anti)quark--gluon channels are
open at LO, also (anti)quark--photon tree level channels contribute.
However the latter involve a relative  suppression factor $\alpha/\alphaS$.
Similar considerations hold also for gluon--photon 
induced processes that involve $q\bar q$ pairs in the final state.

As an alternative power-counting approach, one can handle the photon PDF as
an $\ord(\alpha)$ contribution.  This is justified by the fact that, 
in the typical kinematic range of LHC collisions, the ratio of the photon to gluon PDFs 
is of order $10^{-2}$. 
In this case, $\gamma$--hadron and $\gamma\gamma$-induced processes enter
only at NLO and NNLO, respectively.  Thus at NLO only tree level
$\gamma$--hadron induced processes need to be included, if they contribute
at all to the considered order in $\alphaS$ and $\alpha$.  Such
$\gamma$--hadron tree processes enter at the same perturbative order as NLO
bremsstrahlung contributions associated with initial-state $q\to q\gamma^*$
and $\bar q\to \bar q\gamma^*$ splittings, thereby ensuring the consistent factorisation of the related
collinear singularities into the photon PDF.

For particular processes and kinematic regions where $\gamma$-induced
contributions turn out to be enhanced one should either 
include all NLO terms by counting the photon density as $\ord(1)$ PDF,
or stick to the $\ord(\alpha)$ photon PDF approach and include those 
photon-induced contributions that are formally of NNLO 
in this counting scheme, but quantitatively important.

\subsection{Democratic jet clustering, quark fragmentation and photon recombination}
\label{se:democlstering}

\def\Drgq{\Delta R_{\gamma,q}}
\def\Drrec{R_{\gamma q}^{\mathrm{rec}}}
\def\zgamma{z_\gamma}
\def\zgammathr{z_{\mathrm{thr}}}
\def\ngamma{N_\gamma}
\def\nqcd{N_{g+q}}

In order to guarantee the cancellation of infrared
(soft and collinear) singularities in perturbative QCD, jet observables 
need to be defined through infrared-safe jet algorithms.  In particular,
jets must be insensitive to radiative processes that involve the emission of
massless QCD partons in the soft and collinear limits, i.e.~emission and
no emission of soft or collinear partons must be indistinguishable at the
level of jet observables.
In presence of NLO EW corrections, it is clear that the requirement of IR safeness needs 
to be extended to the singularities associated with $q\to
q\gamma$, $\bar q\to \bar q\gamma$ and $\gamma\to q\bar q$ QED splittings.  In
principle, this can be easily achieved through the so-called democratic
jet clustering approach~\cite{Glover:1993xc,GehrmannDeRidder:1997wx,GehrmannDeRidder:1998ba}, 
where photons and QCD partons are handled on the same footing at each clustering step.
Jets resulting from democratic clustering contain photons, quarks and
gluons, and their four-momenta are determined by the sum of all
jet constituents, including photons.  

While the cancellation of collinear singularities of QCD and QED
type is automatically ensured by democratic jet clustering, such a combined treatment 
of collinear quark--photon and gluon--photon pairs can
hamper the cancellation of soft-gluon singularities.  This is due to the
fact that democratic jets are completely inclusive with respect to 
collinear photon emission, i.e.~the photon energy fraction
inside a jet, $\zgamma=E_{\gamma}/E_{\mathrm{jet}}$, extends over the whole range 
$0\le \zgamma\le 1$.
This inclusiveness is crucial for 
the cancellation of collinear singularities associated with 
(anti)quark--photon pairs. However, in the case of 
gluon--photon pairs, in the region $\zgamma\to 1$, where the jet consists of an almost pure photon,
the gluon emission inside the jet becomes arbitrarily soft, 
thereby giving rise to IR QCD singularities.

The consistent cancellation of this kind of singularities can be achieved in
two different ways.  The first solution is to adopt a democratic
treatment of photons and QCD partons also in the definition of processes
that involve final-state jets.  This implies that, at tree level, a jet can
consist of either a QCD parton or a photon, while $N$-jet production 
receives tree level contributions from subprocesses 
with a variable number of final state
QCD partons, $\nqcd$, and final state photons, $\ngamma=N-\nqcd$, depending
on the actual order $\alphaS^n\alpha^m$.
In this approach, the related NLO EW photon bremsstrahlung
at $\ord(\alphaS^n\alpha^{m+1})$
involves processes with
$\ngamma+1$ photons and $\nqcd$ final-state partons, and since photons
count as jets, the requirement of $N$ hard jets does not guarantee that 
all $\nqcd$ partons are hard and well separated. In fact, the radiated photon
can play the role of the $N^{\mathrm{th}}$ jet, thereby allowing one of the QCD partons 
to become soft and/or collinear to a photon or to another parton.
Nevertheless, in this approach, all resulting QCD singularities are 
cancelled by the virtual QCD corrections to the production of $\ngamma+1$ photons 
plus $\nqcd-1$ QCD partons, 
which are automatically included in the democratic definition of $N$-jet final states.

Alternatively, one can adopt an approach aimed at 
preserving the distinction
between QCD jets and photons, in such a way that processes with different numbers of
QCD jets and photons do not mix.  In this case, in order to avoid the soft QCD
singularities that arise from jets with $\zgamma\to 1$, 
the notion of QCD jets needs to be restricted to clusters of partons and photons 
where the photon-energy fraction does not exceed a certain threshold 
$\zgammathr<1$,  while jets with
$\zgamma> \zgammathr$ have to be considered as photons.
As for IR singularities of QED type, a strict implementation of the condition $\zgamma<\zgammathr$ 
implies a fully exclusive description of collinear photon emissions off quarks,
which hampers the cancellation of the related collinear singularity.
A rigorous solution to this problem requires the 
factorisation of the collinear QED singularity
in a non-perturbative quark-fragmentation
function~\cite{Glover:1993xc,Buskulic:1995au,Bourhis:1997yu,Fontannaz:2001ek,Klasen:2002xb,GehrmannDeRidder:2006vn,Denner:2014ina}.  
However, as a pragmatic alternative to the fragmentation formalism,
the cancellation of the collinear singularity can be enforced 
by recombining (anti)quark--photon pairs
in a tiny cone around the singular region.  As discussed in the following,
this latter solution provides a quite reliable approximation to the rigorous fragmentation
approach. Its algorithmic formulation, at NLO parton level,
is as follows:
\begin{enumerate}

\item Collinear (anti)quark--photon pairs with rapidity--azimuthal separation
$\Drgq\le\Drrec\ll 1$ are recombined and are
treated as (anti)quarks, so that collinear photons remain unresolvable in
all subsequent steps of the algorithm.
\label{jetalgo:1st}
 
\item A jet-clustering algorithm is applied, where 
photons and QCD partons are treated on equal footing at 
each recombination step.
\label{jetalgo:2nd}

\item Jets that contain resolvable photons, i.e.~photons that have not been recombined
in step~\ref{jetalgo:1st}, are considered as QCD jets 
only if the photon-energy fraction $\zgamma=E_\gamma/E_{\mathrm{jet}}$ 
does not exceed a certain threshold $\zgammathr<1$.
\label{jetalgo:3rd}

\end{enumerate}

Is is clear that step~\ref{jetalgo:1st} ensures the cancellation of
collinear QED singularities.  At the same time, the fact that the condition
$\zgamma<\zgammathr$ is not applied to recombined photons represents an
approximate treatment of step~\ref{jetalgo:3rd}.  Since this approximation
is restricted to a tiny cone around the collinear region, its quality can be
easily assessed in a process independent way.  To this end, let us consider a
collinear $q\to q\gamma$ splitting, where a quark with transverse momentum
$p_\rT$ gives rise to a photon and a quark with momenta $\zgamma p_\rT$ and
$(1-\zgamma)\,p_\rT$, respectively.  Combining the perturbative contribution
associated with the splitting function $P_{q\gamma}(z)=[1+(1-z)^2]/z$ with
the non-perturbative fragmentation function extracted from ALEPH
data~\cite{Buskulic:1995au,GehrmannDeRidder:2006vn} leads to the following
expression for the probability to find a photon with energy fraction
$\zgamma>\zgammathr$ within a cone of radius $R$~\cite{Kuhn:2007cv}: 
\beqar
\epsilon_q(\zgammathr,R,p_\rT)&=&\int_{\zgammathr}^1 \rd z\, {\mathcal D}_{q
\gamma}(z,R,p_\rT), 
\eeqar 
with 
\beqar 
{\mathcal D}_{q \gamma}(z,R,p_\rT)
&=& \frac{\alpha Q_q^2}{2\pi}\left[ 2\,P_{q\gamma}(z) \ln\left(
\frac{z R p_{\rT}}{\mu_0}\right) +z-C\right], 
\label{eq:recombb}
\eeqar 
where $Q_q$ is the electromagnetic charge of the quark, 
while the scale $\mu_0=0.14$\,GeV and the 
parameter $C=13.26$ enter through the fit of the fragmentation component 
to ALEPH data.  This quantity corresponds to the probability that a
photon-like jet is misinterpreted as a QCD jet due to the 
photon--quark recombination prescription, i.e.~it represents the relative
uncertainty inherent in the first step of the
above jet definition~\ref{jetalgo:1st}--\ref{jetalgo:3rd}.
Its quantitative impact in the case 
of up-type quarks is illustrated in~\reffi{fig:recomb}
for a wide range of photon-energy thresholds and jet transverse momenta.
For realistic threshold values $\zgammathr\ge 0.5$,
it is clear that the error induced by the recombination prescription is 
at the permil level. Moreover, in realistic jet-production processes 
this error is further suppressed since the treatment of gluon--photon pairs is exact,
while for down-quark--photon pairs eq.~\refeq{eq:recombb} involves a smaller charge factor,
$Q^2_d=Q_u^2/4$.
We thus conclude that the error inherent in the above recombination prescription can hardly 
exceed the few permil level in a very broad kinematic range.

\begin{figure*}[t]
\centering
\includegraphics[width=0.5\textwidth]{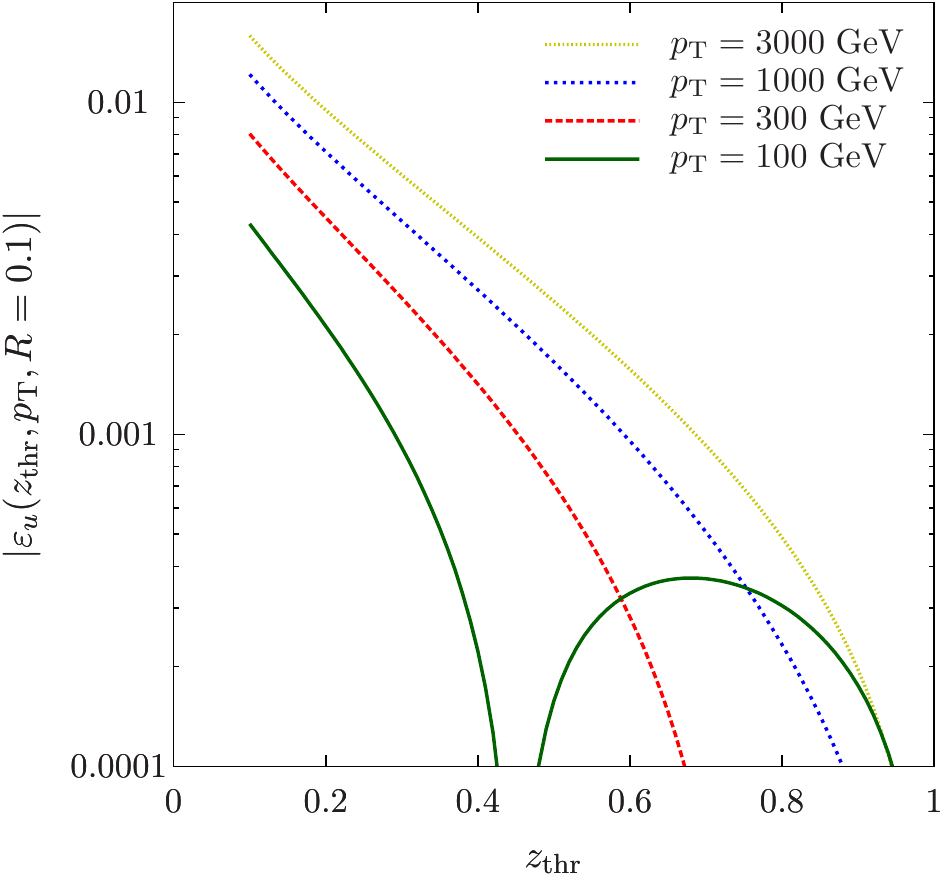}
\caption{Probability $\epsilon_u(\zgammathr,p_\rT,R)$ 
 of $u\to u\gamma$ fragmentation in a cone of radius $R=0.1$ 
as a function of the photon-energy threshold $\zgammathr$
for different values of the jet transverse momentum, 
$p_\rT=100, 300, 1000, 3000$\,GeV.}
\label{fig:recomb}
\end{figure*}

\section{Automation of electroweak corrections in \OpenLoops, \Munich and \Sherpa}
\label{se:automation}

In this section we discuss the fully automated implementation of
NLO QCD+EW corrections in 
\OpenLoops~\cite{Cascioli:2011va,hepforge}, 
\Munich~\cite{munich} and \Sherpa~\cite{sherpaqedbrems,Gleisberg:2007md,Gleisberg:2008ta}.
In this computing framework, the \OpenLoops program generates the relevant one-loop and, if needed, tree
matrix elements, while the \Munich and \Sherpa Monte Carlo programs take care of all complementary NLO tasks,
i.e.~the bookkeeping of partonic processes, the subtraction of IR singularities, 
and phase-space integration.
For what concerns Born and real-emission matrix elements,
in \Sherpa they are provided by the two internal tree-level 
generators \Amegic \cite{Krauss:2001iv} and \Comix \cite{Gleisberg:2008fv},
while \Munich takes them from \OpenLoops.
The present implementation supports parton-level NLO QCD+EW simulations in 
a fully automated way, and any hadron-collider observable can be generated 
in a few simple steps upon specification of the desired hadronic process 
and the relevant input parameters.
In the following we will focus our attention on nontrivial
aspects that had to be addressed in order to extend the functionality of the 
various tools from NLO QCD to NLO EW.
The automation of NLO EW calculations will be available
in future public releases of \OpenLoops, \Munich and \Sherpa.

\subsection{Tree and one-loop amplitudes with \OpenLoops}
\label{se:virtrad}

The \OpenLoops program is a fully automated generator of tree and one-loop
scattering amplitudes within the Standard Model.  Matrix elements are built
with a recursive numerical algorithm~\cite{Cascioli:2011va}, which is
flexibly applicable to any desired process and guarantees high CPU
performance up to high particle multiplicity.  The first public version of
\OpenLoops was released very recently~\cite{hepforge}.  It supports NLO QCD
calculations for a wide range of processes up to four
final-state particles.  The code is available as a set of
compact libraries that cover more than one hundred different processes at
hadron colliders, and the number of supported processes is continuously
growing.  The various process libraries contain all relevant ingredients for
NLO QCD calculations: tree amplitudes, renormalised one-loop amplitudes, and
colour- and helicity-correlated matrix elements for the subtraction of IR
singularities.
\OpenLoops provides easy to use {\sc Fortran} and {\sc C++} interfaces,
as well as a standard interface based on the Binoth Les Houches Accord~\cite{Alioli:2013nda},
and can therefore be easily integrated within any Monte Carlo framework.
Moreover, {\sc Sherpa}~\cite{Gleisberg:2008ta} 
and {\sc Herwig}'s {\sc MatchBox}~\cite{Bellm:2013lba}
as well as \Munich~\cite{munich} dispose of generic built-in interfaces to \OpenLoops.

In \OpenLoops tree and one-loop amplitudes are computed in terms of
individual colour-stripped Feynman diagrams.  While the reduction of colour
factors, colour interferences and colour sums are performed with
algebraic techniques, the construction of colour-stripped diagrams
is entirely numerical.  The tree algorithm is based on subtrees, which
correspond to pieces of individual colour stripped tree diagrams that
result from cutting an internal propagator.  Tree amplitudes are generated
via recursive merging of subtrees, and the systematic exploitation of
relations between diagrams that share common subtrees allows one to evaluate
multi-particle amplitudes with high CPU efficiency.

One-loop amplitudes in \OpenLoops are constructed by means of a hybrid
tree--loop recursion that generates cut-open loops as functions of the circulating
loop momentum~\cite{Cascioli:2011va}. The basic building blocks are 
individual colour-stripped one-loop diagrams of the form
\begin{eqnarray}
\label{eq:npointloop}
\vcenter{\hbox{\includegraphics[height=30mm]{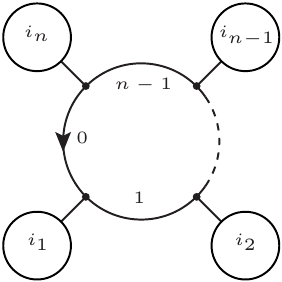}}} 
\hspace{3mm}
=
\hspace{3mm}
\int
\rd^Dq\;
\frac{ \calN(\calI_n;q)}{D_0D_1\dots D_{n-1}}\;,
\end{eqnarray}
where $D_i=(p_i-q)^2-m_i^2+\ri\epsilon$, the blobs $i_1,\dots, i_n$ represent external subtrees,
and the numerator $\calN(\calI_n;q)$ is a polynomial in the loop momentum $q$.
\begin{figure*}[t]
\centering
\includegraphics[width=0.55\textwidth]{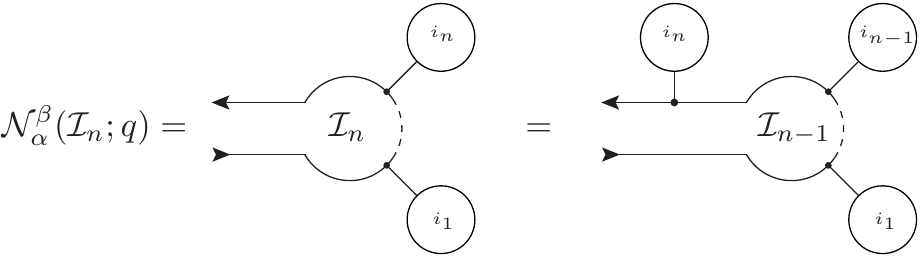}
\caption{Schematic representation of the open-loops 
recursion: $n$-point open loops are
constructed by merging $(n-1)$-point open loops
and external subtrees.
}
\label{fig:looprec}
\end{figure*}
Cut-opening the internal line associated with the $D_0$ propagator, 
converts the loop into a tree structure and promotes the numerator
to a tensor, $\calN^\beta_\alpha(\calI;q)$, whose two indices are associated 
with the spin or vector degrees of freedom of the cut propagator.
As sketched in~\reffi{fig:looprec}, these objects can be constructed in a similar way 
as tree amplitudes, by recursively merging the external subtrees that are attached to the loop.
Formally, this corresponds to the recurrence relation
\begin{eqnarray}
\label{eq:looprecb}
\mathcal{N}_\alpha^\beta(\calI_n;q)
=
\X_{\gamma\delta}^{\beta}(\calI_n,i_{n},\calI_{n-1})
\;\calN^\gamma_\alpha(\calI_{n-1};q)
\; w^\delta(i_n) \,,
\;
\end{eqnarray}
where $w^\delta(i_n)$ represents the $n$-th external subtree, while
the tensor $\X_{\gamma\delta}^{\beta}$, which describes the interaction of the
$i_n$-th subtree with the rest of the cut-open diagram, 
depends only on the flavour and the momenta of the involved particles 
in a way that is dictated by the Feynman rules
of the theory.
In contrast to conventional tree algorithms of type~\refeq{eq:looprecb},
in \OpenLoops all ingredients are handled as polynomials in the circulating
loop momentum. The numerator assumes the form
\begin{eqnarray}
\label{eq:openloop}
\calN^\beta_\alpha(\calI_n;q)= \sum_{r=0}^R \calN^\beta_{\mu_1\dots \mu_r;\alpha}(\calI_n)\; q^{\mu_1}\dots q^{\mu_r},
\end{eqnarray}
where $R\le n$ is the maximum rank of tensor integrals that contribute to the actual loop
diagram, while the interaction term is expressed as\footnote{Here we restrict ourselves to 
a linear $q$-dependence, assuming renormalisable interactions, but the generalisation to 
an arbitrary polynomial degree is straightforward.
Also the formulation of quartic and higher-point interactions is obvious.
}
\begin{eqnarray}
\label{eq:coeffvert}
&&\X_{\gamma\delta}^{\beta}
=
\Y_{\gamma\delta}^{\beta}
+
q^\nu\; \Z_{\nu;\gamma\delta}^{\beta}.
\end{eqnarray}
The one-loop algorithm is formulated as a recurrence relation 
for the direct construction of the $q$-polynomial coefficients:
\begin{eqnarray}
\label{eq:coefflooprec}
&&\calN_{\mu_1\dots\mu_r;\alpha}^\beta(\calI_n)=
\left[
\Y_{\gamma\delta}^{\beta}
\; 
\calN^\gamma_{\mu_1\dots\mu_r;\alpha}(\calI_{n-1})
+\Z_{\mu_1;\gamma\delta}^{\beta}
\; 
\calN^\gamma_{\mu_2\dots\mu_r;\alpha}(\calI_{n-1})
\right]\; w^\delta(i_n).
\end{eqnarray}
This type of algorithm was originally proposed in the framework 
of a Dyson--Schwinger recursion for colour-ordered gluon-scattering 
amplitudes~\cite{vanHameren:2009vq}.
The fact that loop numerators are
directly constructed as {\it functions} of the loop momentum
represents a great advantage for the speed of the algorithm. 
The actual implementation of~\refeq{eq:coefflooprec} in \OpenLoops 
employs fully symmetrised tensors. Its CPU efficiency is further augmented 
by means of parent--child relations 
and thanks to further tricks that exploit the systematic factorisation of colour-, helicity-, and
$q$-dependent objects~\cite{Cascioli:2011va}.

In order to extend \OpenLoops to EW one-loop corrections,
all EW Feynman rules for fermions, vector bosons, scalars and ghosts
have been implemented in the form of numerical routines 
corresponding to the generic recursion relation~\refeq{eq:coefflooprec}.
Each interaction term described by~\refeq{eq:coeffvert} 
is associated
with three lines\footnote{In the case of quartic vertices there is a fourth line that 
enters as additional external wave function.} 
that play different roles: external subtree,
inflowing and outflowing loop line. Thus, in general, each vertex in the Feynman rules 
requires three numerical routines of type~\refeq{eq:coefflooprec}.
Once implemented, these universal routines
are applicable to any one-loop amplitude within the QCD+EW Standard Model.
Moreover, they can be easily extended to BSM interactions.

The numerical polynomial representation~\refeq{eq:openloop} of loop
numerators provides full information on the functional $q$-dependence of the
integrand, thereby allowing for great flexibility in the reduction
of~\refeq{eq:npointloop} to scalar integrals.  On the one hand, the
reduction can be performed at the level of individual tensor integrals
associated with the monomials $q^{\mu_1}\dots{q^{\mu_r}}$ 
in~\refeq{eq:openloop}.  To this end, \OpenLoops is interfaced with
the~\Collier library~\cite{Denner:2014gla}, which implements the
Denner--Dittmaier reduction techniques~\cite{Denner:2002ii,Denner:2005nn}
and the scalar integrals of~\cite{Denner:2010tr}.  Sophisticated analytic
expansions~\cite{Denner:2002ii,Denner:2005nn} render this approach very
robust against numerical instabilities in exceptional phase-space regions. 
Alternatively, the reduction of~\refeq{eq:npointloop} to scalar integrals
can be performed at the integrand level using the OPP
method~\cite{Ossola:2006us} as implemented in~{\sc
CutTools}~\cite{Ossola:2007ax} or {\sc Samurai}~\cite{Mastrolia:2010nb},
which both rely on the {\sc OneLOop} library~\cite{vanHameren:2010cp}
for the evaluation of scalar integrals.

The evaluation of one-loop QCD amplitudes with \OpenLoops is very
fast~\cite{Cascioli:2011va}, both in combination with tensor integral reduction and
OPP reduction.  In this context it was observed that CPU timings grow only
linearly with the number of Feynman diagrams, which guarantees a fairly
favourable scaling with the external-particle multiplicity.  We find that
this property holds also for one-loop EW calculations.  More precisely, the
dependence of CPU timings on the number of Feynman diagrams per process
is roughly universal, i.e.~approximately the same for 
QCD and EW corrections.

Within \OpenLoops, ultraviolet (UV) and infrared (IR) divergences are
dimensionally regularised and take the form of 
poles in $(4-D)$. However, all ingredients of 
the numerical recursion~\refeq{eq:openloop}--\refeq{eq:coefflooprec}
are handled in four space-time dimensions.
The missing $(4-D)$-dimensional contributions---called $R_2$
rational terms---are universal and can be restored from process-independent 
effective counterterms~\cite{Ossola:2008xq,Binoth:2006hk,Bredenstein:2008zb}.
Corresponding Feynman rules have been derived for
QED in \cite{Ossola:2008xq}, for QCD in \cite{Draggiotis:2009yb}
and for the complete EW Standard Model in
\cite{Garzelli:2009is,Garzelli:2010qm,Garzelli:2010fq,Shao:2011tg}.
We implemented all QCD and EW $R_2$ counterterms in
\OpenLoops and validated them against 
independent algebraic results in 
$D=4-2\epsilon$ dimensions.

For the renormalisation of UV divergences
we adopted the on-shell scheme~\cite{Denner:1991kt}
and implemented all relevant $\ord(\alpha)$ counterterm Feynman rules
and related renormalisation constants for the full Standard Model,
including the option of the complex mass scheme~\cite{Denner:2005fg}
for unstable gauge bosons and top quarks.
In NLO QCD calculations the strong coupling constant is
renormalised in the $\overline{\text{MS}}$ scheme, and
heavy quark contributions can be decoupled in a flexible way, depending on the 
number of active flavours in the evolution of $\alphaS$.
For the renormalisation of the electroweak couplings we implemented the 
$\gmu$ scheme, where the fine-structure 
constant $\alpha=e^2/4\pi$ and the weak mixing angle 
$\theta_{\mathrm{w}}$ are given by
\beq\label{eq:GFscheme}
\alpha=\frac{\sqrt{2}}{\pi}\GF\/\MW^2\left(1-\frac{\MW^2}{\MZ^2}\right),\qquad
\cos\theta_{\mathrm{w}}=\frac{M_W}{M_Z}
\eeq
This requires a redefinition of the renormalisation constant associated with the
electromagnetic coupling, 
\beq
\delta Z_e \vert_{\gmu} = \delta Z_e \vert_{\alpha(0)} - \frac{1}{2}\Delta r \,,
\eeq
where $\Delta r$ is defined in~\cite{Denner:1991kt}, 
and $\alpha(0)$ denotes the standard on-shell renormalisation 
prescription in the Thompson limit.

For the cancellation of the remaining IR singularities in the virtual 
QCD and EW corrections, \OpenLoops provides dedicated routines
that implement the so-called $I$-operator in the dipole subtraction 
formalism~\cite{Catani:1996vz,Catani:2002hc} and its extension to 
QED corrections~\cite{Dittmaier:1999mb,Dittmaier:2008md,Gehrmann:2010ry}.
In this context also colour-correlated and charge-correlated Born matrix elements 
at any desired order in $\alpha$ and $\alphaS$ are supported. 
Their content can be schematically represented as 
\beqar
\label{eq:QCDcc}
g_\rS^2\,\langle M_0|\,T^a(i)\,T^a(j)\,|M_0\big\rangle\bigg|_{\alphaS^{n+1}\alpha^m}
&=&
g_\rS^2\sum_{p,p',q,q'}\, \big\langle M_0^{(p,q)}|\,T^a(i)\,T^a(j)\,|M_0^{(p',q')}\big\rangle\,
\delta_{2n,p+p'}\,\delta_{2m,q+q'},\nonumber\\\\
\label{eq:QEDcc}
e^2\,\langle M_0|\,Q(i)\,Q(j)\,|M_0\big\rangle\bigg|_{\alphaS^n\alpha^{m+1}}
&=&
e^2\sum_{p,p',q,q'}\, \big\langle M_0^{(p,q)}|\,Q(i)\,Q(j)\,|M_0^{(p',q')}\big\rangle\,
\delta_{2n,p+p'}\,\delta_{2m,q+q'},\nonumber\\
\eeqar
where $T^a(i)$ denotes the usual colour-insertion operator acting on the $i^{\mathrm{th}}$
external leg, and $Q(i)$ is the corresponding electromagnetic charge operator.
The usual bra--ket notation is used for Born matrix elements and their complex conjugates, 
and sums over external-leg colours are implicitly understood.
Born matrix elements of $\ord(g_\rS^pe_{\phantom{\rS}}^q)$ are denoted as $M_0^{(p,q)}$, and all relevant contributions
to a predefined overall order are included in a fully automated way.
Furthermore, \OpenLoops provides extra routines 
to calculate gluon- and photon-helicity correlated 
Born amplitudes, which are needed by Monte Carlo programs to construct 
IR subtraction terms for real-emission matrix elements.
 
As far as the bookkeeping of the perturbative orders in $\alphaS$ and
$\alpha$ is concerned, all relevant LO and NLO virtual contributions 
are generated and combined in a similar way as 
in~\refeq{eq:QCDcc}--\refeq{eq:QEDcc}, i.e.~the following
colour-summed Born--Born and 
Born--virtual interference terms that contribute to a given 
order are automatically combined,
\beqar
\label{eq:BBint}
\langle M_0|M_0\big\rangle\bigg|_{\alphaS^n\alpha^m}
&=&
\sum_{p,p',q,q'}\, \big\langle M_0^{(p,q)}|M_0^{(p',q')}\big\rangle\,
\delta_{2n,p+p'}\,\delta_{2m,q+q'},\\
\label{eq:BVint}
\langle M_0|M_1\big\rangle\bigg|_{\alphaS^n\alpha^m}
&=&
\sum_{p,p',q,q'}\,\big\langle M_0^{(p,q)}|M_1^{(p',q')}\big\rangle\,
\delta_{2n,p+p'}\,\delta_{2m,q+q'}.
\eeqar
Here, the inclusion of all counterterm contributions
of UV and $R_2$ kind is implicitly understood.
All nontrivial EW--QCD interference contributions
described in section~\ref{se:powercounting} are thus automatically taken into account. 
From the user viewpoint, specifying the desired order $\alphaS^n\alpha^m$ at LO and
the type of correction, NLO QCD or NLO EW, is sufficient in order
to obtain all relevant NLO terms of $\ord(\alphaS^{n+1}\alpha^{m})$ 
or $\ord(\alphaS^{n}\alpha^{m+1})$, respectively.
Also the calculation of the complete NLO Standard Model corrections, including
all relevant contributions of $\ord(\alphaS^{n-k+1}\alpha^{m+k})$ 
with $-m\le k\le n+1$ is possible.
This flexible power counting is fully supported by 
the available \OpenLoops interface~\cite{hepforge}. 

The entire implementation of NLO EW virtual contributions in \OpenLoops, including
the finite parts of the UV renormalisation, has been checked for several
processes.  To this end we implemented NLO EW corrections in a second and fully independent 
in-house generator, which was originally 
developed for NLO QCD calculations~\cite{Bredenstein:2010rs,Denner:2012yc}.
Detailed checks have been performed for all 
building blocks that enter the NLO QCD+EW corrections for
$W$-boson production in association with jets presented in this paper.

\subsection{Real radiation and QCD+QED subtraction with \Sherpa and \Munich}
\label{se:realrad}

This section deals with the automated calculation of real-emission contributions at
NLO QCD+EW level in \Munich and \Sherpa.  In this context, the first key task is the fully
automated bookkeeping of the real-emission channels that contribute 
to any user-defined process with a certain number of jets, photons, leptons and
additional heavy particles at Born level.
More precisely, the programs generate the full list of contributing
partonic processes organised according to their orders in $\alphaS$ and $\alpha$,
together with the ones that involve one extra massless object in the final state,
i.e.~an extra gluon, a quark pair instead of a gluon, an extra photon, or a fermion 
pair instead of a photon. 
As discussed in \refse{se:democlstering}, jets and photons can be handled on
the same footing or as separate physics objects, and the list of contributing 
subprocesses depends on the details of the photon/jet definition.\footnote{Note that \refse{se:democlstering}
deals only with the infrared-safe definition of jets in processes
with hard jets and no resolved photons, while the issue of IR safeness 
for processes with resolved photons at NLO QCD+EW 
is not addressed in this paper.}
However, the process bookkeeping can adapt to the
above two options in a fully flexible way.

In connection with the generation of the real radiation the  
main task of \Munich and \Sherpa is the consistent subtraction of IR singularities.
To this end, both programs implement the Catani--Seymour formalism~\cite{Catani:1996vz,Catani:2002hc}.
Light quarks and leptons are treated as massless particles, 
and the related singularities are regularised in $D$ dimensions.
All relevant subtraction terms in the real-emission phase
space are obtained from the convolution of QCD and QED
Catani--Seymour splitting kernels with reduced Born contributions.
Their integrated counterparts factorise into reduced Born matrix elements 
times the so-called $I$, $K$, and $P$ operators~\cite{Catani:1996vz,Catani:2002hc}.
In this context, starting from existing implementations of dipole subtraction 
at NLO QCD, all process-independent building blocks, i.e.~splitting kernels
and $I+K+P$ operators, have been extended to NLO QCD+QED.\footnote{The construction 
of QED dipole-subtraction  terms has been discussed in \citeres{Dittmaier:1999mb,Dittmaier:2008md,Gehrmann:2010ry}.}
In particular, all contributions associated with 
$f \to f\gamma$, $\bar f \to \bar f\gamma$, and
$\gamma\to f\bar f$ QED splittings
can be obtained 
from the related QCD contributions by applying the substitutions
\beq\label{eq:subtraction_replacements}
  \alpha_s\longrightarrow\alpha,\qquad
  C_F \longrightarrow Q_f^2,\qquad
  T_R \longrightarrow N_{c,f} Q_f^2,\qquad
  T_R N_f \longrightarrow \sum_f N_{c,f} Q_f^2,\qquad
  C_A\longrightarrow 0\,,
\eeq
and the following additional replacements for the colour-correlation operators associated with an emitter $ij$ and a spectator $k$,
\beq\label{eq:QEDcorrel}
  \frac{\mathbf{T}_{ij}\cdot\mathbf{T}_k}{\mathbf{T}_{ij}^2}
  \longrightarrow 
\left\{\begin{array}{l @{\quad}l}
\frac{Q_{ij}Q_k}{Q^2_{ij}} & \mbox{if the emitter $ij$ is a (anti)fermion}\\
\kappa_{ij,k} &  \mbox{if the emitter $ij$ is a photon\,,}\\
\end{array}\right.\qquad
\mbox{with}\qquad
\sum_{k\neq ij} \kappa_{ij,k} = -1\,.
\eeq
In practice, for the case of a photon emitter, one can restrict oneself
to a single spectator particle $e_{ij}$ different from 
the fermion--antifermion emitter $ij$, i.e. $\kappa_{ij,k}=-\delta_{e_{ij},k}$.
Alternatively any sum over spectators different from $ij$ can be chosen 
as long as the last constraint in \refeq{eq:QEDcorrel} is fullfilled.
While the colour-insertion operators are reduced to multiplicative 
scalars in \refeq{eq:QEDcorrel}, the spin correlators of the real-subtraction terms associated with 
$\gamma\to f\bar f$ splittings preserve the same form as 
for $g\to q\bar q$ splittings in QCD.

Besides singularities of pure QED type,
processes with external on-shell $W$ bosons
involve additional singularities associated with 
$W\to W\gamma$ splittings.
In this case, due to the large $W$-boson mass, 
no collinear singularity or 
logarithmic enhancement is present, and only the soft-photon singularity 
has to be subtracted. Exploiting the universal nature of 
soft singularities, in this publication this is achieved by using the 
heavy-fermion or 
heavy-scalar splitting function of~\cite{Catani:2002hc}, and, after the 
replacements of \refeq{eq:subtraction_replacements}, identifying the 
heavy particle with the external $W$ boson.

As discussed in \refse{se:powercounting}, NLO QCD
and EW corrections have to be understood, respectively, as the full set of
$\ord(\alphaS)$ and $\ord(\alpha)$ corrections relative to a certain
tree-level order $\alpha_S^n\alpha^m$. Moreover, in general, NLO QCD and EW corrections 
are not uniquely associated with the emission of corresponding
(strongly or electroweakly interacting) particles.
Actually, given a certain correction order, $\alphaS^{n+1}\alpha^m$ or $\alphaS^{n}\alpha^{m+1}$,
each of the contributing real-emission processes can comprise various 
types of unresolved massless particles (gluons, photons, quark or lepton 
pairs) and IR singularities. In particular,
NLO QCD (EW) corrections can involve singularities associated with both
order $\alphaS$ ($\alpha$) splittings times order
$\alphaS^{n}\alpha^m$ Born terms, and with 
order $\alpha$ ($\alphaS$) splittings times order
$\alphaS^{n+1}\alpha^{m-1}$ ($\alphaS^{n-1}\alpha^{m+1}$) Born terms.
Therefore, \Munich and \Sherpa implement a
fully general bookkeeping of perturbative orders and singularities.
The relevant dipole terms, to account for all possible 
QCD and QED splittings in a generic real-correction process, are selected in 
a fully automated way.  
Inevitably, the associated reduced Born matrix elements
are allowed to be at a different order than the original
Born configuration. 
For the integrated subtraction terms, a similarly general bookkeeping is
applied, where all relevant QED and QCD contributions to the $I+K+P$
operators are combined with factorised Born matrix elements at the
appropriate orders in $\alpha$ and $\alphaS$.  
This requires nontrivial
combinations of charge/colour insertion operators and interferences of Born
amplitudes at different orders, similarly as in~\refeq{eq:QCDcc}--\refeq{eq:QEDcc}.

For phase-space integration, both \Munich and \Sherpa employ adaptive
multi-channel techniques.  In \Sherpa, dipole subtraction terms can be
restricted by means of the so-called $\alpha$-dipole 
parameter~\cite{Nagy:1998bb,Nagy:2003tz,Campbell:2004ch,Bevilacqua:2009zn,
Campbell:2005bb,Frederix:2010cj}, while
\Munich constructs extra phase-space mappings based on the dipole
kinematics, and automatically adds them to the generic set of the real-emission based
phase-space parametrisations used in the multi-channel approach.

The \Sherpa and \Munich  implementations have been validated with standard
self-consistency checks, such as the local cancellation of singularities in
the real-emission phase space, the cancellation of the
$\alpha$-dipole  dependence in \Sherpa and the equivalence of
fermion and scalar splitting kernels for the subtraction of $W\to W\gamma$
soft singularities.

All involved colour-, charge- and spin-correlated matrix elements are
provided by the \OpenLoops generator in case of \Munich, whereas they are
supplied by \Amegic and \Comix within the \Sherpa implementation.  Apart
from the contributions that involve charge/colour insertions of
type~\refeq{eq:QCDcc}, which are still under construction within \Sherpa, for
all other building blocks the two programs have been validated against each
other on a point-wise basis as well as for integrated cross sections for a
wide range of processes, giving rise to full agreement on the level of
machine precision and statistical precision, respectively.  
The point-wise agreement for the $I$-operator provided
by \OpenLoops, \Munich and \Sherpa was also checked.
The results presented in~\refse{se:results} have been obtained with \MunichOpenLoops.

\section{Electroweak and QCD corrections to $\boldsymbol{pp\to W^++1,2,3}$\,jets}
\label{se:Wjets}

To demonstrate the flexibility and the performance of NLO automation in
\OpenLoops together with \Sherpa and \Munich, as a first application we
consider the NLO QCD+EW corrections to $W$-boson production in association
with up to three jets at the LHC.  In this paper we focus on the production of
stable $W^+$ bosons, while the case of $W^-$ production as well as $W$-boson
decays will be addressed in a subsequent publication.
In the following we discuss the building blocks of our calculation and technical 
subtleties related to the on-shell treatment of final-state $W$ bosons at NLO EW.

\subsection{Partonic channels}

The level of automation of the employed tools is such that, to generate and
evaluate all relevant contributions to a desired hadronic cross section, it
is sufficient to specify the desired final state and the perturbative order
in $\alphaS$ and $\alpha$.  Thus, from the user viewpoint, there is no need
to worry about the detailed content of the simulation in terms of partonic channels,
scattering amplitudes and Feynman diagrams.  Nevertheless, a basic knowledge 
of these ingredients plays an important role for the
understanding of the physics content of the simulation and for the 
interpretation of the phenomenological results.

At tree level, the only crossing-independent partonic process
that contributes to $pp\to W^+\jet$ is
\beq
u_i\bar d_i  \to W^+ g, \label{eq:processes_w1j}
\eeq
where $u_i=(u,c)$ and $d_i=(d,s)$. 
All other relevant channels can be obtained from~\refeq{eq:processes_w1j} 
through permutations of initial- and final-state partons.
For $pp\to W^+2\jet$ there are two crossing-independent 
subprocesses:
\beqar
u_i \bar d_i&\to& W^+ q\bar q, \label{eq:processes_w2j_1}\\
u_i \bar d_i&\to& W^+ gg, \label{eq:processes_w2j_2}
\eeqar
and the relevant crossing-independent
subprocesses for $pp\to W^+3\jet$ are obtained form~\refeq{eq:processes_w2j_1} 
and~\refeq{eq:processes_w2j_2} by adding an extra gluon:
\beqar
u_i \bar d_i&\to& W^+ q\bar qg, \label{eq:processes_w3j_1}\\
u_i \bar d_i&\to& W^+ ggg. \label{eq:processes_w3j_2}
\eeqar
The above processes can 
be categorised into two-quark and four-quark channels, 
according to the total number of external (anti)quarks.
In the case of the four-quark channels,~\refeq{eq:processes_w2j_1}
and~\refeq{eq:processes_w3j_1},
the additional $q\bar q$ system can consist of any light-quark pair
with $q\in\{u,d,s,c,b\}$.
All light quarks are treated as massless particles in our calculation.

The main focus of this paper is on the NLO QCD
and NLO EW corrections with respect to the dominant 
$\ord(\alphaS^n\alpha)$ tree-level contributions to $pp\to W^++n$\,jets. 
With other words we will consider NLO contributions of $\ord(\alphaS^{n+1}\alpha)$ 
and $\ord(\alphaS^n\alpha^2)$, respectively.
In both cases, $W^++n$-jet production
receives NLO bremsstrahlung contributions from 
tree-level amplitudes involving an extra parton. The relevant partonic channels
are obtained from \refeq{eq:processes_w1j}--\refeq{eq:processes_w3j_2} either by
replacing an external gluon by a $q\bar q$ pair, or by adding and external gluon 
or an external photon.
At Born level, in the following
we will discuss also mixed EW--QCD contributions of 
$\ord(\alphaS^{n-1}\alpha^2)$, pure EW contributions of 
$\ord(\alphaS^{n-2}\alpha^3)$, the tower of photon--proton induced contributions of $\ord(\alphaS^{n-1}\alpha^2)$, $\ord(\alphaS^{n-2}\alpha^3)$ and $\ord(\alphaS^{n-3}\alpha^4)$, 
and photon--photon induced contributions of $\ord(\alphaS^{n-2}\alpha^3)$.

Table~\ref{tab:diagnumber} summarises the number of $\ord(\alphaS^n\alpha)$
tree and corresponding QCD and EW one-loop Feynman diagrams that contribute to the various 
parton-level processes in $pp\to W+1,2,3$\,jets. This gives an impression of 
the complexity of the calculation and its dependence on the
jet multiplicity. We observe that the number of one-loop EW diagrams 
is from 30\% to 3 times higher as compared to the case of one-loop QCD.
Moreover, as discussed below, the NLO EW corrections to 
four-quark processes require both one-loop EW and one-loop QCD diagrams.
The number of one-loop EW diagrams increases by 
about one order of magnitude for each extra jet, similarly as in the one-loop QCD case,
and for $W+3\jet$ production it ranges from about 1000 to 2600 per partonic 
subprocess.

\begin{table}[t]
\setlength{\tabcolsep}{1.2ex}%
\renewcommand{\arraystretch}{1.2}
\begin{center}
\begin{tabular}{lllll}\hline
Channel & QCD trees & EW trees & QCD 1-loop & EW 1-loop\\
\hline
$u_i\bar d_i\to W^+ g$                   & 2 & - & 11 & 32\\
$u_i \bar d_i\to W^+ q\bar q$            & 2 (4) & 7 (14) & 33 (66) & 105 (210) \\
$u_i \bar d_i\to W^+ gg$                 & 8 & - & 150 & 266 \\
$u_i \bar d_i\to W^+ q\bar qg$           & 12 (24) & 33 (66) & 352 (704)& 1042 (2084) \\
$u_i \bar d_i\to W^+ ggg$                & 54 & - & 2043 & 2616\\
\hline
\end{tabular}
\caption{Number of tree and one-loop Feynman diagrams in the various
$pp\to W^++n$-jet partonic subprocesses: QCD trees of $\ord(g_\rS^n e)$,
EW trees of $\ord(g_\rS^{n-2} e^3)$, 1-loop QCD diagrams  
 of $\ord(g_\rS^{n+2} e)$, 
and 1-loop EW diagrams of $\ord(g_\rS^{n} e^3)$.
Numbers in parenthesis refer to the case of 
four-quark processes with same flavour, $q=u_i$ or $q=d_i$.
In the \OpenLoops framework individual contributions associated with the three 
independent colour structures of four-gluon vertices
count as separate diagrams.
}
\label{tab:diagnumber}
\end{center}
\end{table}


\subsection[Two-quark contributions to $pp\to W^++n$\,jets]{Two-quark contributions to $\boldsymbol{pp\to W^++n}$\,jets}
\label{se:twoquarks}
Due to the presence of a single quark pair,
the $u_i \bar d_i\to W^++n$-gluon channels
feature a rather simple structure
from the viewpoint of EW interactions:
the $W$ boson is necessarily coupled to the $u_i \bar
d_i$ quark line, while gluons can be produced
only through strong
interactions. 
Representative tree diagrams for processes with
$n=1,2,3$ gluons are depicted in~\reffi{fig:udwtrees}.
For each of these two-quark channels, tree-level amplitudes 
are characterised by a unique order, $g_\rS^n e$.
Thus at NLO each one-loop Feynman diagram can be uniquely 
assigned either to the QCD or to the EW corrections, depending on its order in 
$g_\rS$ and $e$. 
\begin{figure*}[t]
\centering
\subfloat[][]{   
   \includegraphics[width=0.21\textwidth]{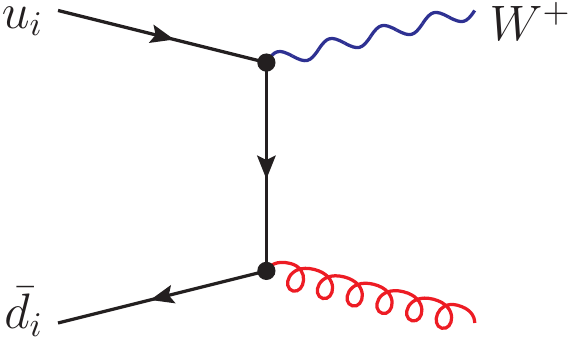}
   \label{fig:wgtree}
}
\quad
\subfloat[][]{   
   \includegraphics[width=0.21\textwidth]{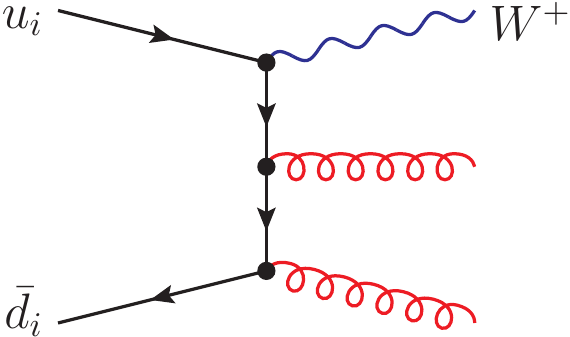}
   \label{fig:wggtree}
}
\quad
\subfloat[][]{   
   \includegraphics[width=0.21\textwidth]{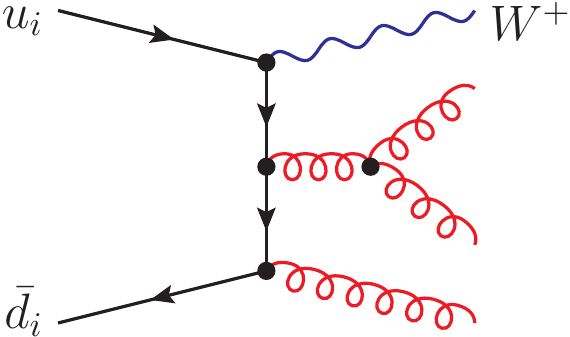}
   \label{fig:wgggtree}
}
\quad
\caption{Representative tree diagrams for 
$u_i \bar d_i\to W^++n$-gluon matrix elements 
at $\ord(g_\rS^n e)$. 
}
\label{fig:udwtrees}
\end{figure*}
\begin{figure*}[t]
\centering
\subfloat[][]{   
   \includegraphics[width=0.21\textwidth]{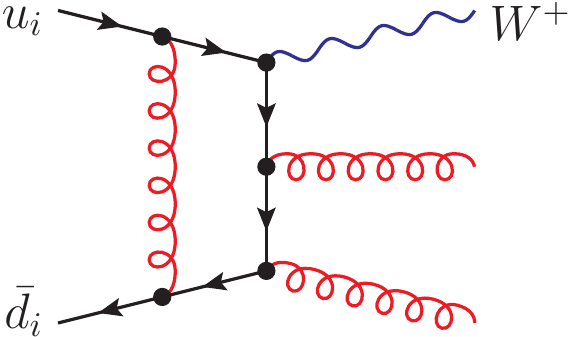}
   \label{fig:wggQCDvirtI}
}
\quad
\subfloat[][]{   
   \includegraphics[width=0.21\textwidth]{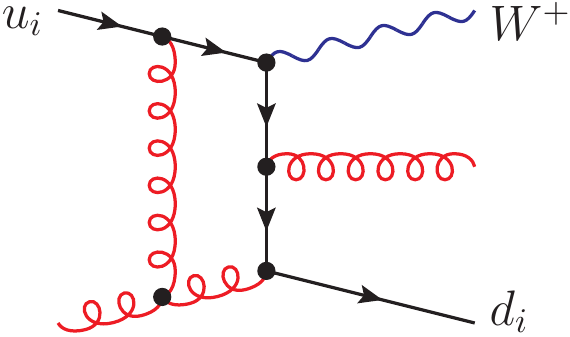}
   \label{fig:wggQCDvirtII}
}
\quad
\subfloat[][]{   
   \includegraphics[width=0.21\textwidth]{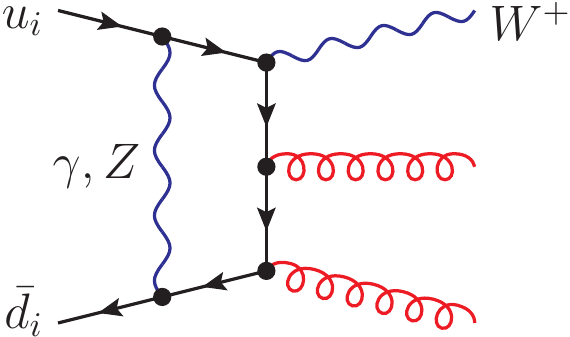}
   \label{fig:wggEWvirtI}
}
\quad
\subfloat[][]{   
   \includegraphics[width=0.21\textwidth]{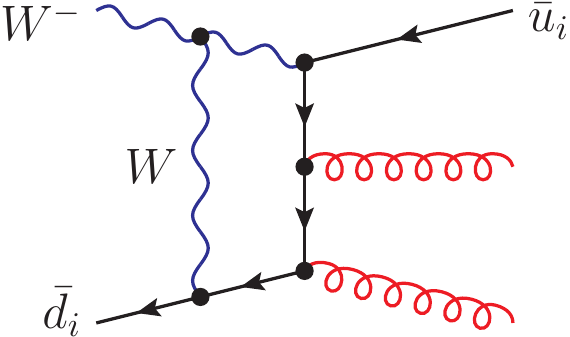}
   \label{fig:wggEWvirtII}
}
\caption{Representative one-loop diagrams for 
$u_i \bar d_i\to W^+gg$ matrix elements 
at $\ord(g_\rS^4 e)$ (\ref{fig:wggQCDvirtI}--\ref{fig:wggQCDvirtII})
and $\ord(g_\rS^2 e^3)$ (\ref{fig:wggEWvirtI}--\ref{fig:wggEWvirtII}).
}
\label{fig:udwggvirt}
\end{figure*}
\begin{figure*}[t]
\centering
\subfloat[][]{   
   \includegraphics[width=0.21\textwidth]{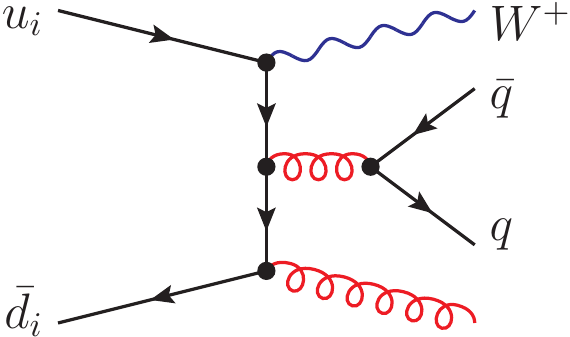}
   \label{fig:wggQCDrealI}
}
\quad
\subfloat[][]{   
   \includegraphics[width=0.21\textwidth]{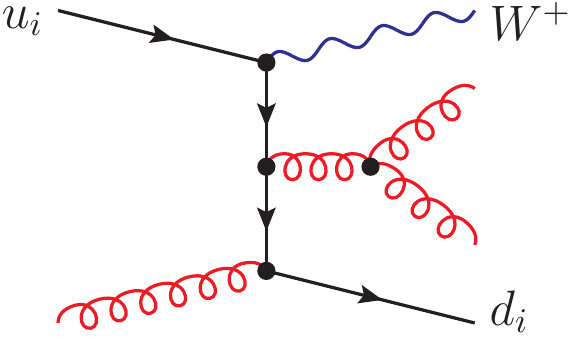}
   \label{fig:wggQCDrealII}
}
\quad
\subfloat[][]{   
   \includegraphics[width=0.21\textwidth]{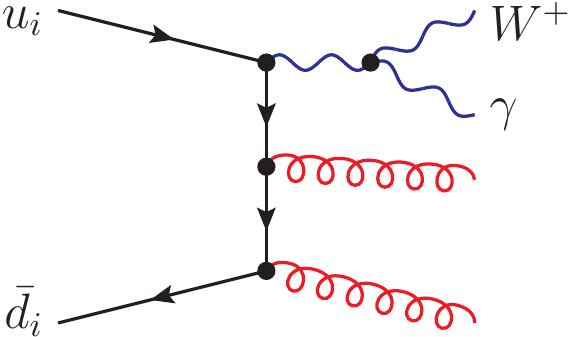}
   \label{fig:wggEWrealI}
}
\quad
\subfloat[][]{   
   \includegraphics[width=0.21\textwidth]{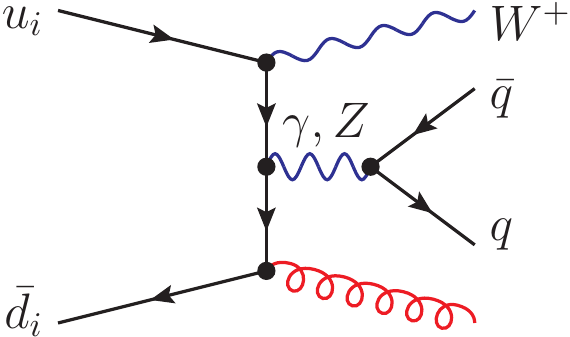}
   \label{fig:wggEWrealII}
}
\caption{
Representative diagrams for the real corrections to
$u_i \bar d_i\to W^+gg$:
contributions to the $\ord(g_\rS^3 e)$ QCD emission amplitudes 
(\ref{fig:wggQCDrealI}--\ref{fig:wggQCDrealII}), the 
$\ord(g_\rS^2 e^2)$ QED emission amplitudes 
(\ref{fig:wggEWrealI}) and the $\ord(g_\rS e^3)$
$q\bar q$ emission amplitudes  (\ref{fig:wggEWrealII}). 
}
\label{fig:udwggreal}
\end{figure*}

Examples of one-loop and real-emission diagrams
that contribute to the NLO QCD+EW corrections to $u_i \bar d_i\to W^+gg$
are displayed in~\reffis{fig:udwggvirt} and~\ref{fig:udwggreal}, respectively.
Corresponding diagrams for 
$pp\to W^++1\jet$ and $pp\to W^++3\jet$
are obtained by removing or adding an external gluon, or for the case of $W^++3\jet$
by replacing an external gluon with a $q\bar q$ pair. 
The $\ord(\alphaS^3\alpha)$ NLO QCD corrections to $u_i \bar d_i\to W^+gg$
receive contributions from  the interference of $\ord(g_\rS^2 e)$ trees (\ref{fig:wggtree}) 
with $\ord(g_\rS^4 e)$ loop diagrams (\ref{fig:wggQCDvirtI}--\ref{fig:wggQCDvirtII}), and 
from squared $\ord(g_\rS^3 e)$ QCD emission amplitudes
(\ref{fig:wggQCDrealI}--\ref{fig:wggQCDrealII}), while
the $\ord(\alphaS^2\alpha^2)$ NLO EW corrections to the same process receive contributions from  
the interference of $\ord(g_\rS^2 e)$ trees (\ref{fig:wggtree}) with 
$\ord(g_\rS^2 e^3)$ loop diagrams (\ref{fig:wggEWvirtI}--\ref{fig:wggEWvirtII}),  
from squared $\ord(g_\rS^2 e^2)$ QED emission amplitudes (\ref{fig:wggEWrealI}) and 
from the interference of $\ord(g_\rS^3 e)$ (\reffi{fig:wggQCDrealI}) and 
$\ord(g_\rS e^3)$ (\reffi{fig:wggEWrealII}) $q\bar q$ emission diagrams.

\subsection[Four-quark contributions to $pp\to W^++n$\,jets]{Four-quark contributions to $\boldsymbol{pp\to W^++n}$\,jets}
\label{se:fourquarks}

\begin{figure*}[t]
\centering
\subfloat[][]{   
   \includegraphics[width=0.21\textwidth]{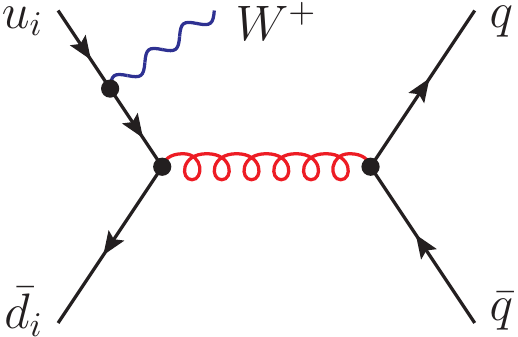}
   \label{fig:wqqQCDtreeI}
}
\quad
\subfloat[][]{   
   \includegraphics[width=0.21\textwidth]{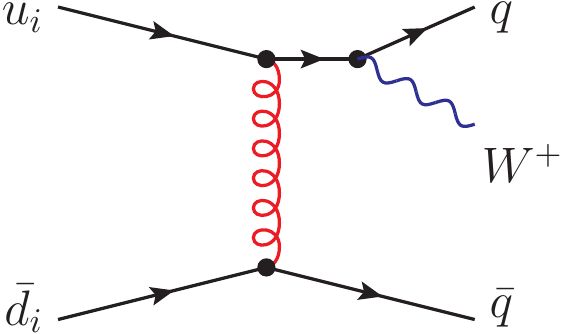}
   \label{fig:wqqQCDtreeII}
}
\quad
\subfloat[][]{   
   \includegraphics[width=0.21\textwidth]{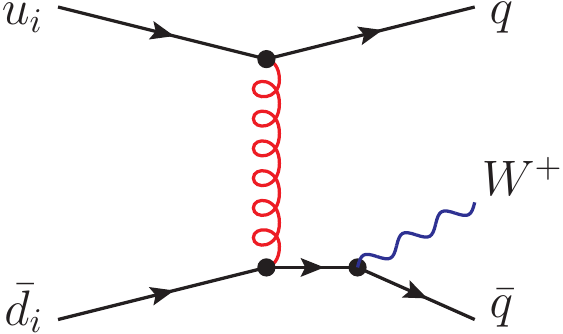}
   \label{fig:wqqQCDtreeIII}
}
\caption{
Representative tree diagrams for 
$u_i \bar d_i\to W^+q\bar q$ matrix elements 
at $\ord(g_\rS^2 e)$. While $s$-channel gluon exchange (\ref{fig:wqqQCDtreeI}) contributes 
to any flavour configuration with $q\in\{u,d,s,c,b\}$, $t$-channel topologies 
of type~\ref{fig:wqqQCDtreeII}
and~\ref{fig:wqqQCDtreeIII} contribute only when
$q=d_i$ and $q=u_i$, respectively.
}
\label{fig:wqqQCDtrees}
\end{figure*}
\begin{figure*}[t]
\centering
\subfloat[][]{   
   \includegraphics[width=0.21\textwidth]{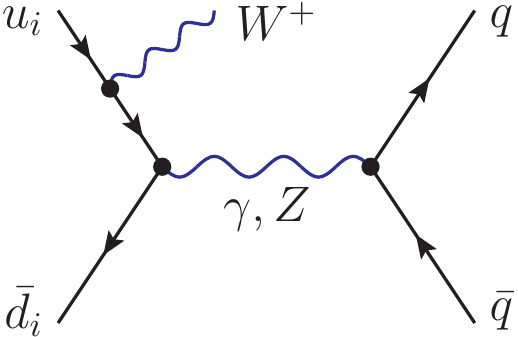}
   \label{fig:wqqEWtreeI}
}
\quad
\subfloat[][]{   
   \includegraphics[width=0.21\textwidth]{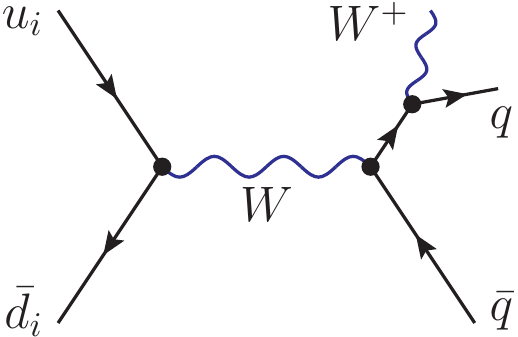}
   \label{fig:wqqEWtreeII}
}
\quad
\subfloat[][]{   
   \includegraphics[width=0.21\textwidth]{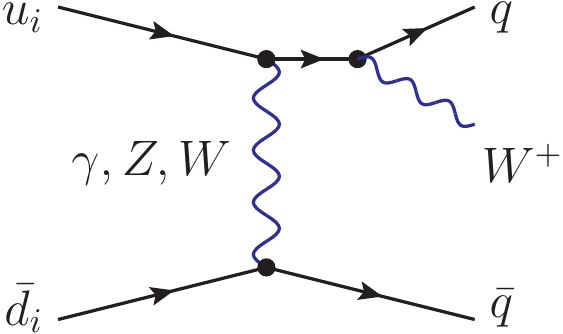}
   \label{fig:wqqEWtreeIII}
}
\quad
\subfloat[][]{   
   \includegraphics[width=0.21\textwidth]{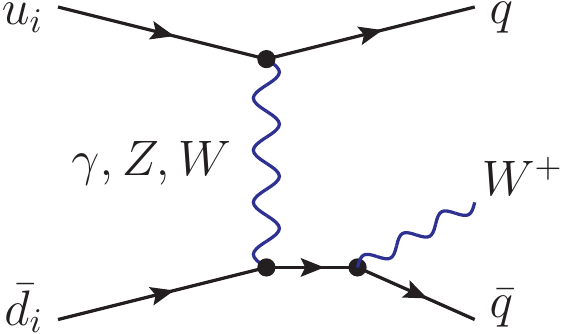}
   \label{fig:wqqEWtreeIV}
}
\caption{
Representative tree diagrams for 
$u_i \bar d_i\to W^+q\bar q$ matrix elements 
at $\ord(e^3)$. While $s$-channel exchange of EW bosons 
(\ref{fig:wqqEWtreeI}--\ref{fig:wqqEWtreeII}) 
contributes to any flavour configuration with $q\in\{u,d,s,c,b\}$, 
processes with $q=d_i$ ($q=u_i$) receive also contributions
from topologies of type~\ref{fig:wqqEWtreeIII} with $t$-channel exchange of 
neutral (charged) EW bosons
and topologies of type~\ref{fig:wqqEWtreeIV} with $t$-channel 
exchange of charged (neutral) EW bosons.
}
\label{fig:wqqEWtrees}
\end{figure*}
\begin{figure*}[t]
\centering
\subfloat[][]{   
   \includegraphics[width=0.21\textwidth]{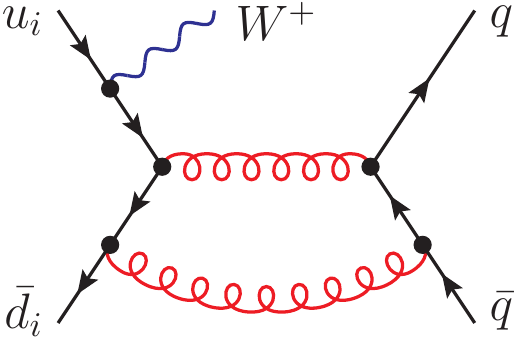}
   \label{fig:wqqQCDvirtI}
}
\quad
\subfloat[][]{   
   \includegraphics[width=0.21\textwidth]{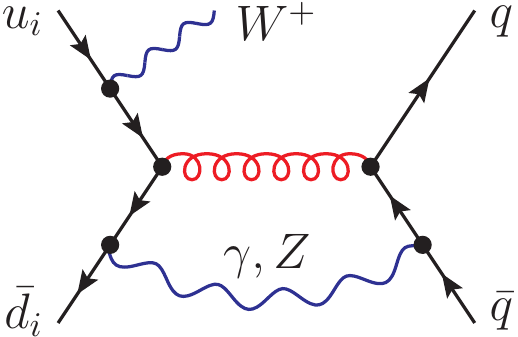}
   \label{fig:wqqEWvirtI}
}
\quad
\subfloat[][]{   
   \includegraphics[width=0.21\textwidth]{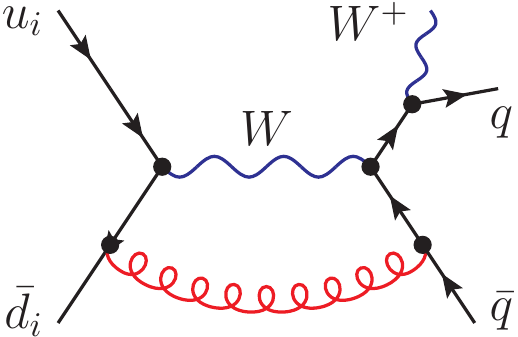}
   \label{fig:wqqEWvirtII}
}
\quad
\subfloat[][]{   
   \includegraphics[width=0.21\textwidth]{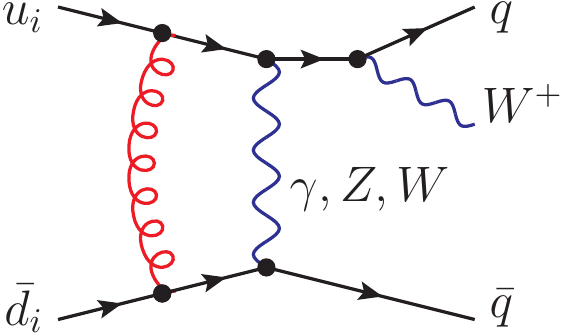}
   \label{fig:wqqEWvirtIII}
}
\caption{
Representative one-loop diagrams for 
$u_i \bar d_i\to W^+q\bar q$ matrix elements 
at $\ord(g_\rS^4 e)$ (\ref{fig:wqqQCDvirtI})
and $\ord(g_\rS^2 e^3)$ (\ref{fig:wqqEWvirtI}--\ref{fig:wqqEWvirtIII}). 
The $s$-channel topologies (\ref{fig:wqqQCDvirtI}--\ref{fig:wqqEWvirtII}) 
contribute to any process with $q\in\{u,d,s,c,b\}$. For $q=b$, diagrams of 
type~\ref{fig:wqqEWvirtII} involve resonant top-quark propagators.
Diagrams of type~\ref{fig:wqqEWvirtIII}, with $t$-channel exchange of neutral 
(charged) EW bosons contribute only to 
processes with $q=d_i$ ($q=u_i$).
}
\label{fig:wqqvirt}
\end{figure*}
\begin{figure*}[t]
\centering
\subfloat[][]{   
   \includegraphics[width=0.21\textwidth]{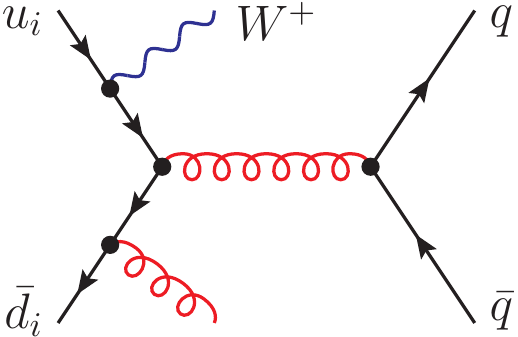}
   \label{fig:wqqQCDrealI}
}
\qquad
\subfloat[][]{   
   \includegraphics[width=0.21\textwidth]{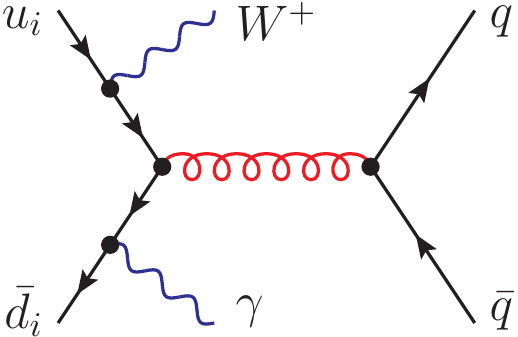}
   \label{fig:wqqEWrealI}
}
\qquad
\subfloat[][]{   
   \includegraphics[width=0.21\textwidth]{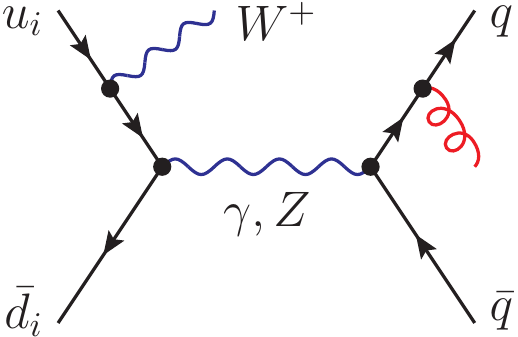}
   \label{fig:wqqMIXrealI}
}
\caption{
Representative diagrams for the real corrections to
$u_i \bar d_i\to W^+q\bar q$:
contributions to the $\ord(g_\rS^3 e)$ QCD emission amplitudes 
(\ref{fig:wqqQCDrealI}), 
$\ord(g_\rS^2 e^2)$ QED emission amplitudes (\ref{fig:wqqEWrealI}),
and $\ord(g_\rS e^3)$ QCD emission amplitudes (\ref{fig:wqqMIXrealI}).
}
\label{fig:wqqreal}
\end{figure*}

The production of $W$ bosons in association with two and three
jets involves
also the four-quark processes~\refeq{eq:processes_w2j_1} and
\refeq{eq:processes_w3j_1}, respectively.  In this case, the possibility to couple the two
quark lines either through gluons or EW bosons gives rise
to a nontrivial interplay between QCD and EW interactions already at tree-level.
In the following we will discuss such effects 
in the context of the NLO QCD+EW corrections to $u_i \bar d_i\to W^+q\bar q$.
Representative
LO and NLO Feynman diagrams for this process are displayed in~\reffis{fig:wqqQCDtrees}--\ref{fig:wqqreal}, while
corresponding diagrams for $u_i \bar d_i\to W^+q\bar qg$ 
are easily obtained by adding an external gluon or, in the case of 
NLO emissions, by converting a gluon into an additional $q\bar q$ pair.

At tree level, $u_i \bar d_i\to W^+q\bar q$ scattering amplitudes receive
QCD contributions of $\ord(g_\rS^2 e)$ (\reffi{fig:wqqQCDtrees}) as well as EW
contributions of $\ord(e^3)$ (\reffi{fig:wqqEWtrees}). 
Squared QCD amplitudes, mixed EW--QCD amplitudes, and
squared EW amplitudes, result in cross section 
contributions of $\ord(\alphaS^2\alpha)$,
$\ord(\alphaS\alpha^2)$, and $\ord(\alpha^3)$, respectively.
In this paper we mainly focus on the leading QCD contributions of
$\ord(\alphaS^2\alpha)$ and related NLO QCD+EW corrections.  Nevertheless,
in~\refse{se:results} we will discuss also the impact of mixed Born
contributions of $\ord(\alphaS\alpha^2)$ arising in the four-
quark channel. In general, all Born contributions are relevant, and their 
simulation is in principle straightforward. 
However, the EW contributions of type~\ref{fig:wqqEWtreeI}--\ref{fig:wqqEWtreeIV} 
involve various unstable particles that can 
give rise to resonances:
besides topologies where
an external quark--antiquark pair is coupled to a $Z$ or $W$ boson propagator
(\reffis{fig:wqqEWtreeI}--\ref{fig:wqqEWtreeIV}), 
in the case of $u_i \bar d_i\to W^+b\bar b$ and crossing related channels
also top-quark propagators coupled to external $Wb$ pairs can occur (\reffi{fig:wqqEWtreeII}).
As a consequence, pure EW $\ord(\alpha^3)$ contributions to $W+2\jet$ production 
involve $Z$, $W$, and top resonances that need to be
regularised in a consistent way by means of the relevant widths, $\Gamma_{Z,W,t}$.
These resonant contributions correspond to $WZ$, $WW$, and $t\jet$ production with
$Z\to \jet\jet$, $W\to \jet\jet$ and $t\to Wb$ decays, respectively. 
However, $W+2\jet$ production at $\ord(\alpha^3)$
contains also non-resonant contributions to the same final states, and
interferences between resonant and non-resonant amplitudes.  Therefore,
contributions of $\ord(\alpha^3)$ can not unambiguously be assigned to
either $WW,WZ, t\jet$ or to $W+2\jet$ production.  
As far as the EW--QCD interference contributions of $\ord(\alphaS\alpha^2)$ are concerned, 
due to the interference with QCD diagrams,\footnote{For $pp\to W+2\jet$, such Born
interferences are possible only in presence of the colour flow associated
with $t$-channel contributions of
type~\ref{fig:wqqQCDtreeII}--\ref{fig:wqqQCDtreeIII}
and~\ref{fig:wqqEWtreeIII}--\ref{fig:wqqEWtreeIV}, i.e.~only for
same-flavour quark combinations with $q=u_i$ or $q=d_i$.  If $u_i\bar d_i\to
W^+q\bar q$ amplitudes are dressed with an extra (virtual or real) gluon,
then EW--QCD interferences contribute to all flavour combinations
$q=u,d,s,c,b$.} the $Z,W,$ and top propagators in the EW
amplitudes do not lead to any Breit--Wigner peak in  $pp\to W+\jet\jet$ 
distributions.

Examples of one-loop and real emission diagrams that contribute to 
$u_i \bar d_i\to W^+ q\bar q$ at NLO QCD+EW are presented 
in~\reffis{fig:wqqvirt} and~\ref{fig:wqqreal}. 
The $\ord(\alphaS^3\alpha)$ NLO QCD corrections 
receive contributions from  the interference of $\ord(g_\rS^2 e)$ 
trees~(\ref{fig:wqqQCDtreeI}-\ref{fig:wqqQCDtreeIII})
with $\ord(g_\rS^4 e)$ loop diagrams (\ref{fig:wqqQCDvirtI}), and 
from the squared $\ord(g_\rS^3 e)$ QCD emission amplitudes
(\ref{fig:wqqQCDrealI}).
The $\ord(\alphaS^2\alpha^2)$ NLO EW corrections involve 
contributions that arise from the interference of QCD trees of 
$\ord(g_\rS^2 e)$~(\reffis{fig:wqqQCDtreeI}--\ref{fig:wqqQCDtreeIII})
with loop diagrams of $\ord(g_\rS^2e^3 )$  (\reffis{fig:wqqEWvirtI}--\ref{fig:wqqEWvirtIII}), and 
from the squared QED emission amplitudes of $\ord(g_\rS^2e^2)$ 
(\reffi{fig:wqqEWrealI}). In addition, the NLO EW corrections involve also
interferences of EW trees of 
$\ord(e^3)$~(\reffis{fig:wqqEWtreeI}--\ref{fig:wqqEWtreeIV})
with $\ord(g_\rS^4 e)$  loop diagrams (\ref{fig:wqqQCDvirtI}), as well as
interferences between QCD real emission amplitudes of 
$\ord(g_\rS^3 e)$ (\reffi{fig:wqqQCDrealI})
and $\ord(g_\rS e^3)$ (\reffi{fig:wqqMIXrealI}).
Similarly as in the Born case, EW--QCD interference terms
at NLO EW order $\alphaS^2\alpha^2$ do not give rise to Breit--Wigner 
resonances in $W+2\jet$ production.
The same holds for EW--QCD interference terms
of order $\alphaS^3\alpha^2$ in $pp\to W+3\jet$.

\subsection{Photon-induced processes}
\label{sec:gamma-induced}

At tree level, if one treats the photon density as a quantity of
$\ord(1)$ as discussed in~\refse{sec:photons}, 
$W$+multijet  production receives
Born contributions from $\gamma p \to W+n$\,jets at $\ord(\alphaS^{n-1} \alpha^2)$,
i.e.~at the same order as EW--QCD interference terms,
as well as $\gamma \gamma \to W+n$-jet contributions
at $\ord(\alphaS^{n-2}\alpha^3)$, which is 
the order of pure EW Born terms. Photon--photon channels 
start contributing at $n=2$.
More explicitly, $W^+\jet$ production 
receives $\gamma p\to W^+\jet$ contributions 
of $\ord(\alpha^2)$ from the partonic process 
\beq
\gamma u_i \to W^+d_i,
\label{eq:gammaprocesses_w1}
\eeq
and crossing-related channels. Hadro-production of $W+2$\,jets involves the following 
single-photon induced processes
of $\ord(\alphaS\alpha^2)$ and
$\gamma\gamma$-induced processes of $\ord(\alpha^3)$,
\beqar
\label{eq:gammaprocesses_w2a}
&&\gamma u_i \to W^+d_i g,\\
\label{eq:gammaprocesses_w2b}
&&\gamma\gamma \to W^+d_i\bar u_i,
\eeqar
while $W+3\jet$ production involves the following 
$\ord(\alphaS^2\alpha^2)$  single-photon induced 
and $\ord(\alphaS\alpha^3)$ $\gamma\gamma$-induced channels,
\beqar
\label{eq:gammaprocesses_w3a}
&&\gamma u_i \to W^+d_i gg,\\
\label{eq:gammaprocesses_w3b}
&&
\gamma\gamma \to W^+d_i\bar u_i g,
\eeqar
together with the following channel contributing at $\ord(\alphaS^2\alpha^2)$, $\ord(\alphaS\alpha^3)$ and $\ord(\alpha^4)$,
\beqar
\gamma u_i \to W^+d_i q\bar q.
\label{eq:gammaprocesses_w3c}
\eeqar
All $\gamma p$- and $\gamma\gamma$-induced processes enter at a
different (lower) order in $\alphaS$ as compared to the NLO EW corrections
of $\ord(\alphaS^n\alpha^2)$ presented in this paper. 
Photon-induced contributions are
thus irrelevant for the cancellation of collinear
initial-state singularities at $\ord(\alphaS^n\alpha^2)$,
and can be handled as separate processes. From the formal power-counting 
perspective, leading $\gamma p$-induced processes are actually more important than
NLO EW corrections, but in most of the phase space
they are strongly suppressed by the small photon PDF.
However, as we will see in~\refse{se:results},
in the very high-energy tails of distributions
photon-induced processes can have a sizable impact 
on $W+$multijet production.
As is well known, this is due to the relative enhancement 
of the photon density at large $x$. At the same time, 
the poor knowledge of the photon PDF in this kinematic region represents
a large source of theoretical uncertainty~\cite{Ball:2013hta}.

\subsection{Technical aspects of the on-shell approximation}
\label{sec:onshell}
\def\Greg{\Gamma_{\mathrm{reg}}}

In this paper we consider $W+$multijet production with
stable on-shell $W$ bosons, and the inclusion of $W\to\ell\nu$ decays will
be addressed in a subsequent publication.  Implementing $W$ boson decays at NLO EW
does not represent a dramatic source of extra complexity 
as long as $W$ bosons are kept on-shell, such that 
$W+$multijet production and $W$ decays can be factorised using the
narrow-width approximation (NWA).  In contrast, a full description of $pp\to
\ell\nu+n$\,jets, including off-shell contributions at NLO EW, would be at
least one order of magnitude more CPU expensive.  
This is simply due to the
fact that, if the $W$ boson is replaced by a $\ell\nu$ pair,
the number of external particles that can enter EW loops increases
by one.\footnote{Note that
off-shell $W\to\ell\nu$ decays are trivial at NLO QCD as they do not
increase the number of external lines that enter QCD loops.} Keeping
the external $W$ boson on-shell---either as stable particle or as decaying
particle in NWA---is thus essential in order to 
be able to push $W+$multijet NLO EW calculations up to the highest
possible jet multiplicity while keeping the 
complexity at a manageable level.
 
Unfortunately, the simplifications that arise from the on-shell
(or narrow-width) approximation are accompanied by some technical
complications at NLO EW.  The key problem is that the on-shell treatment of
external $W$ bosons implies that the $W$ boson width is set to zero,
while EW corrections give rise to internal $W$ propagators that can produce
physical resonances, which requires a non-zero width.  
In presence of physical resonances, it is clear that all
$W$ bosons must be consistently handled as unstable particles with non-zero
width, and in order to preserve gauge
invariance the complex-mass scheme~\cite{Denner:2005fg} has to be used, which means that 
the on-shell description of $W$+multijet production has to be abandoned 
(or improved in a nontrivial way that preserves gauge invariance).
However, as discussed in \refse{se:fourquarks}, internal $W$, $Z$, and top propagators that enter the
 EW corrections to $pp\to W+n$\,jets cannot give rise to Breit--Wigner resonances at
$\ord(\alphaS^n\alpha^2)$. 
At this perturbative order, resonant $Z,W$, and $t$ propagators
appear in the EW Born amplitudes of $\ord(g_\rS^{n-2} e^3)$ (see \ref{fig:wqqEWtreeI}--\ref{fig:wqqEWtreeIII}),
in the EW virtual amplitudes of $\ord(g_\rS^n e^3)$ (see \ref{fig:wqqEWvirtI}--\ref{fig:wqqEWvirtIII}),
and in the QCD emission amplitudes of $\ord(g_\rS^{n-1} e^3)$ (see \ref{fig:wqqMIXrealI}),
but they contribute to the physical cross section
only through interference with non-resonant QCD amplitudes. 
As illustrated in Figure~\ref{fig:resdiags}, also two-quark processes 
involve EW 1-loop topologies with potentially resonant particles, including
Higgs bosons. In any case,
as a result of the interference with QCD amplitudes, none of these 
contributions can give rise to a physical resonance.

\begin{figure*}[t]
\centering
\subfloat[][]{   
   \includegraphics[height=0.14\textwidth]{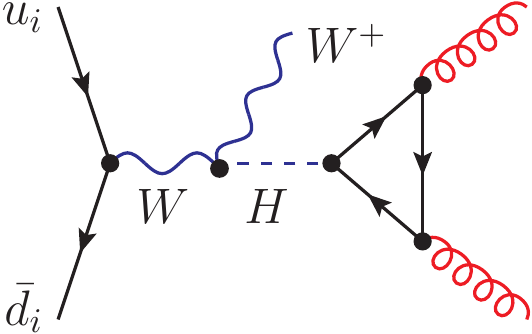}
   \label{fig:wggEWresI}
}
\quad
\subfloat[][]{   
   \includegraphics[height=0.14\textwidth]{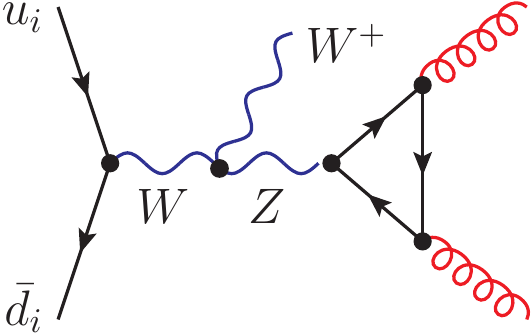}
   \label{fig:wggEWresII}
}
\quad
\subfloat[][]{   
   \includegraphics[height=0.14\textwidth]{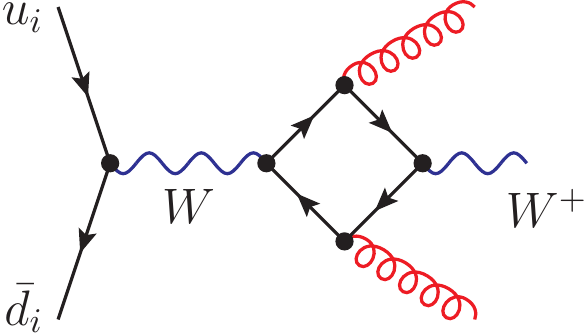}
   \label{fig:wggEWresIII}
}
\caption{
Examples of 1-loop $u_i \bar d_i\to W^+gg$  matrix elements 
at $\ord(g_\rS^2 e^3)$ that involve potentially resonant
Higgs-boson (\ref{fig:wggEWresI}), $Z$-boson (\ref{fig:wggEWresII})
and $W$-boson (\ref{fig:wggEWresIII}) propagators, where the last 
example diagram can only become resonant in the $gg \to W^+ \bar u_i d_i$ crossing.
}
\label{fig:resdiags}
\end{figure*}

Since $pp\to W+n$\,jets at $\ord(\alphaS^n \alpha^2)$ 
is free from Breit--Wigner resonances, in principle
the $W$ width can be set to zero in all scattering amplitudes,
consistently with the on-shell treatment of external 
$W$ bosons. However, the interference of resonant and 
non-resonant contributions gives rise to spikes 
that can disturb the numerical stability of the 
phase-space integration in the vicinity of the 
``pseudo-resonance''.  
An optimal treatment of these regions
can be achieved by introducing an ad-hoc technical width $\Greg$
in the potentially resonant propagators, in such a way that 
the pseudo-resonant contributions behave as
\beq
\label{eq:pseudoresonance}
\lim_{Q^2\to M^2}\frac{\rd \sigma}{\rd Q^2} \,\,\propto\,\, \frac{Q^2-M^2}{(Q^2-M^2)^2+\Greg^2M^2}.
\eeq
The idea is that the $1/\Greg^2$ enhancement at $Q^2\sim M^2$
cancels upon integration over $Q^2$, and 
the overall dependence on the technical
regulator should be $\ord(\Greg/M)$ suppressed, 
while all contributions should formally behave smoothly
when $\Greg\to 0$. If these conditions are fulfilled, 
then the calculation should consistently converge 
towards the correct on-shell limit, and  using a sufficiently small value for
$\Greg$ should guarantee a negligible numerical impact of $\ord(\Greg/M)$ effects 
and related violations of gauge invariance.

In this context, 
due to the presence of IR
singularities that arise from (virtual and real) soft photons coupled to
external $W$ bosons, the smooth convergence of the $\Greg\to 0$ limit represents
a nontrivial requirement. In fact, a naive introduction of
$\Greg>0$ in all $W$ propagators would turn 
such soft-photon singularities into $\ln(\Greg)$ terms that do not converge towards 
the correct $1/(D-4)$ poles when $\Greg\to 0$.
Fortunately, all diagrams that involve real or virtual 
photons are free from potential resonances.
Therefore, in order to guarantee a smooth $\Greg\to 0$ behaviour, 
one can simply restrict the $\Greg>0$ regulator
to those diagrams that are free from photons, 
and evaluate all photonic corrections at zero width. 
More precisely, we will adopt the following approach,
which is applicable at $\ord(\alphaS^n \alpha^2)$ 
for the case of stable $W$ bosons as well as for
decaying $W$ bosons in NWA:
\bit
\item the physical width of all unstable particles ($W,Z,t,H$)
is never included in the corresponding propagators, and the 
corresponding masses, as well as the related mixing angles and Higgs couplings,
are treated as real parameters, i.e.~the complex mass scheme is not used;
\item external $W$ bosons are kept on their mass shell, $p_W^2=M_W^2$;
\item in diagrams that do not involve photons,
possible $W$, $Z$, $H$, and top-quark  propagators are regularised 
as $1/(Q^2-M^2+i\Greg M)$ with a small technical 
width $\Greg$.
\eit
The dependence of physical observables on the value of $\Greg$
must be regarded as a small uncertainty associated with a gauge-dependent 
$\ord(\Greg)$ deformation around the exact gauge-invariant limit
$\Greg\to 0$ limit. In this respect, it should be stressed that, thanks to the 
smooth convergence of the $\Greg\to 0$ limit, these
violations of gauge invariance are controllable, in the sense that 
they can be quantified and systematically reduced by chosing 
an appropriate $\Greg$ value.\footnote{This is completely different with respect to 
violations of gauge invariance in process that involve 
physical resonances.}

Results presented in \refse{se:results} have been obtained using
$\Greg=1$\,GeV, which turns out to guarantee good numerical stability 
and negligible $\Greg$ dependence.
More precisely, we have checked that for all
integrated and differential results presented in section \ref{se:results} 
the shift resulting from variations of $\Greg$ between 0.1 and 1\,GeV
is well below one percent.

\section{Setup of the simulation}
\label{se:setup}

As input parameters to 
simulate $W+$\,multijet production 
at NLO QCD+EW we use the 
gauge-boson, Higgs-boson, and top-quark masses
\beqar
\MZ=91.1876~\GeV,\quad
\MW=80.385~\GeV,\quad
\MH=126~\GeV,\quad
\Mt=173.2~\GeV.
\eeqar
The corresponding Lagrangian parameters are kept strictly real since we
treat all heavy particles as stable.  
The electroweak couplings are derived
from the gauge-boson masses and the Fermi constant,
$\GF=1.16637\times10^{-5}~\GeV^{-2}$, in the 
$G_{\mu}$-scheme~\refeq{eq:GFscheme}.
The CKM matrix is assumed to be diagonal, while colour effects and related
interferences are included throughout, without applying any large-$N_c$
expansion.

For the calculation of hadron-level cross sections we employ the NNPDF2.3
QED parton distributions~\cite{Ball:2013hta}, which include NLO QCD and LO
QED effects, and we use the PDF set corresponding to
$\alphaS(\MZ)=0.118$.\footnote{To be precise we use the
NNPDF23\_nlo\_as\_0118\_qed set interfaced through the LHAPDF library 5.9.1.}
Matrix elements are
evaluated using the running strong coupling supported by the PDFs and,
consistently with the variable flavour-number scheme implemented in the
NNPDFs, at the top threshold we switch from five to six active quark
flavours in the renormalisation of $\alphaS$.  All light quarks, including
bottom quarks, are treated as massless particles, and top-quark loops are included
throughout in the calculation.  The NLO PDF set is used
for LO as well as for NLO QCD and NLO EW predictions. 
Using the same PDFs for LO and NLO predictions exposes matrix-element
correction effects in a more transparent way.  In particular, it guarantees
that NLO EW $K$-factors remain free from QCD effects related to the
difference between LO and NLO PDFs.

The renormalisation scale $\mu_R$ and factorisation scale $\mu_F$ are set to
\beq\label{eq:RFscales} 
\mu_{\rR,\rF}=\xi_{\rR,\rF}\mu_0
\quad\mbox{with}\quad \mu_0= \hat H_{\mathrm{T}}/2, 
\eeq 
where $\hat H_{\mathrm{T}}$ is the scalar sum of the transverse energy of all 
partonic final-state particles, 
\beq\label{eq:BHscale} 
\hat H_{\mathrm{T}} = \sum_{\text{partons}} E_{\mathrm{T}} = \sum_{i} E_{\rT,\jet_i} + E_{\rT,\gamma} + \sqrt{{p^2_{\mathrm{T, W}}}+M_W^2}\,.
\eeq
Our default scale choice corresponds to $\xi_{\rR}=\xi_{\rF}=1$,
and theoretical uncertainties are assessed by
applying the scale variations $(\xi_\rR,\xi_\rF)=(2,2)$,
$(2,1)$, $(1,2)$, $(1,1)$, $(1,0.5)$, $(0.5,1)$, $(0.5,0.5)$.
As shown in
\cite{Berger:2009zg,Ellis:2009zw,KeithEllis:2009bu,Berger:2009ep,
Berger:2010zx, Bern:2013gka} the scale choice~\refeq{eq:RFscales} guarantees
a good perturbative convergence for $W+$\,multijet production over a wide
range of observables and energy scales.

For the definition of jets we employ the anti-$k_{\mathrm{T}}$
algorithm \cite{Cacciari:2008gp} with $R=0.4$. More precisely,
in order to guarantee IR safeness in presence of NLO QCD and EW corrections,
we adopt the democratic clustering approach introduced in~\refse{se:democlstering}, 
treating QCD jets and photons as separate physics objects.
To this end we impose an 
upper bound $\zgammathr=0.5$ to the photon energy fraction
inside jets, and the recombination of collinear (anti)quark--photon pairs
is applied within a cone of radius $\Drrec=0.1$.

\section{NLO QCD+EW predictions for $\boldsymbol{W^++1,2,3}$~jets at the LHC}
\label{se:results}

\def\gammaind{\gamma-\mathrm{ind.}}
\def\interf{\mathrm{interf.}}
\def\relplotwidth{0.46}

In the following we present a series of NLO QCD+EW simulations
for $W^+$ production in association with one, two, and three jets 
in proton--proton collisions at 13\;TeV. 
Events are categorised according to the number of jets 
in the transverse-momentum and pseudo-rapidity region defined by
\beqar\label{eq:JETcuts}
p_{\rT,\jet}&>&30\,\GeV, \qquad {|\eta_{\jet}|<4.5},
\eeqar
and for each $W+n$-jet sample
we present an inclusive analysis, where we do not impose any selection cut apart from
requiring the presence of $n$ (or more) jets.
In addition, to study the high-energy behaviour of EW corrections,
we also consider cross sections and distributions in presence of one of the
following cuts:
\beqar
\label{eq:hardcuts}
p_{\mathrm{T,W}}>1\,\TeV \,, \qquad p_{\rT,\jet_1}>1\,\TeV \,, \qquad \mathrm{or} \qquad \HTtot  > 2\,\TeV \,.
\eeqar
Here $\jet_1$ denotes the first jet, while the total transverse energy
$\HTtot$ is defined in terms of the jet and $W$-boson 
transverse momenta\footnote{Note that at variance with the definition~\refeq{eq:BHscale} of 
$\hat H_{\mathrm{T}}$, here we use transverse momenta and not transverse energies.} as
\beqar
\label{eq:htdefs}
%
%
\HTtot= p_{\mathrm{T,W}}
+\sum_{k} p_{\rT,\jet_k},
\eeqar
where all jets that satisfy~\refeq{eq:JETcuts}
are included.


Our default NLO results are obtained by combining 
QCD and EW predictions,
\beqar
\sigma^{\NLO}_{\QCD} &=& 
\sigma^{\LO}+\delta\sigma^\NLO_\QCD,
\qquad\sigma^{\NLO}_{\EW} =
\sigma^{\LO}+\delta\sigma^\NLO_\EW,
\eeqar
with a standard additive prescription
\beqar
\sigma^{\NLO}_{\QCDpEW} &=& 
\sigma^{\LO}+\delta\sigma^\NLO_\QCD + \delta\sigma^\NLO_{\EW},
\label{eq:qcdplusew}
\eeqar
where $\delta\sigma^\NLO_\QCD$ and
$\delta\sigma^\NLO_\EW$ correspond to $pp\to W+n$-jet contributions of 
$\ord{(\alphaS^{n+1}\alpha})$ and
$\ord{(\alphaS^{n}\alpha^2})$, respectively.
As LO contributions, 
in \refses{se:wjsec}--\ref{se:wjjjsec} only the leading-QCD terms of
$\ord(\alphaS^n\alpha)$ will be included, while 
subleading Born contributions and photon-induced terms 
will be discussed in \refse{se:lomixsec}.
In order to identify potentially large effects due to the interplay of 
EW and QCD corrections beyond NLO, 
we will also consider the following factorised 
combination of EW and QCD corrections,
\beqar
\sigma^\NLO_\QCDtEW &=& 
\sigma^\NLO_\QCD\left(1 + \frac{\delta\sigma^\NLO_{\EW}}{\sigma^\LO}\right)
=
\sigma^\NLO_\EW\left(1 + \frac{\delta\sigma^\NLO_{\QCD}}{\sigma^\LO}\right).
\label{eq:qcdtimesew}
\eeqar
If this approach can be justified by a 
clear separation of scales---such as in situations 
where QCD corrections are dominated by soft interactions well below the EW scale---the 
factorised formula \refeq{eq:qcdtimesew} can be regarded as an improved prediction. 
Otherwise, the difference between~\refeq{eq:qcdplusew} 
and~\refeq{eq:qcdtimesew} should be considered 
as an estimate of unknown higher-order corrections.

In the following sections, we will present QCD+EW and QCD$\times$EW NLO corrections 
relative to $\sigma^\NLO_\QCD$, which corresponds to the ratios
\beqar
\frac{\sigma^\NLO_\QCDpEW}{\sigma^\NLO_\QCD} &=& \left(1 + \frac{\delta\sigma^\NLO_{\EW}}{\sigma^\NLO_\QCD}\right),
\label{eq:qcdplusewratio}\\
\frac{\sigma^\NLO_\QCDtEW}{\sigma^\NLO_\QCD} &=& 
\left(1 + \frac{\delta\sigma^\NLO_{\EW}}{\sigma^\LO}\right).
\label{eq:qcdtimesewratio}
\eeqar
Note that the QCD$\times$EW ratio~\refeq{eq:qcdtimesewratio}
corresponds to the 
usual NLO EW correction relative to LO, which is free from NLO QCD effects,
while the QCD+EW ratio~\refeq{eq:qcdplusewratio} depends on 
$\sigma^\NLO_\QCD$. In particular, 
for observables that receive 
large NLO QCD corrections, the relative QCD+EW correction 
can be drastically suppressed as compared to the QCD$\times$EW one. 
This feature is typically encountered 
in observables that receive huge QCD corrections
of real-emission type. In such situations,
NLO QCD+EW predictions for $pp\to W+n$\,jets 
are dominated by tree-level contributions with one extra jet,
and the inclusion of NLO QCD+EW corrections for
$pp\to W+(n+1)$\,jets becomes mandatory.

Thanks to the high efficiency of the employed tools,
the simulation of $W$+multijet production at NLO QCD+EW 
requires a moderate amount of computing resources. The runtime needed 
 to achieve very high statistical accuracy, 
at the level of 0.1\%, in the NLO QCD+EW integrated cross section amounts 
to about 
13, 210 and 6300
CPU hours for $pp\to W^++1,2,3$\,jets, 
respectively.\footnote{
The stated runtimes 
refer to a single core and
are estimated from runs on a cluster based on
Intel$^{\textregistered}$ Xeon$^{\textregistered}$ E5-2660 (20MB Cache, 2.20GHz)
processors by means of an extrapolation to an overall statistical error 
of 0.1\% wrt.\ $\sigma^{\NLO}_{\QCDpEW}$.
} 
In order to obtain 0.1\% statistical accuracy also in the phase-space region 
with $\HTtot>2\TeV$, where the cross section is suppressed by about 4 orders of 
magnitude, less than a factor 10 of extra CPU time is needed (without using any 
generation cut). For all processes under consideration, the evaluation of the 
NLO EW corrections consumes a subleading part of the total CPU budget.

\subsection[$W^++1$ jet]{$\mathbf{W^++1}$ jet}
\label{se:wjsec}

\def\arraystretch{1.3}
\begin{table}[t]
\renewcommand{\arraystretch}{1.5}%
\setlength{\tabcolsep}{1.2ex}%
\begin{center}
\begin{tabular}{lllllll}\hline
\input{XStable.wj.tex}
\end{tabular}
\caption{Integrated $pp\to W^++1\jet$ 
cross sections with inclusive cuts~\refeq{eq:JETcuts}
and in presence of additional cuts. 
Born cross sections ($\sigma^\LO$) include only the leading QCD contributions of
$\ord(\alphaS\alpha)$.
}
\label{tab:XSwj}
\end{center}
\end{table}

Among the various $W+$(multi)jet production processes, 
the inclusive production of a $W$ boson in association with 
(at least) one jet is the one that features the strongest sensitivity to 
NLO QCD radiation. This is clearly illustrated by
the results shown in \reffis{fig:wj_pTall}--\ref{fig:wj_HTtot} and \refta{tab:XSwj}.
In particular, large NLO QCD effects arise in the tails of the inclusive
distributions in the $W$-boson and in the jet transverse momenta, shown in the left plot of
\reffi{fig:wj_pTall}.
For the $W$-boson $\pT$ distribution NLO QCD corrections exceed 50\%, 
while in the case of the jet-$p_\rT$ distribution they become as 
large as 400\% at 1\;TeV.
As is well known, this extreme behaviour is due to the fact that $W$-boson
production in association with very hard jets is dominated by $W+$multijet
events where two or more high-$p_\rT$ jets recoil against each other, while
the $W$ boson tends to be rather
soft~\cite{Arnold:1988dp,Arnold:1989ub,Campbell:2002tg,Denner:2009gj}.  In
this kinematic regime, inclusive NLO simulations of $pp\to W+1\jet$  are dominated by
tree-level contributions with two jets, which results in large scale uncertainties. 
The inclusion of NLO corrections for $W+$multijet production is thus
mandatory for a well-behaved theoretical prediction 
of the inclusive jet-$\pT$ spectrum.
Predictions for $pp\to W+1\jet$ at NLO QCD are perturbatively stable only 
in presence of ad-hoc cuts that separate one-jet configurations form 
the bulk of the extra jet emission. As shown in the right plot of
\reffi{fig:wj_pTall}, this can be achieved by means of a veto
against dijet configurations with azimuthal separation
$\Delta\phi_{\jet\jet}>3\pi/4$.
Thanks to this cut, which avoids hard events characterised by 
a  back-two-back dijet system with $\Delta\phi_{\jet\jet}\to\pi$, 
NLO QCD correction become acceptably small and reasonably stable, 
even at very large jet $\pT$.

As discussed in the following, the behaviour of NLO EW effects is strictly connected to the one of 
NLO QCD corrections.
The shape of NLO EW corrections to the inclusive $W$-boson $p_\rT$ distribution
(\reffi{fig:wj_pTall}, left) 
is consistent with the expected presence of negative
Sudakov logarithms that grow as $\ln^2(\hat s/M_W^2)$. 
However, in the tail we observe a large gap between
QCD+EW and QCD$\times$EW predictions,
which points to the presence of 
sizable EW higher-order effects that are not captured 
by the NLO QCD+EW approximation. This is clearly due 
to the fact that the NLO QCD cross section 
involves large radiative contributions 
that are effectively described at LO EW accuracy only.
In any case, it is clear that 
NLO EW effects are large. 
Noteworthy, already for $p_{\rT,W}\gtrsim 300\,\GeV$
they become larger than the NLO QCD uncertainties. 
For the inclusive jet-$\pT$ distribution, due to the huge impact 
of QCD radiation, NLO EW corrections behave in a pathological way. The expected
Sudakov suppression is completely absent, and above 1\;TeV one observes 
a strong enhancement. This 
can be attributed to $\ord(\alphaS\alpha^2)$  mixed EW--QCD 
contributions to hard-dijet plus soft-$W$ events~\cite{Denner:2009gj}, 
which result from the interference between 
diagrams of type 
\ref{fig:wqqQCDtreeI}--\ref{fig:wqqQCDtreeIII}
and \ref{fig:wqqEWtreeI}--\ref{fig:wqqEWtreeIV}.
The increase at large $\pT$ can be understood as a PDF effect 
at large Bjoerken $x$.

\begin{figure*}[t]
\centering
   \includegraphics[width=\relplotwidth\textwidth]{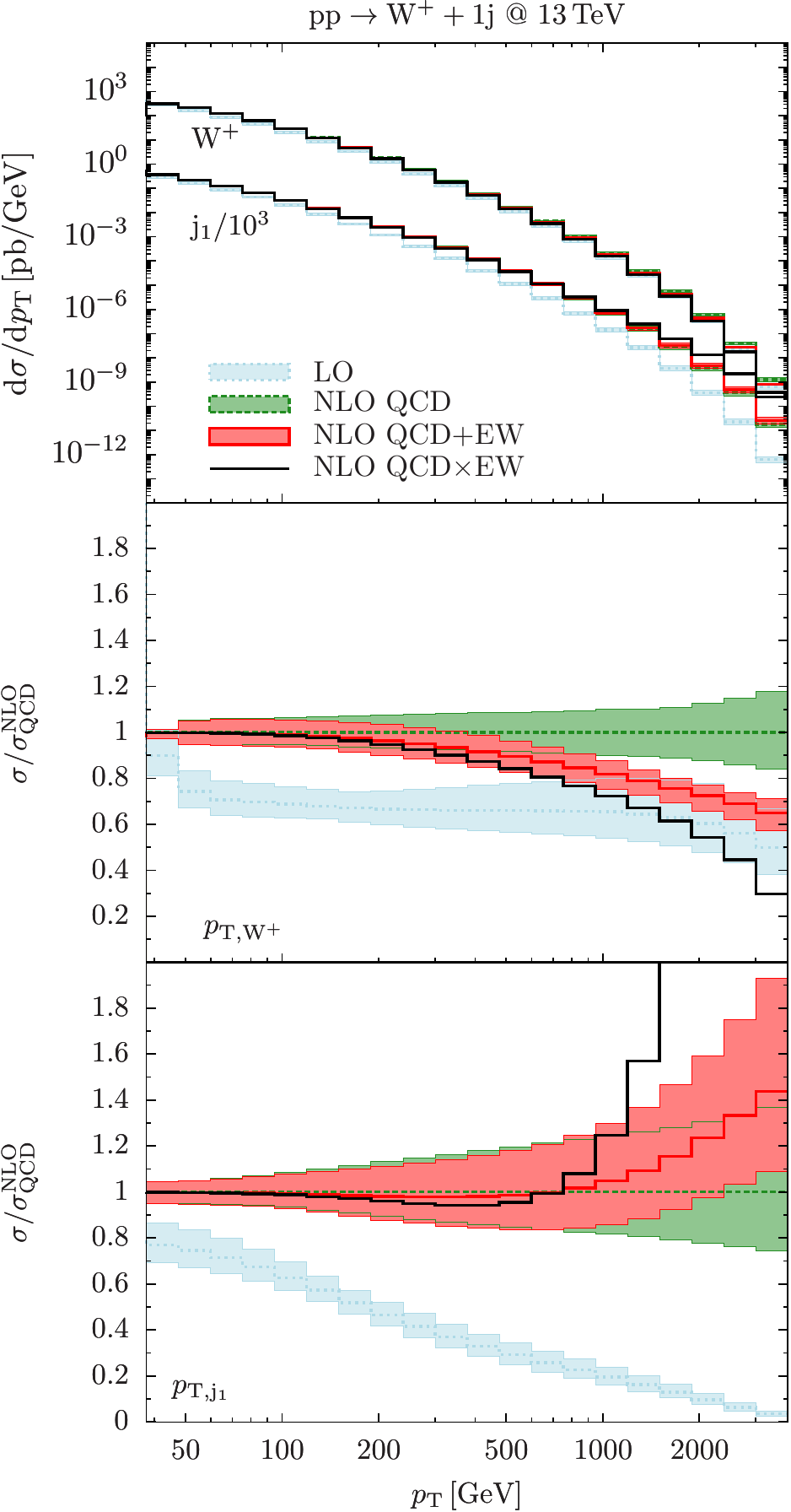}
         \qquad 
   \includegraphics[width=\relplotwidth\textwidth]{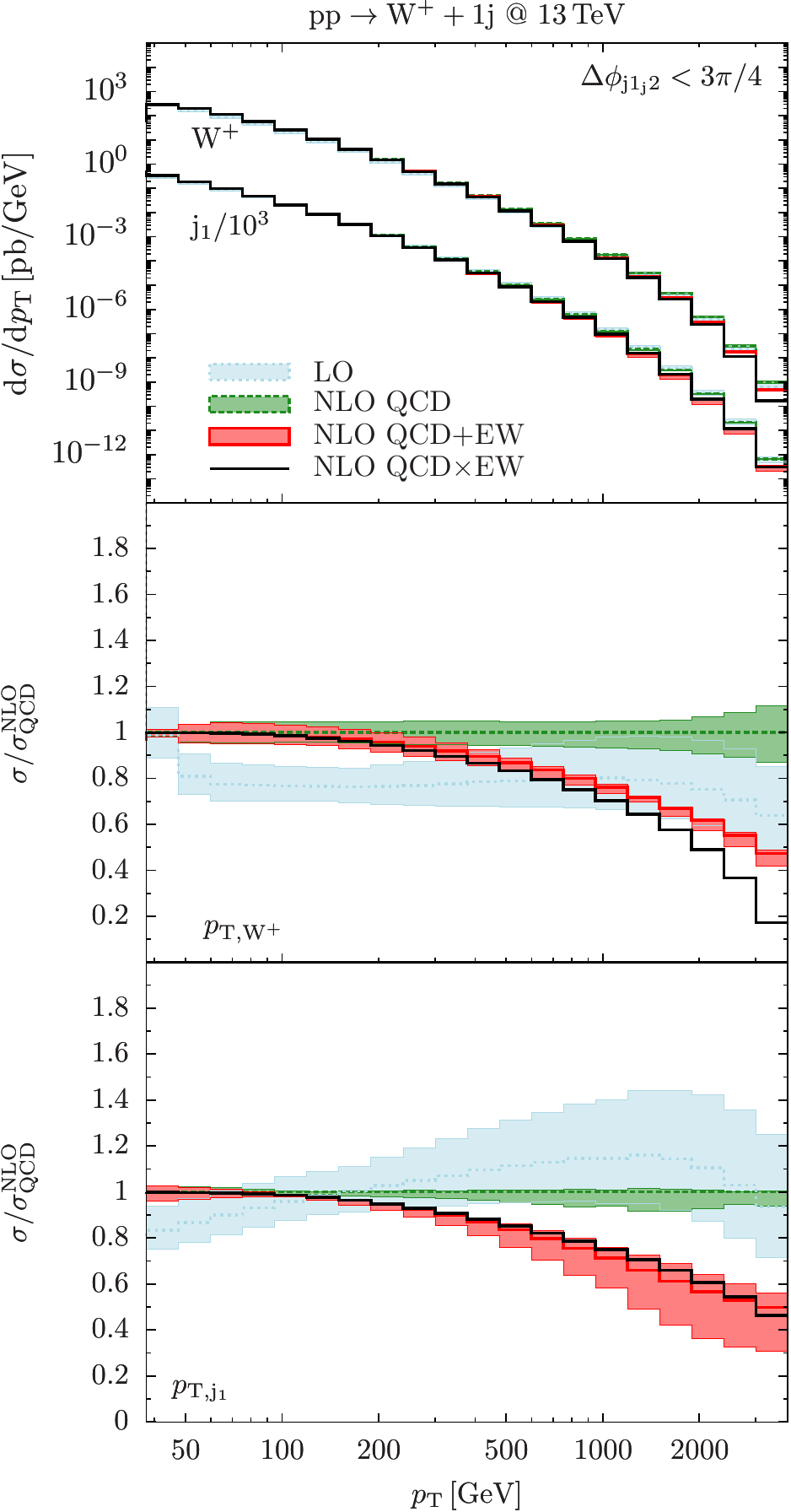}
\caption{
Distributions in the transverse momenta of the $W$ boson and of the first jet
for inclusive (left) \mbox{$W^++1\jet$} production and with a cut
$\deltajj <3\pi/4$ (right).
Absolute LO (light blue), NLO QCD (green), NLO QCD+EW (red) and NLO
QCD$\times$EW (black) predictions (upper panel) and relative corrections
with respect to NLO QCD (lower panels).  
The bands correspond to scale variations, 
and in the case of ratios only the numerator is varied.
The distribution in $\pTjone$ is
rescaled by a factor $10^{-3}$.
}
\label{fig:wj_pTall}
\end{figure*}

\begin{figure*}[t]
\centering
   \includegraphics[width=\relplotwidth\textwidth]{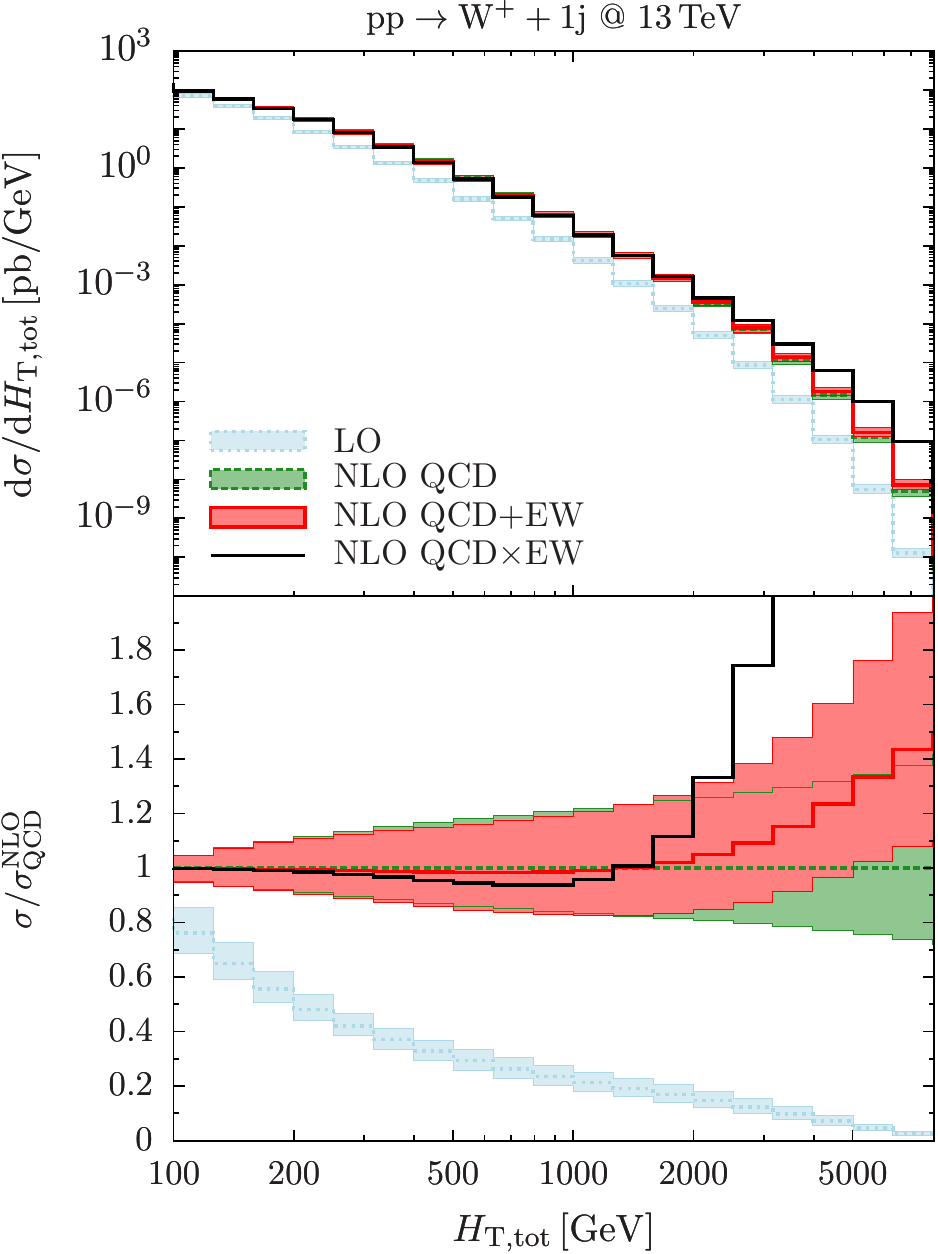}
         \qquad 
   \includegraphics[width=\relplotwidth\textwidth]{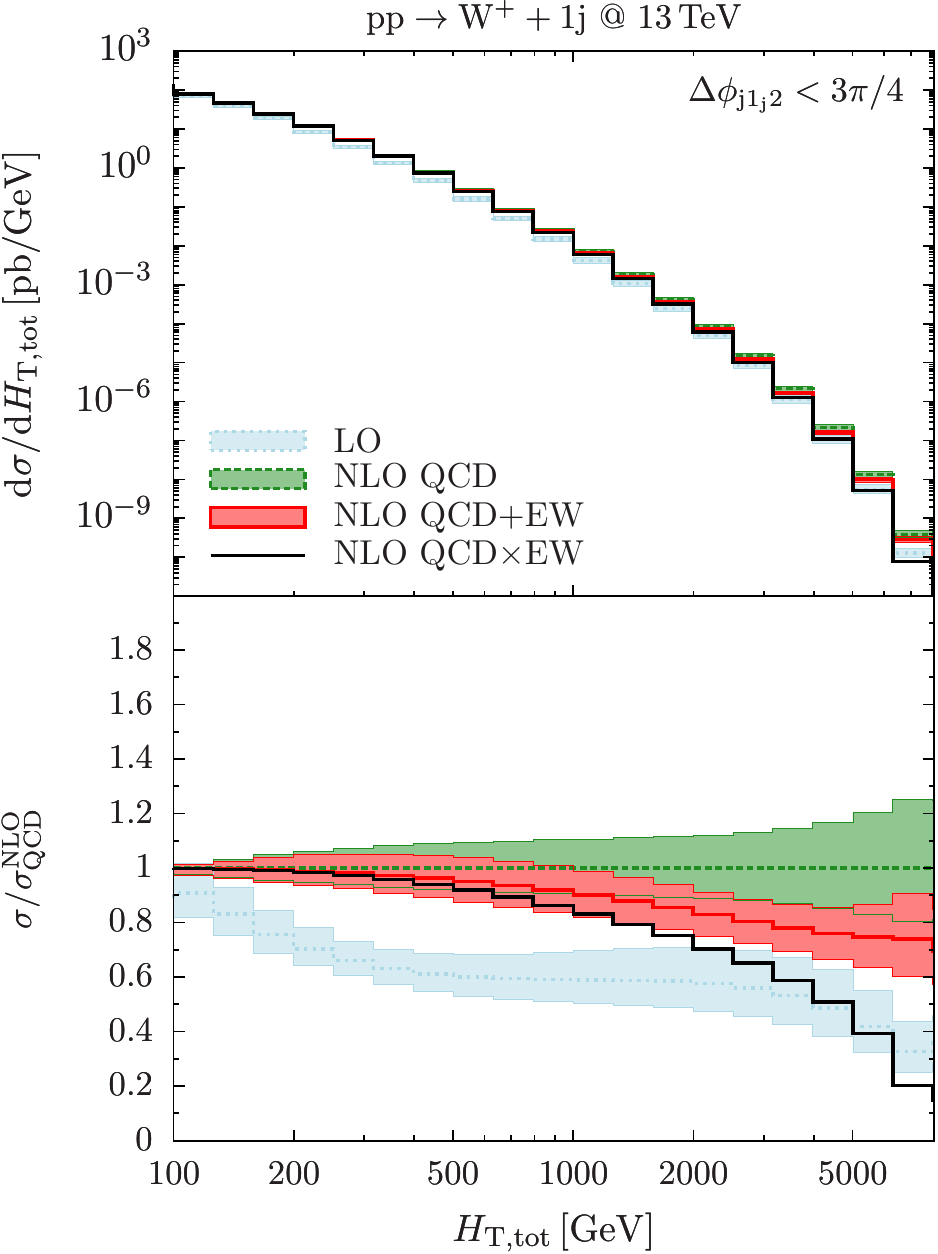}
\caption{
Distribution in $\HTtot$
for inclusive (left) \mbox{$W^++1\jet$} production and with a cut
\mbox{$\deltajj < 3\pi/4$} (right). 
Curves and bands as in \reffi{fig:wj_pTall}.
}
\label{fig:wj_HTtot}
\end{figure*}

As can be seen in the right plot of \reffi{fig:wj_pTall}, in presence of the
cut on $\Delta\phi_{\jet\jet}$, the improved perturbative QCD convergence
leads to a consistent Sudakov behaviour for the NLO EW corrections to the
$W$- and jet-$p_\rT$ distributions.
These two observables behave in a quite similar way, as expected 
for exclusive $W+1\jet$ events, 
where the jet and the $W$ boson recoil against each 
other, and the  size of the corrections is around $-40\%$ at $\pT=2~\TeV$.
Note that, in presence of the cut on $\Delta\phi_{\jet\jet}$,
EW corrections exceed NLO QCD scale variations 
already at $\pT \sim 200\,\GeV$.
%
%
The gap between the EW+QCD and EW$\times$QCD curves completely disappears
in the case of the jet-$\pT$ distribution, while for the
$W$-boson $\pT$ it remains problematic, due to the persistence 
of sizable QCD effects.

The distribution in $\HTtot$, shown in~\reffi{fig:wj_HTtot}, behaves in a
qualitatively similar way as the jet-$\pT$ distribution.  However, also in presence
of the $\Delta \phi_{\jet\jet}$ cut, this observable remains very sensitive to NLO
QCD radiation, and the QCD$\times$EW curve indicates that the observed 
NLO QCD+EW correction of $-25\%$ at $\HTtot=4\,\TeV$ might be underestimated by up to a
factor two.

In summary, the strong sensitivity of $W+1$\,jet production to NLO QCD real
emission, which is effectively described at LO accuracy, leads to a sizable
scale dependence and to an underestimate of EW correction effects in various
observables.  This calls for the calculation of NLO QCD+EW corrections to 
$W+2\jet$ and $W+3\jet$ production that we are going to present in the following sections.
Numerical results for $pp\to W^++1\jet$ cross sections with
different cuts are collected in \refta{tab:XSwj}.

\subsection[$W^++2$ jets]{$\mathbf{W^++2}$ jets}
\label{se:wjjsec}

\begin{table}[t]
\renewcommand{\arraystretch}{1.5}%
\setlength{\tabcolsep}{1.2ex}%
\begin{center}
\begin{tabular}{lllllll}\hline
\input{XStable.wjj.tex}
\end{tabular}
\caption{
Integrated $pp\to W^++2\jet$ 
cross sections with inclusive cuts~\refeq{eq:JETcuts}
and in presence of additional cuts. 
Born cross sections ($\sigma^\LO$) include only the leading QCD contributions of
$\ord(\alphaS^2\alpha)$.
}
\label{tab:XSwjj}
\end{center}
\end{table}

\begin{figure*}[t]
\centering
   \includegraphics[width=\relplotwidth\textwidth]{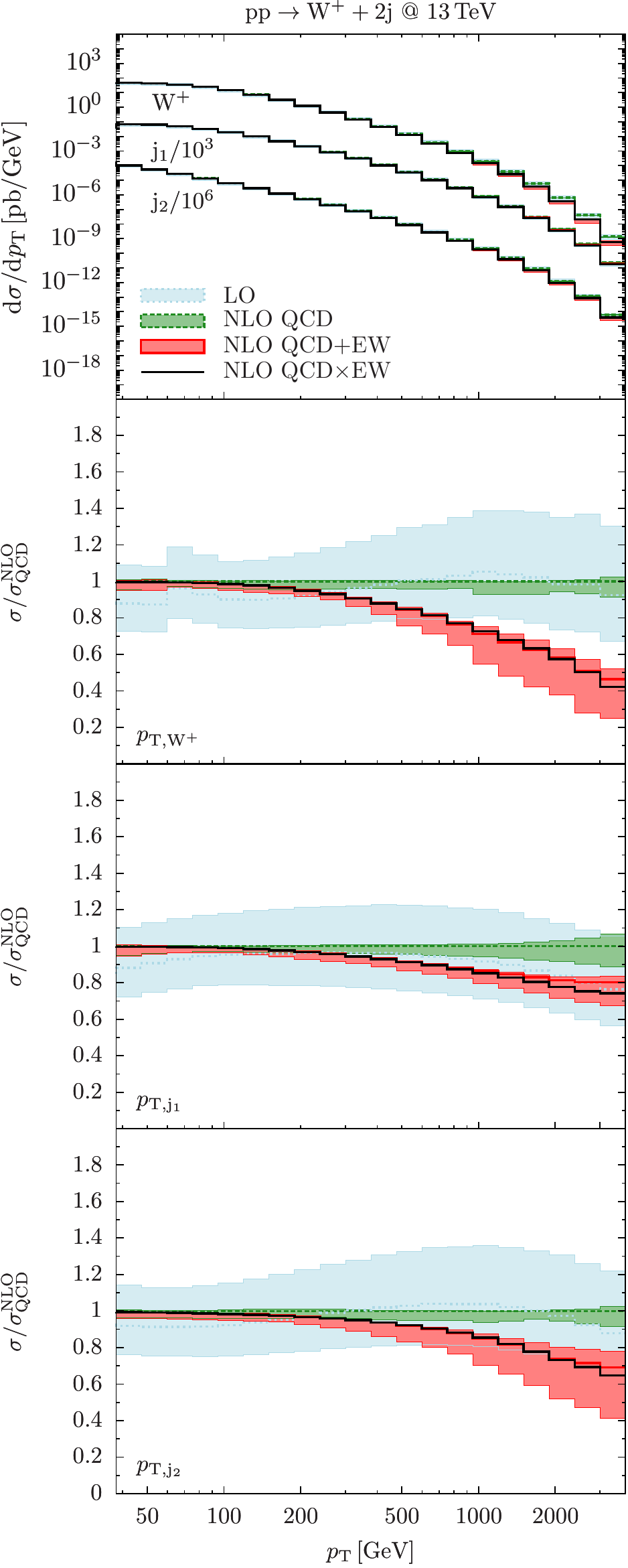}
   \qquad
   \includegraphics[width=\relplotwidth\textwidth]{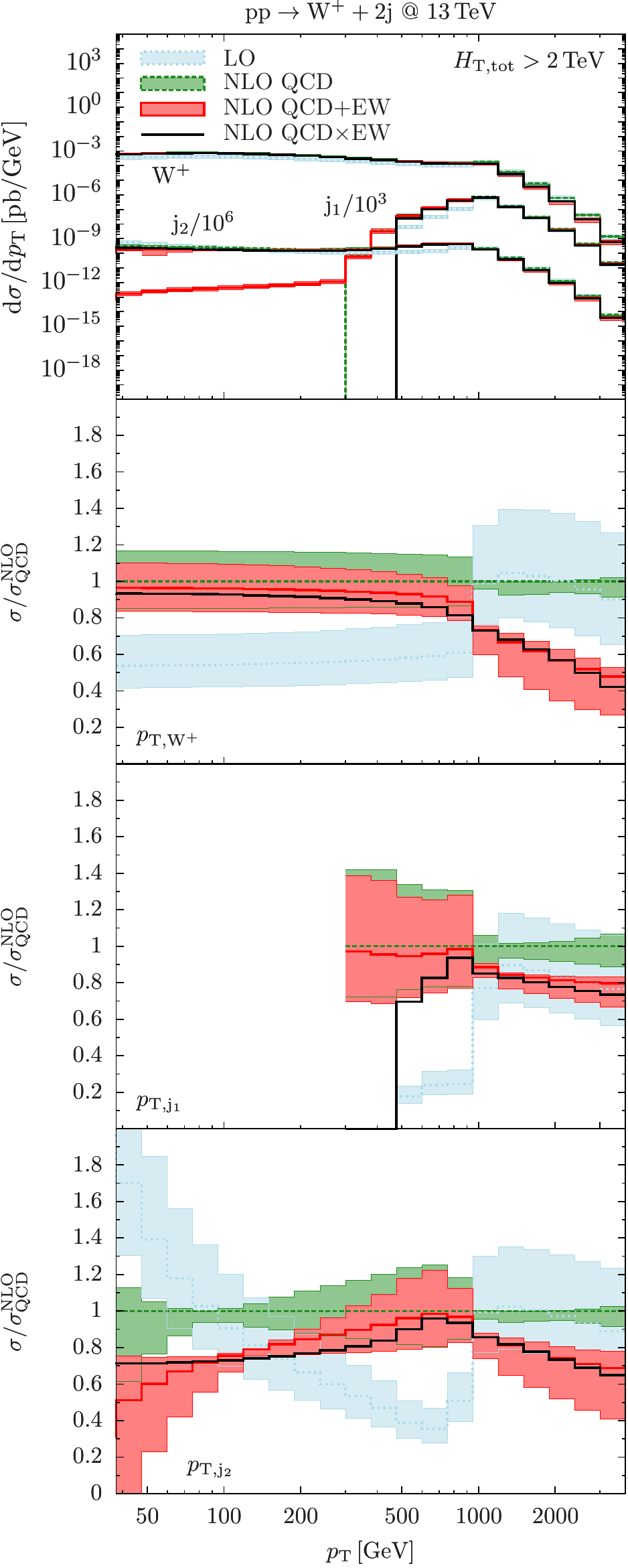}
\caption{
Distributions in the transverse momenta of the 
$W$ boson and the first two jets
for inclusive (left) \mbox{$W^++2\jet$} production and with a cut
$\HTtot > 2\,\TeV$ (right).
The distributions in the $n$-th jet $\pT$ are rescaled by factors $10^{-3n}$.
Curves and bands as in \reffi{fig:wj_pTall}.
}
\label{fig:wjj_pTall}
\end{figure*}

\begin{figure*}[t]
\centering
   \includegraphics[width=\relplotwidth\textwidth]{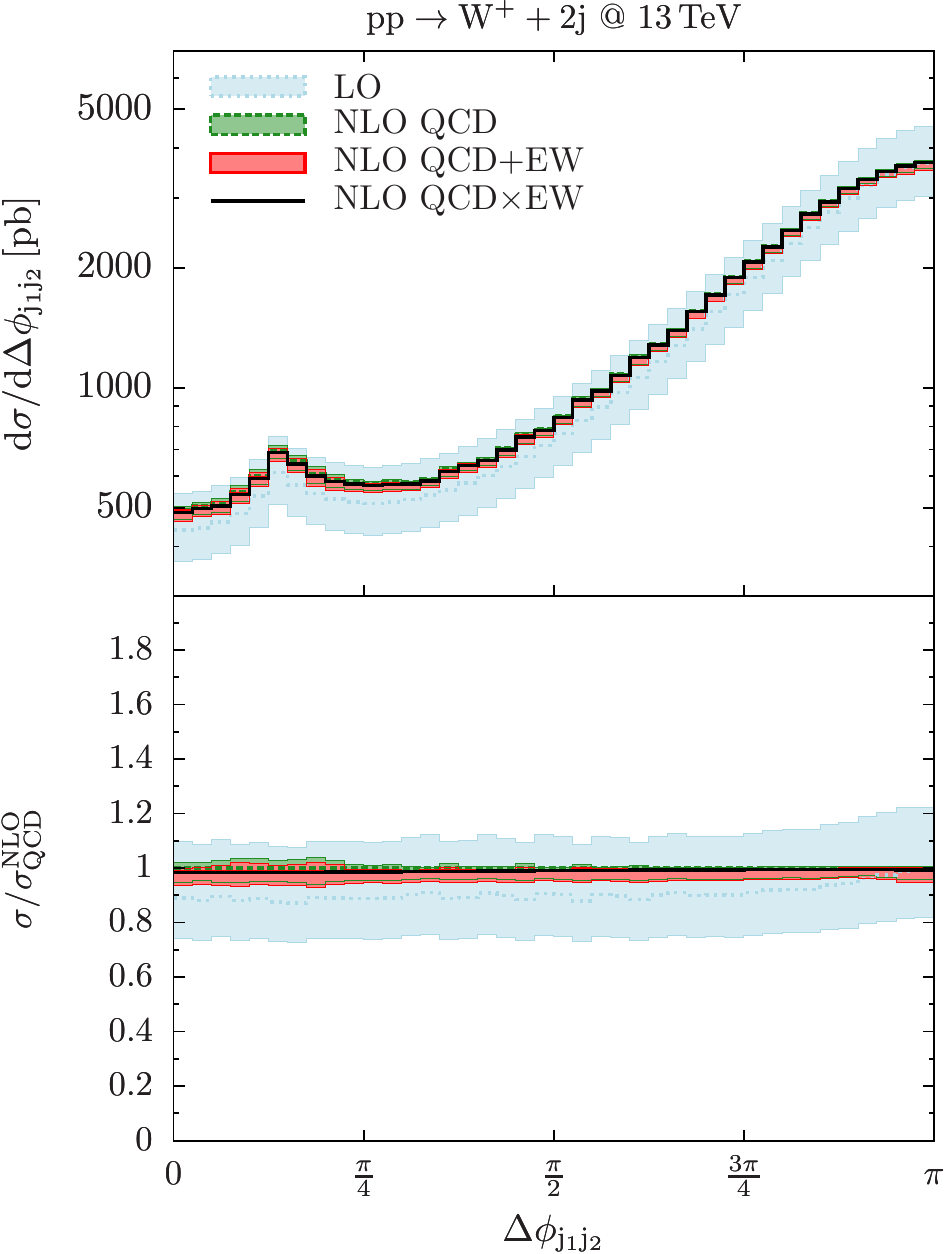}
      \qquad   
   \includegraphics[width=\relplotwidth\textwidth]{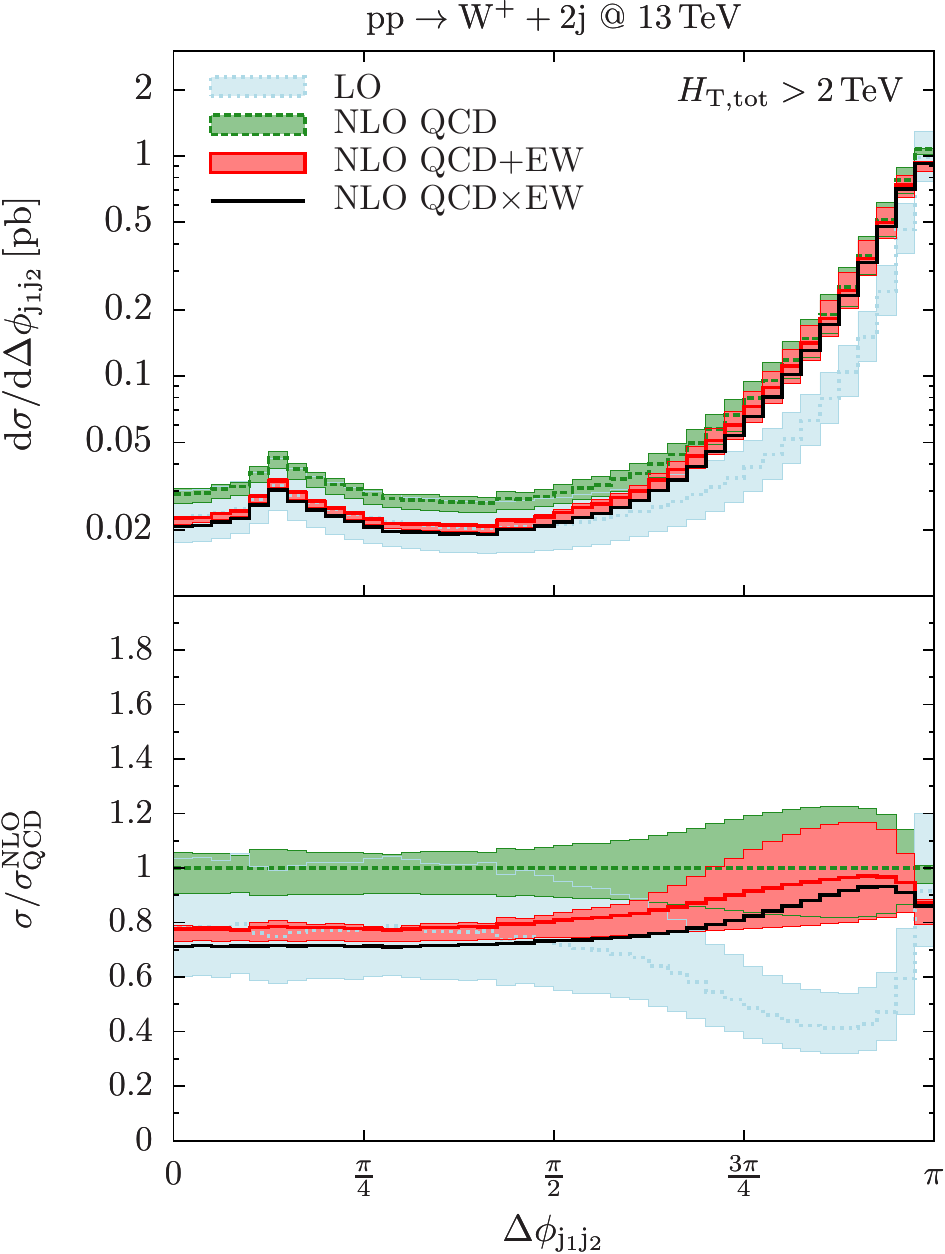}
\caption{
Distributions in the azimuthal separation of the 
first two jets for inclusive (left) \mbox{$W^++2\jet$} production
and with a cut $\HTtot > 2\,\TeV$ (right).
Curves and bands as in \reffi{fig:wj_pTall}.
 }
\label{fig:wjj_dphi}
\end{figure*}

Distributions and integrated cross sections for $pp\to W^++2\jet$ 
are presented in \reffis{fig:wjj_pTall}--\ref{fig:wjj_Httot_minv} and \refta{tab:XSwjj}, 
respectively.
When the $W$ boson is accompanied by two jets,
all one-particle inclusive $\pT$ distributions,
shown in the left plot of \reffi{fig:wjj_pTall},
are quite stable with respect to NLO QCD corrections.
Scale uncertainties at NLO QCD are generally very small, 
and even in the tails they hardly exceed $10\%$.
The NLO EW corrections show a standard Sudakov behaviour
and exceed NLO QCD uncertainties already at 
a few hundred~GeV. For the
$W$-boson $\pT$ distribution they behave very similarly as in the case of $pp\to
W+1\jet$, reaching $-40\%$ at 2\;TeV. The EW corrections to the jet-$\pT$ distributions
are significantly smaller. At 2\;TeV they 
are around $-20\%$, both
for the first and for the second jet.  
Moreover, in the tails of the jet-$\pT$ distributions, the trend of
increasingly negative Sudakov corrections gets suppressed due to positive
contributions from mixed EW--QCD bremsstrahlung, which result from 
interferences between
diagrams of type~\ref{fig:wqqQCDrealI} and~\ref{fig:wqqMIXrealI}.

In \reffi{fig:wjj_pTall}, results for inclusive $\pT$ distributions (left)
are compared to the same observables
in presence of a cut $\HTtot > 2\,\TeV$ (right).  
The region of high $\HTtot$ plays a central role 
for BSM searches, and the upper right plot in \reffi{fig:wjj_pTall}
provides insights into the
interplay between $W$-boson and jet transverse momenta in this kinematic region.
The interesting part of the plot is the $\pT$-range below 
$\HTtotcut/2$, where the $\HTtot$ cut is not trivially fulfilled,
and the distributions behave in a significantly different way from the inclusive case.
The shape of the various distributions can be understood in terms of
a hard-$W$ regime---where the $W$ boson carries
$\pT\gtrsim \HTtotcut/2$ and recoils against all jets---and a soft-$W$ 
regime---where the hardness of the event is driven by two
back-to-back jets with $\pT\gtrsim \HTtotcut/2$.
The transition between these two regimes is controlled by the
$W$-boson $\pT$, whose distribution features a sharp change around $p_{\rT,W}=\HTtotcut/2$.
When the $W$-boson $\pT$ enters the region below $\HTtotcut/2$
and approaches the soft regime, we observe
that the growth of the cross section is drastically reduced 
as compared to the hard-$W$ regime. 
This indicates that, at large $\HTtot$,   
hard dijet signatures with soft $W$ bosons tend to be favoured,
but $W$-boson emissions are distributed in a rather smooth way 
from low to high $\pT$.
This is consistent 
with the flatness of the $\pT$-distribution of
the second jet, which shares the 
first-jet recoil with the $W$ boson 
when the latter is not the hardest object.
As for the first-jet $\pT$, 
the peak at $\pT=\HTtotcut/2$ indicates that
the cross section is dominated by events where
the $W$ boson and the second jet recoil against the 
first jet, while the region 
$\HTtotcut/4<\pT<\HTtotcut/2$ corresponds to the hard-$W$
regime, where the two jets recoil against a $W$ boson
with $\pT>\HTtot/2$.

\begin{figure*}[t]
\centering
   \includegraphics[width=\relplotwidth\textwidth]{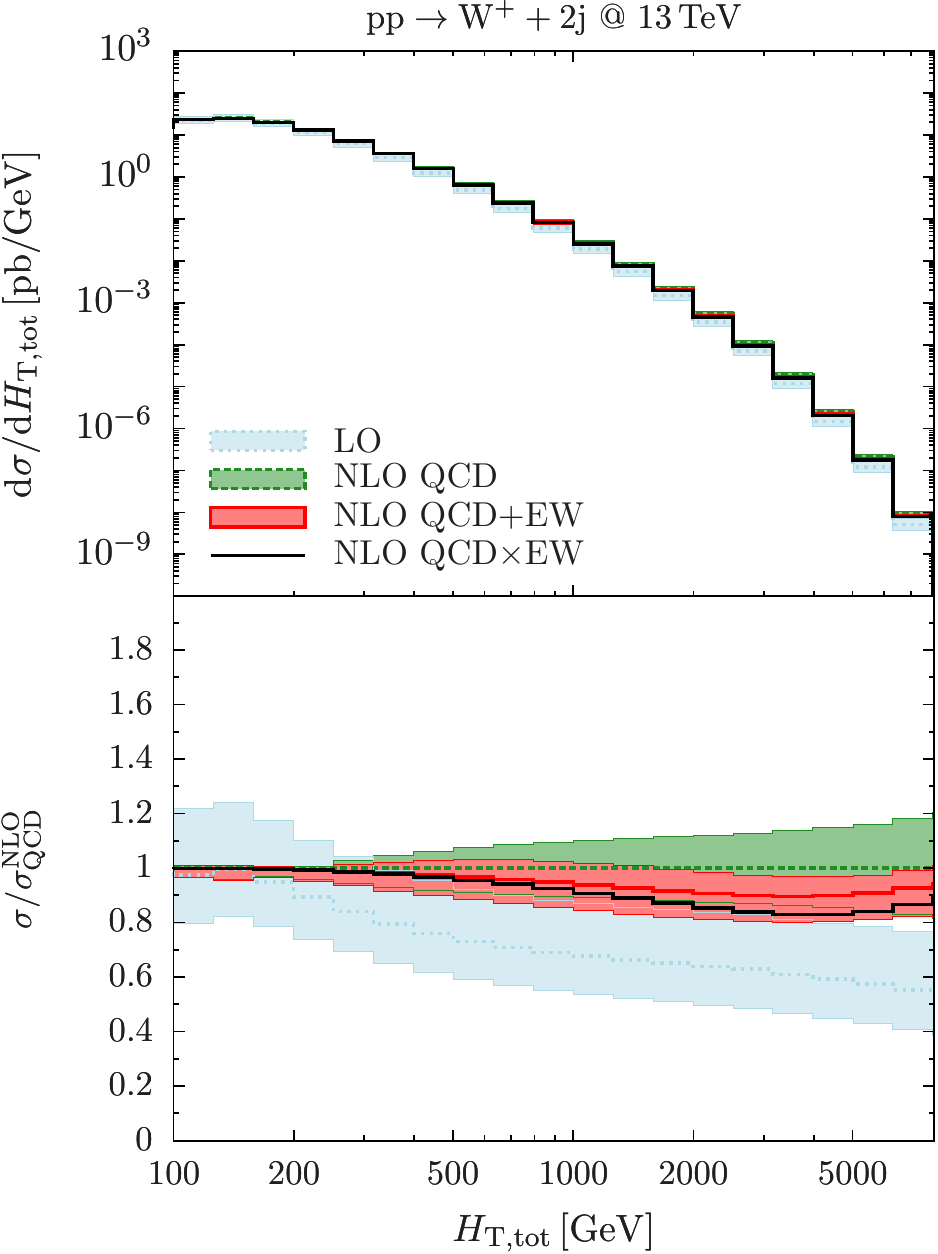}
      \qquad   
   \includegraphics[width=\relplotwidth\textwidth]{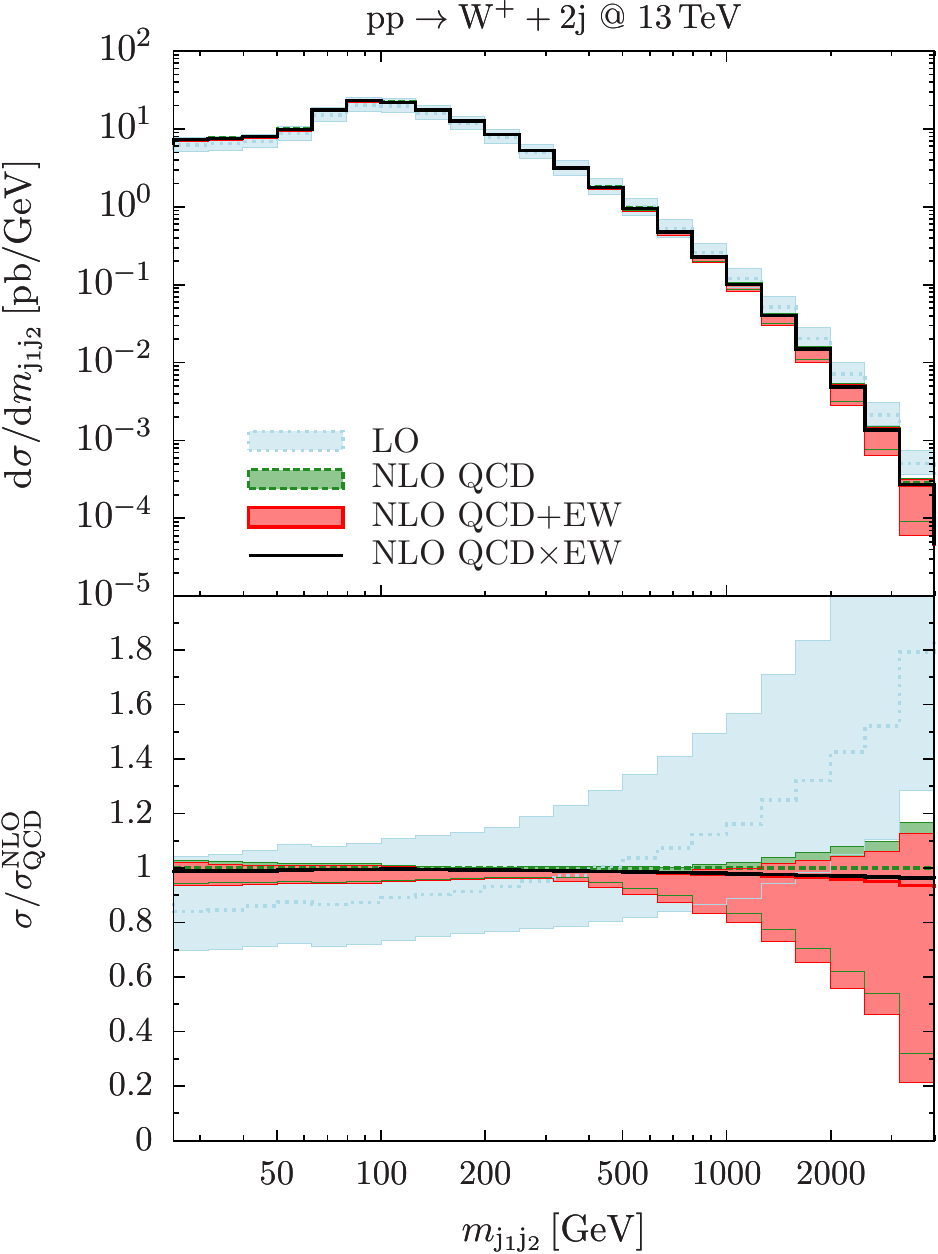}
\caption{
Distributions in $\HTtot$ (left)
and in the invariant mass of the first two jets (right)
for inclusive \mbox{$W^++2\jet$} production.
Curves and bands as in \reffi{fig:wj_pTall}.
 }
\label{fig:wjj_Httot_minv}
\end{figure*}

For what concerns the behaviour of NLO corrections,
we observe that, in contrast to the inclusive case,
in the $\pT$-region sensitive to the $\HTtotcut$, i.e.~below
$\HTtotcut/2$, 
the various $\pT$ distributions involve 
very strong NLO QCD corrections of $\ord(100\%)$ and correspondingly
large NLO scale uncertainties. This can be attributed to the fact that the 
cut on $\HTtot$ can be efficiently saturated through 
QCD real-emission processes. Again, the fact that extra jet emission 
is effectively described at LO accuracy leads to underestimated
EW correction effects.
These features are clearly visible in the transition 
region around $\HTtotcut/2$, where 
the relative QCD+EW correction jumps by about a factor two
as compared to the smooth QCD$\times$EW prediction.
Apart from the problematic interplay of NLO QCD and EW corrections, the latter grow continuously with 
$p_{\rT,W}$ as expected from EW Sudakov logarithms.
The opposite trend in the second-jet $\pT$ distribution is 
due to the fact that, below $\HTtotcut/2$,
large $p_{\rT,\jet_2}$ corresponds to small
$p_{\rT,W}$ and vice versa.
Let us point out that the behaviour of the $W$-boson $\pT$ distribution in
the right plot of \reffi{fig:wjj_pTall} is relevant for BSM searches
that require very large $\HTtot$ without a correspondingly high cut on the
leptonic and/or missing transverse energy.  In this case, the $W+2\jet$
background is clearly dominated by the region $p_{\rT,W}<\HTtotcut/2$, where the $\HTtot$
cut leads to a bad perturbative QCD behaviour. This calls for higher-order
corrections to $W+3\jet$ production and, ultimately, for matching to the parton
shower and multi-jet merging at NLO.
 
\reffi{fig:wjj_dphi} presents the distribution in the 
azimuthal separation between the first two jets, $\deltajj$, i.e.~the variable
used to isolate hard dijet configurations 
in \refse{se:wjsec}. The left plot shows that 
inclusive $\PWp+2$-jet production
is dominated by back-to-back dijet configurations,
while the collinear peak around $\pi/6$ remains 
clearly subdominant. For such an inclusive observable,
NLO EW corrections are essentially negligible, 
and QCD corrections are rather small and stable.
In presence of a cut $\HTtot>2$\;TeV (right plot),
$\deltajj$ allows to discriminate the hard-$W$ regime---where 
all jets are in the hemisphere opposite to the $W$ boson---from 
the soft-$W$ regime---where the jets are back-to-back.
As expected, the largest EW corrections are observed in the
hard-$W$ regime (small $\deltajj$), where they amount to about 
$-20\%$, consistently with 
the inclusive result at $\pT=1\,\TeV$.
In the soft-$W$
regime (large $\deltajj$) NLO EW effects are much less pronounced.
This is in part due to the fact 
that hard jets receive smaller EW Sudakov corrections
as compared to hard $W$ bosons. Moreover, the presence of 
$\ord(100\%)$ QCD corrections induces a further strong suppression of NLO EW effects
in this region.

The distribution in $\HTtot$, displayed in \reffi{fig:wjj_Httot_minv},
provides further evidence of the poor stability of this observable with
respect to QCD radiation effects.  In the tail, NLO QCD and EW corrections
approach the 100\% and 10\% level, respectively, and the QCD$\times$EW curve
suggests that the importance of NLO EW corrections is underestimated by a
factor 2 in the NLO QCD+EW prediction.
Finally, the distribution in the invariant mass of the 
first two jets (\reffi{fig:wjj_Httot_minv}, right)
behaves in a very different way: NLO EW corrections
turn out to be very small and almost completely independent 
of the dijet mass, even in the multi-TeV range.
This is explained by the fact that, in absence of an explicit
high-$\pT$ requirement, the region of large dijet mass
is dominated by $t$-channel production at small 
$\pT$. Note that in the tail of the $\minvjj$
distribution QCD corrections become large, which results in 
sizable scale uncertainties.

In summary, NLO QCD+EW predictions for $pp\to W+2\jet$ show a 
significantly improved perturbative stability as compared to the 
$W+1\jet$ case. Nevertheless, the strong sensitivity of 
certain observables---in particular $\HTtot$---to NLO QCD radiation
calls for the extension of NLO QCD+EW 
calculations to $W+3$-jet production.

\subsection[$W^++3$ jets]{$\mathbf{W^++3}$ jets}
\label{se:wjjjsec}

\begin{table}[t]
\renewcommand{\arraystretch}{1.5}%
\setlength{\tabcolsep}{1.2ex}%
\begin{center}
\begin{tabular}{lllllll}\hline
\input{XStable.wjjj.tex}
\end{tabular}
\caption{
Integrated $pp\to W^++3\jet$ 
cross sections with inclusive cuts~\refeq{eq:JETcuts}
and in presence of additional cuts. 
Born cross sections ($\sigma^\LO$) include only the leading QCD contributions of
$\ord(\alphaS^3\alpha)$.
}
\label{tab:XSwjjj}
\end{center}
\end{table}

Numerical results for $pp\to W^++3\jet$ at NLO QCD+EW
are presented in \reffis{fig:wjjj_pTall}--\ref{fig:wjjj_dphi} and in \refta{tab:XSwjjj}.
At variance with the $W+2\jet$ case, for one-particle 
inclusive $\pT$ distributions, shown in \reffi{fig:wjjj_pTall},  
we find stable NLO QCD predictions only for the $W$ boson and 
the third jet, while the distributions in the $\pT$ of the first two jets receive sizable negative QCD corrections 
in the region around 1\,TeV. This suggests that the QCD 
scale choice~\refeq{eq:RFscales} might be suboptimal for 
$W+3\jet$ final states, and, in order to achieve
better perturbative convergence, alternative dynamical scales should be considered.
For instance, \reffi{fig:wjjj_pTall} indicates that 
using $\mu_0=\hat H_{\mathrm{T}}$ instead of 
$\mu_0=\hat H_{\mathrm{T}}/2$, which corresponds to the lower boundary 
of the LO uncertainty band, would already improve the convergence in a significant way.
However, in this paper we will stick to the
standard choice~\refeq{eq:RFscales} that was used in the 
most recent ATLAS analysis~\cite{Aad:2014qxa},
and we defer a detailed study of alternative scale choices to
a future publication.

As far as predictions obtained at the central scale 
$\mu_0=\hat H_{\mathrm{T}}/2$ are concerned, 
NLO EW corrections in \reffi{fig:wjjj_pTall} 
are well behaved: the tails of all $\pT$
distributions feature the expected EW Sudakov suppression, and the quantitative
impact of the corrections is rather consistent with what is observed in the
$W+2\jet$ case.  For the first- and second-jet $\pT$-distributions,  
the QCD$\times$EW curve suggests that 
the NLO QCD+EW approximation 
might overestimate EW correction effects,
as a result of the negative QCD corrections.
For what concerns scale variations, the fact that 
NLO QCD and NLO EW corrections are both very large 
leads to a very strong scale dependence at high $\pT$.  This illustrates, once again, 
that the optimal convergence of QCD predictions plays a key role for the
stability of NLO QCD+EW predictions. 

A very good perturbative convergence is found in \reffi{fig:wjjj_Httot_minv} for the 
$\HTtot$ distribution (left). In presence of three associated 
jets this important observable receives fairly small QCD corrections.
The NLO QCD+EW approximation can thus be regarded as a reliable description 
of EW correction effects, which reach $-20\%$ at $\HTtot=4\,\TeV$.
For the distribution in the invariant mass of the first two jets,
shown in the right plot of \reffi{fig:wjjj_Httot_minv},
a similar picture as in the case of $W+2\jet$ production emerges. In particular
EW corrections remain negligible in the entire $\minvjj$ range.

Also the distribution in the azimuthal angular separation of the two hardest jets,
shown in \reffi{fig:wjjj_dphi}, behaves in a fairly similar way as for
$W+2\jet$ production. In particular, a cut $\HTtot > 2\,\TeV$ (right plot) 
induces EW corrections around $-20\%$ in the hard-$W$ regime (small $\deltajj$) 
while in the soft-$W$ regime (large $\deltajj$)
NLO EW effects are clearly less pronounced.

Numerical results for $pp\to W^++3\jet$ cross sections with
different cuts are collected in \refta{tab:XSwjjj}.

\begin{figure*}[t]
\centering
   \includegraphics[width=\relplotwidth\textwidth]{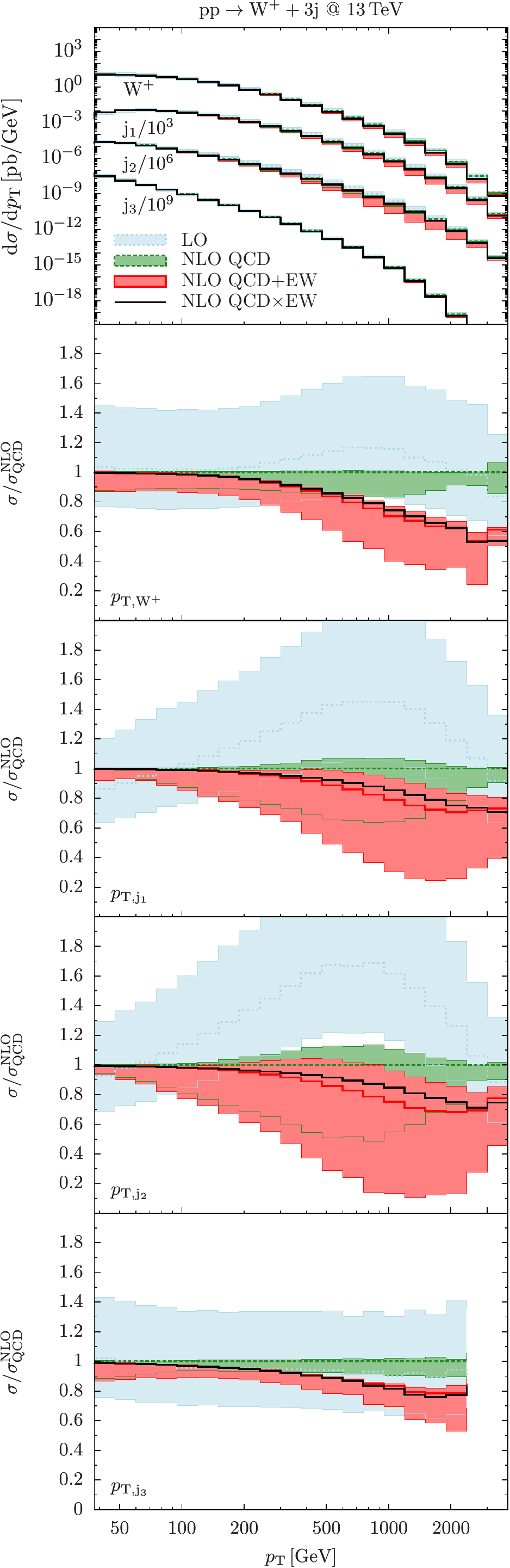}
\caption{
Distributions in the transverse momenta of the 
$W$ boson and of the first three jets
for inclusive \mbox{$W^++3\jet$} production. 
The distributions in the $n$-th jet $\pT$ are rescaled by factors $10^{-3n}$.
Curves and bands as in \reffi{fig:wj_pTall}.
 }
\label{fig:wjjj_pTall}
\end{figure*}

\begin{figure*}[t]
\centering
   \includegraphics[width=\relplotwidth\textwidth]{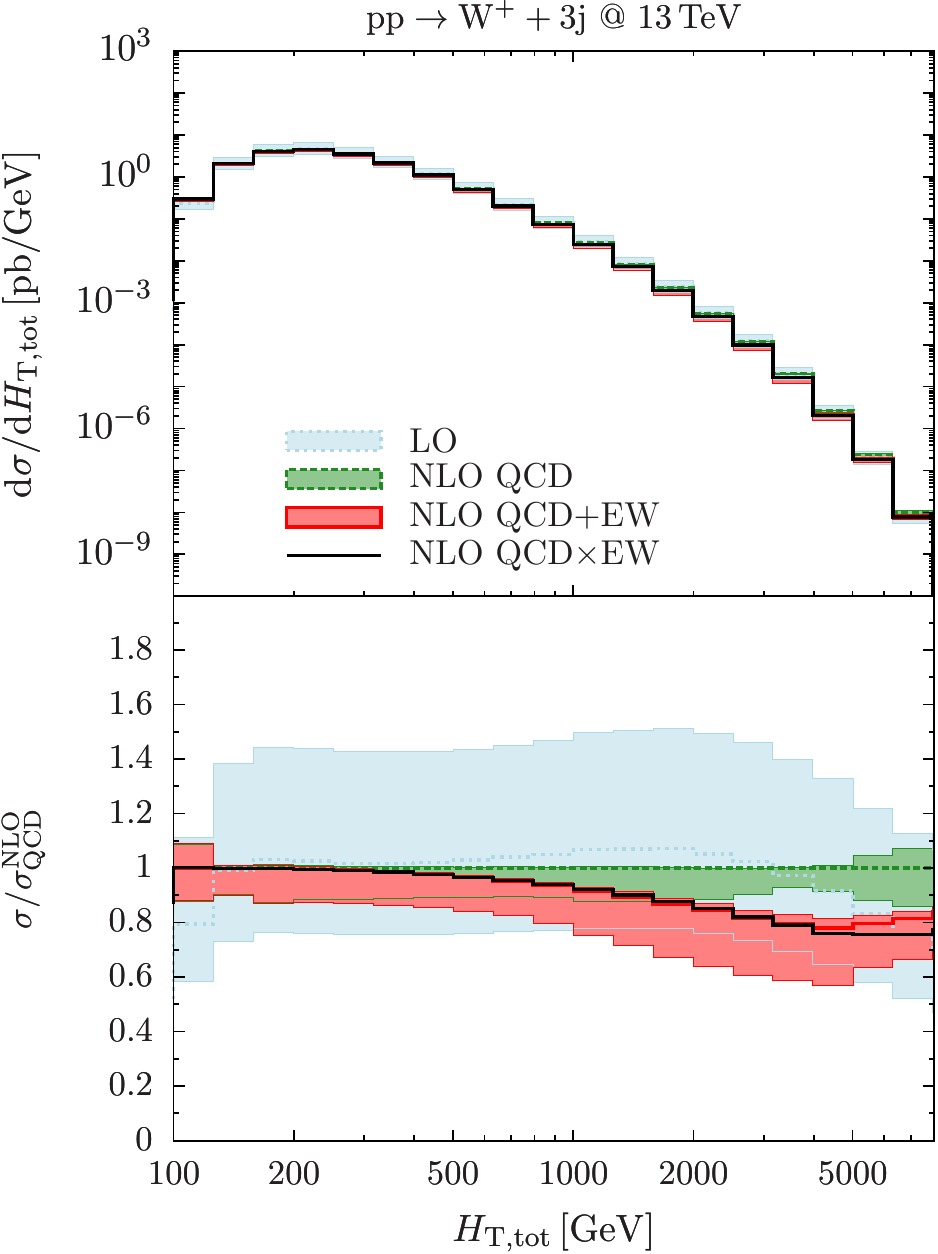}
      \qquad   
   \includegraphics[width=\relplotwidth\textwidth]{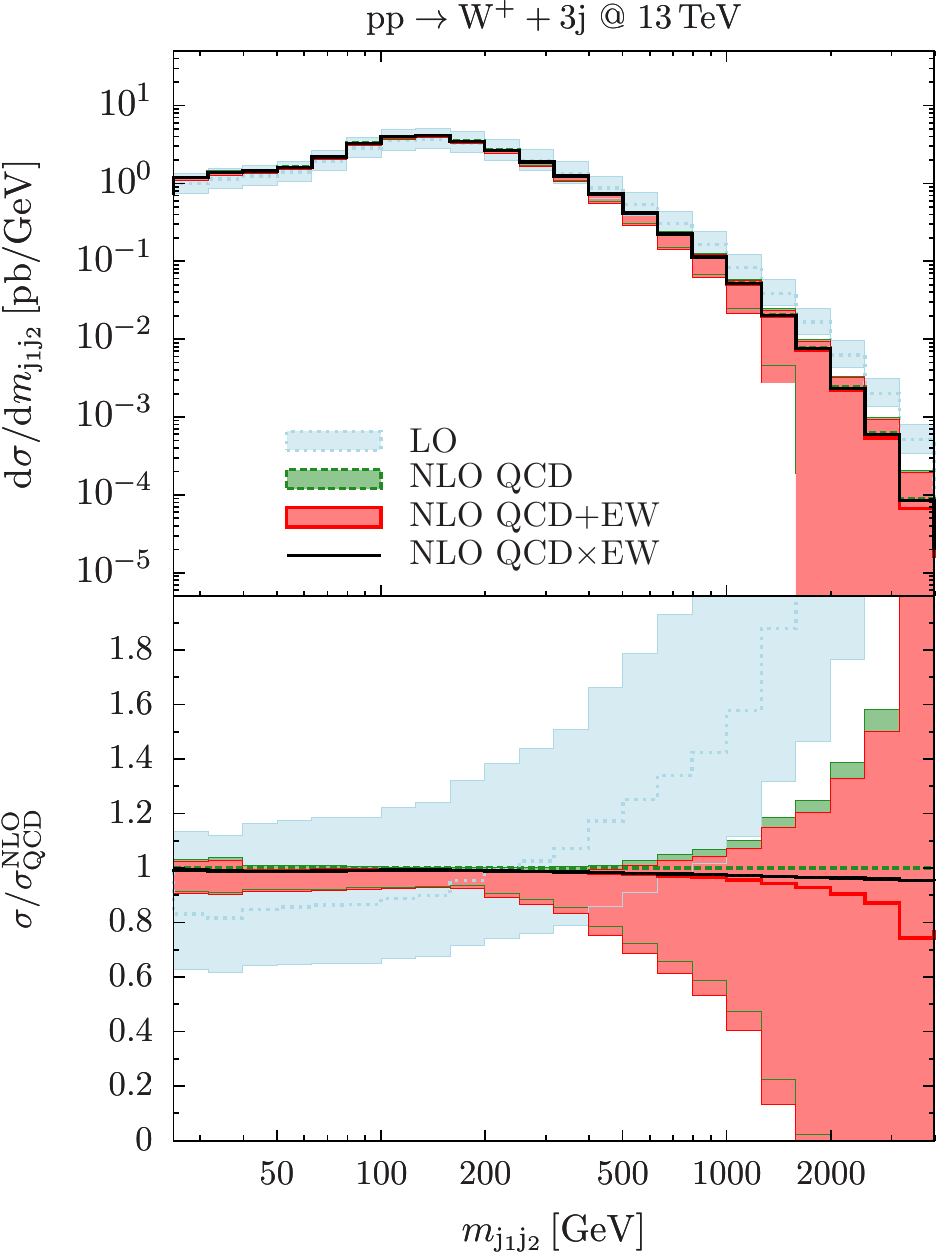}
\caption{
Distributions in $\HTtot$ (left)
and in the invariant mass of the first two jets (right)
for inclusive \mbox{$W^++3\jet$} production.
Curves and bands as in \reffi{fig:wj_pTall}.
 }
\label{fig:wjjj_Httot_minv}
\end{figure*}

\begin{figure*}[t]
\centering
   \includegraphics[width=\relplotwidth\textwidth]{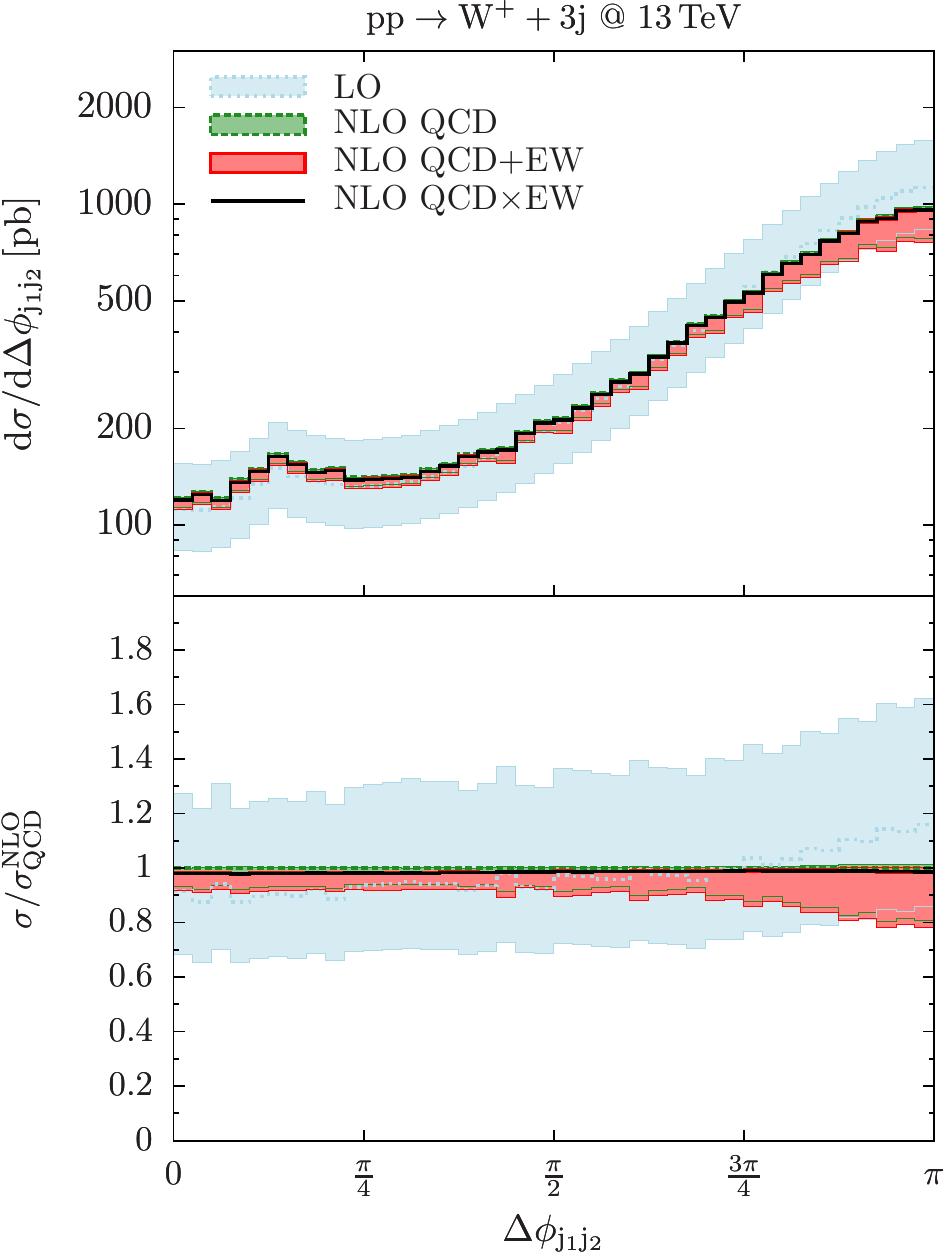}
      \qquad   
   \includegraphics[width=\relplotwidth\textwidth]{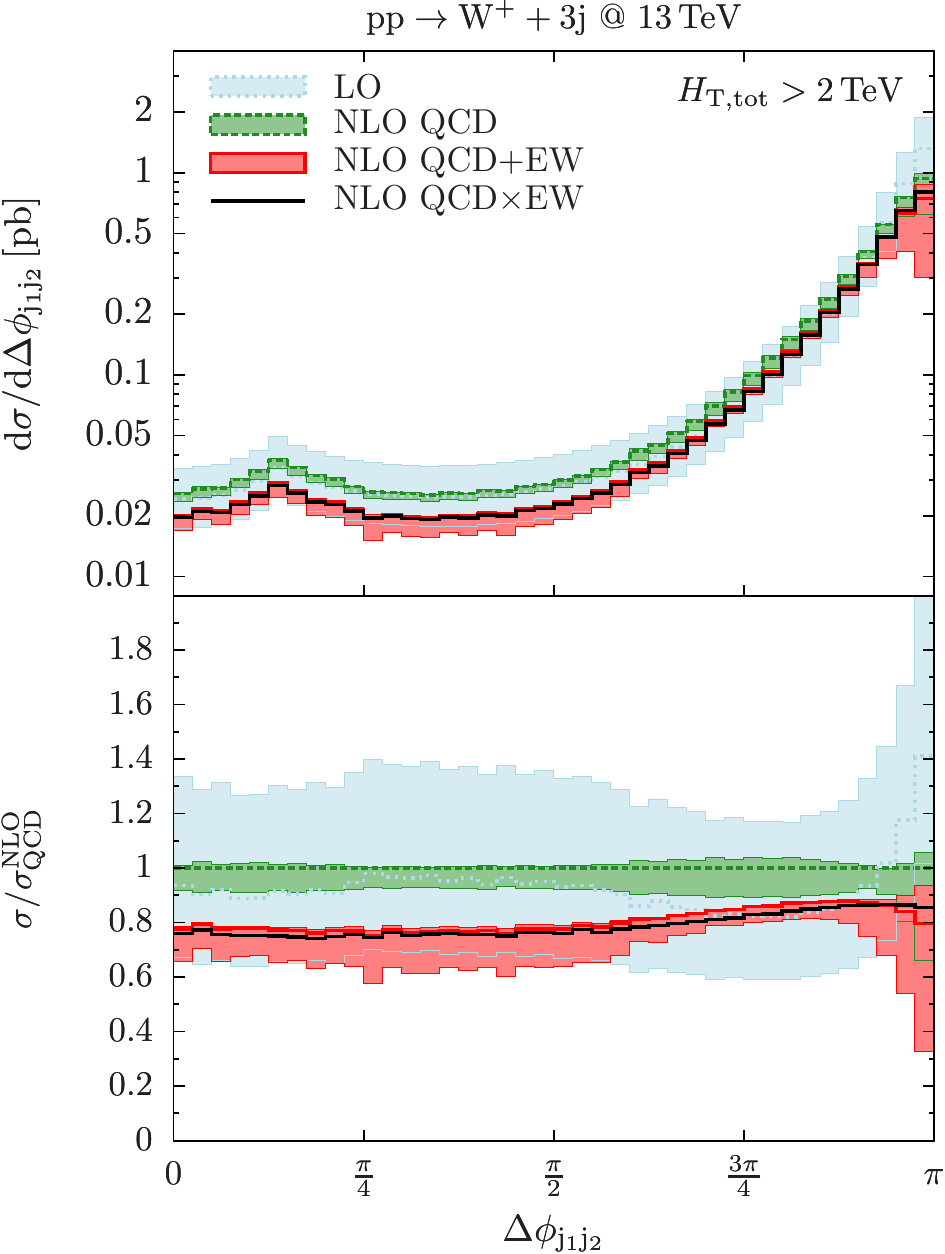}
\caption{
Distributions in the azimuthal separation of the 
first two jets for inclusive (left) \mbox{$W^++3\jet$} production
and with a cut $\HTtot > 2\,\TeV$ (right).
Curves and bands as in \reffi{fig:wj_pTall}.
 }
\label{fig:wjjj_dphi}
\end{figure*}

\clearpage

\subsection{Subleading and photon-induced Born contributions}
\label{se:lomixsec}

In the following we briefly discuss
the numerical impact of subleading and photon-induced Born
contributions to the production of a $W^+$ boson in association with $n=1,2,3$ jets.

As discussed in section~\ref{se:fourquarks}, 
the production of $W+2,3$\;jets receives pure EW contributions of 
$\ord(\alphaS^{n-2}\alpha^3)$ as well as contributions of $\ord(\alphaS^{n-1}\alpha^2)$ 
from interferences of QCD- and EW-type diagrams in the four-quark channels.
In addition, the production of $W+1,2,3$\;jets 
can proceed through different $\gamma$-induced processes, as discussed in
section~\ref{sec:gamma-induced}. 
The pure EW contributions and the resonant
$\gamma\gamma$-induced processes of pure EW-type at $\ord(\alphaS^{n-2}\alpha^3)$
involve physical Z, W, and top-quark resonances, which are regularized by 
their corresponding physical decay widths.\footnote{We use the following values of the relevant particle widths, which are calculated at LO from the parameters stated in \refse{se:setup}, 
$\Gamma_{\mathrm{W}}=2.04544~\GeV$,
$\Gamma_{\mathrm{Z}}=2.44408~\GeV$, and
$\Gt=1.50175~\GeV$.} The impact of the resulting violation of gauge invariance due to the approximation of an
 on-shell W was found to be at the small percent level of the respective contribution. A consistent gauge-invariant treatment for these processes at NLO will require a full SM calculation with decays.

Results for integrated cross sections and distributions 
are listed in \refta{tab:XSBborn} and \reffis{fig:bornwj_pTall}--\ref{fig:bornwjjj_pTall}, respectively.
As far as subleading Born contributions are concerned, in the 
integrated $W+2,3$\;jet cross sections
EW--QCD mixed Born effects of $\ord(\alphaS^{n-1}\alpha^2)$ are at the
permil level, while the pure EW contributions of $\ord(\alphaS^{n-2}\alpha^3)$ 
are one order of magnitude larger. This is due to the presence of resonances that 
correspond to di-boson and single-top production (with hadronic decays of a $W^-$ or $Z$ boson).
The relative importance of EW--QCD Born interference terms grows with 
the jet-$\pT$, and at 1\;TeV these contributions reach 11\%\;(14\%) in $pp\to W^++2\jet\;(3\jet)$.
This enhancement can also be understood as a PDF effect where
the contribution of the four-quark channel increases over the two-quark channel due
to a relative increase of the quark PDFs over the gluon PDFs for large $x$.
In certain phase-space regions, EW--QCD interference contributions
become negative.

Photon-induced effects induce only permil-level contributions to the various inclusive cross sections.
Still, the increasing importance of the photon density at high
Bjorken $x$ leads to an enhancement of $\gamma$-induced cross sections at large
$W$-boson transverse momenta.  
At $\pTwp >1~$TeV, the dominant $\ord(\alphaS^{n-1}\alpha^2)$ photon-induced contributions to
$W^++1,2,3$\;jets are around 8\%, 4\%, and 2\%, respectively, 
and their magnitude grows extremely rapidly up to $\ord(100\%)$
in the multi-TeV range. In this respect, one should keep in mind that
the photon PDF is still very poorly constrained in this regime~\cite{Ball:2013hta}, and
$W$+jets measurements at large transverse momenta
might provide useful input for a better determination of the photon PDF.
We observe that $\gamma\gamma$-induced processes are strongly suppressed in the entire phase space.

%

\begin{table}[t]
\renewcommand{\arraystretch}{1.5}%
\setlength{\tabcolsep}{1.2ex}%
\begin{center}
\begin{tabular}{cllllll}\hline
\input{XStable.born_new.tex}

\end{tabular}
\caption{
Integrated cross sections for $pp\to W^++n$\,jet 
production with $n=1,2,3$ with inclusive cuts~\refeq{eq:JETcuts}
and in presence of additional cuts:
$\ord(\alphaS^{n-1}\alpha^2)$ mixed EW--QCD, $\ord(\alphaS^{n-2}\alpha^3)$ pure EW,   
photon--proton induced Born contributions of $\ord(\alphaS^{n-1}\alpha^2)$ \dots $\ord(\alpha^{n+1})$
and photon--photon induced contributions of $\ord(\alphaS^{n-2}\alpha^3)$.
The various contributions are normalised to corresponding NLO QCD predictions.
}
\label{tab:XSBborn}
\end{center}
\end{table}

\def\relplotwidthBorn{0.44}
\begin{figure*}[t]
\centering
   \includegraphics[width=\relplotwidthBorn\textwidth]{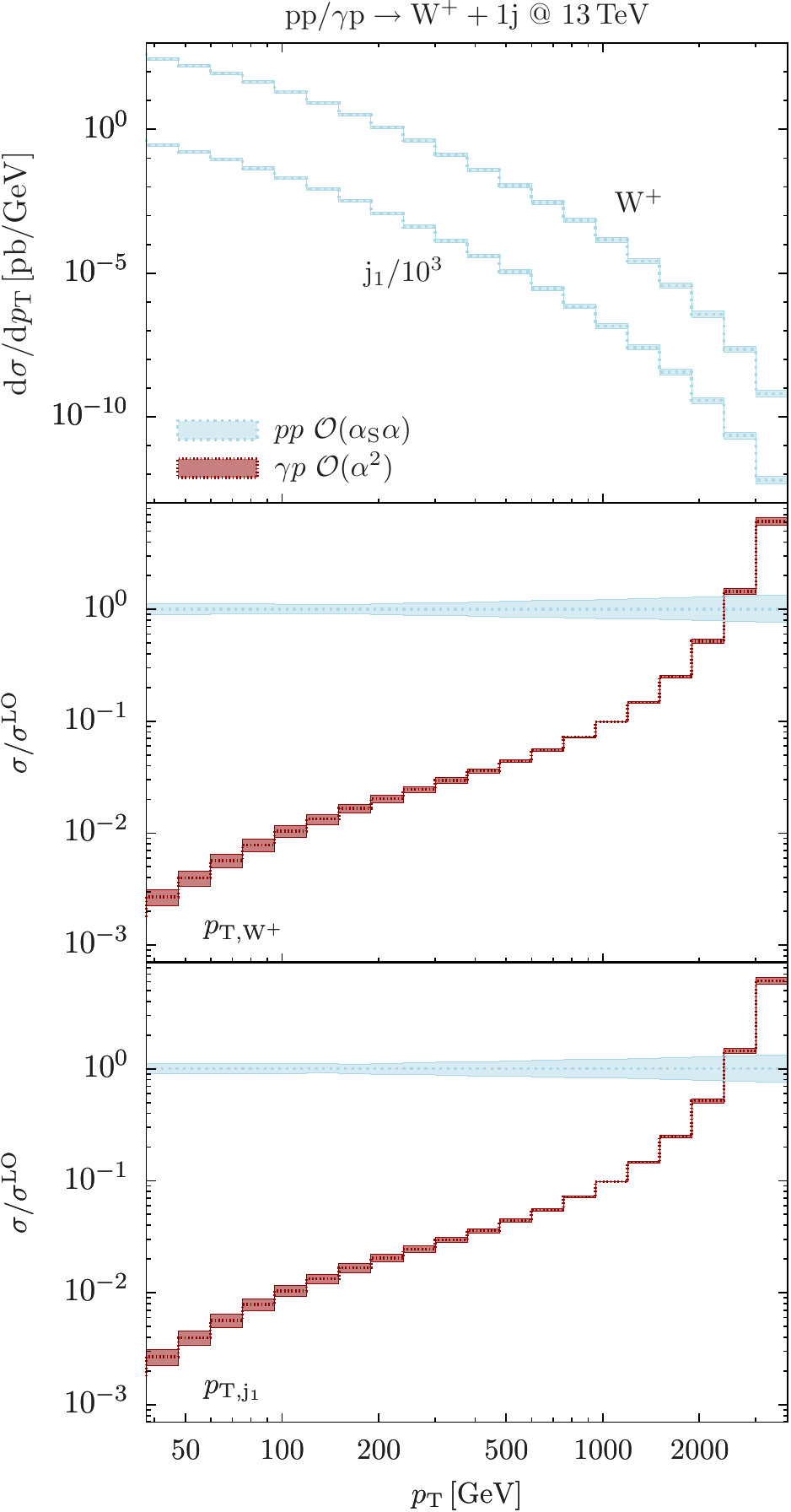}
\caption{
Distributions in the transverse momenta of the $W$ boson and of the first jet
for inclusive \mbox{$W^++1\jet$} production. In the upper panel absolute predictions for the LO Born contribution at $\ord(\alphaS\alpha)$ (light blue) are shown.
The distribution in $\pTjone$ is
rescaled by a factor $10^{-3}$.
In the lower panels photon--proton induced predictions at $\ord(\alpha^2)$ (dark red) are shown relative to the LO Born contribution.
The bands correspond to scale variations, 
and in the case of ratios only the numerator is varied.
}
\label{fig:bornwj_pTall}
\end{figure*}

\def\imagetop#1{\raisebox{-\height}{#1}}
\begin{figure*}[t]
\centering
\includegraphics[width=\relplotwidthBorn\textwidth]{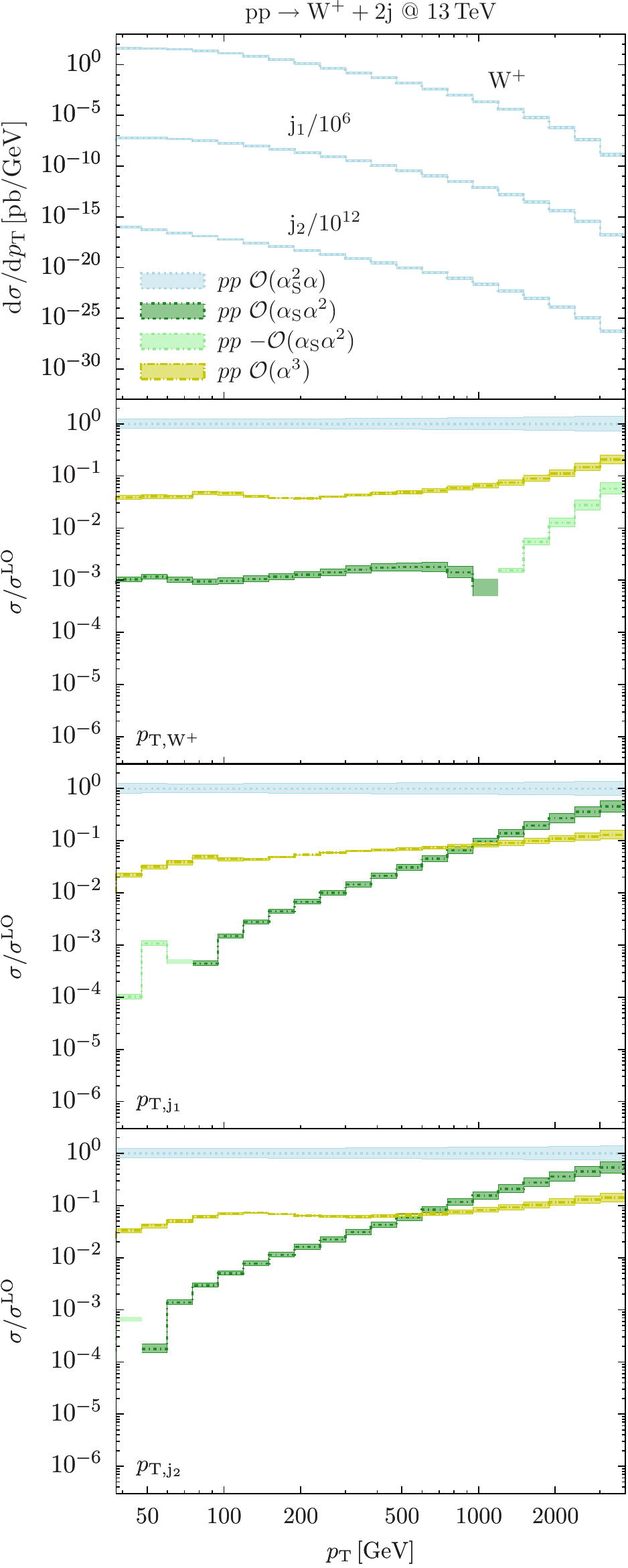}
\qquad
\includegraphics[width=\relplotwidthBorn\textwidth]{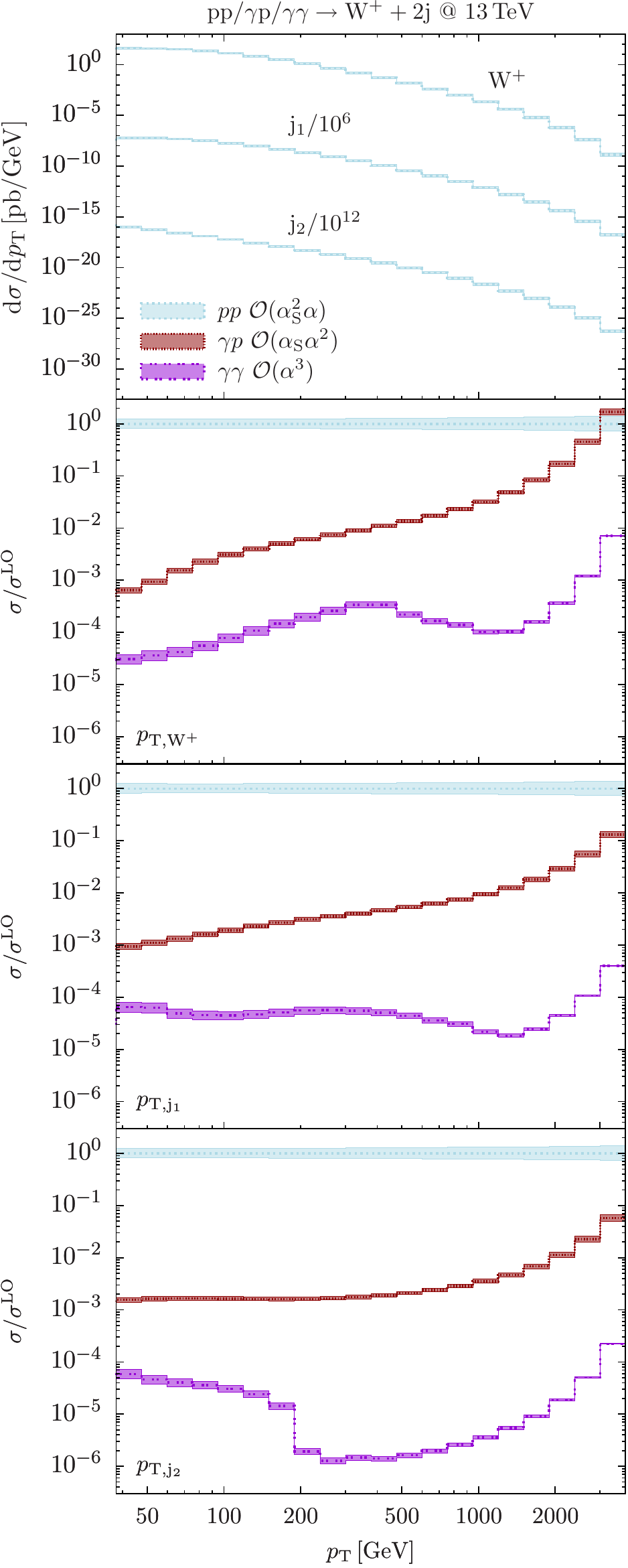}
\caption{
Distributions in the transverse momenta of the $W$ boson and of the first two jets
for inclusive \mbox{$W^++2\jet$} production. In the upper panels absolute predictions for the LO Born contribution at $\ord(\alphaS^2\alpha)$ (light blue) are shown.
The distributions in the $n$-th jet are
rescaled by a factor $10^{-6n}$.
In the lower panels, on the left, proton--proton induced mixed EW--QCD predictions at $\ord(\alphaS\alpha^2)$ (dark/light green depending on the sign) and resonant EW predictions at $\ord(\alpha^3)$  (olive) relative to the LO Born contribution are shown. On the right, photon--proton and photon--photon induced predictions at $\ord(\alphaS\alpha^2)$ (dark red) and $\ord(\alpha^3)$ (violet), respectively, are shown relative to the LO Born contribution.
Bands as in \reffi{fig:bornwj_pTall}.
}
\label{fig:bornwjj_pTall}
\end{figure*}

\def\imagetop#1{\raisebox{-\height}{#1}}
\begin{figure*}[t]
\centering
\imagetop{\includegraphics[width=\relplotwidthBorn\textwidth]{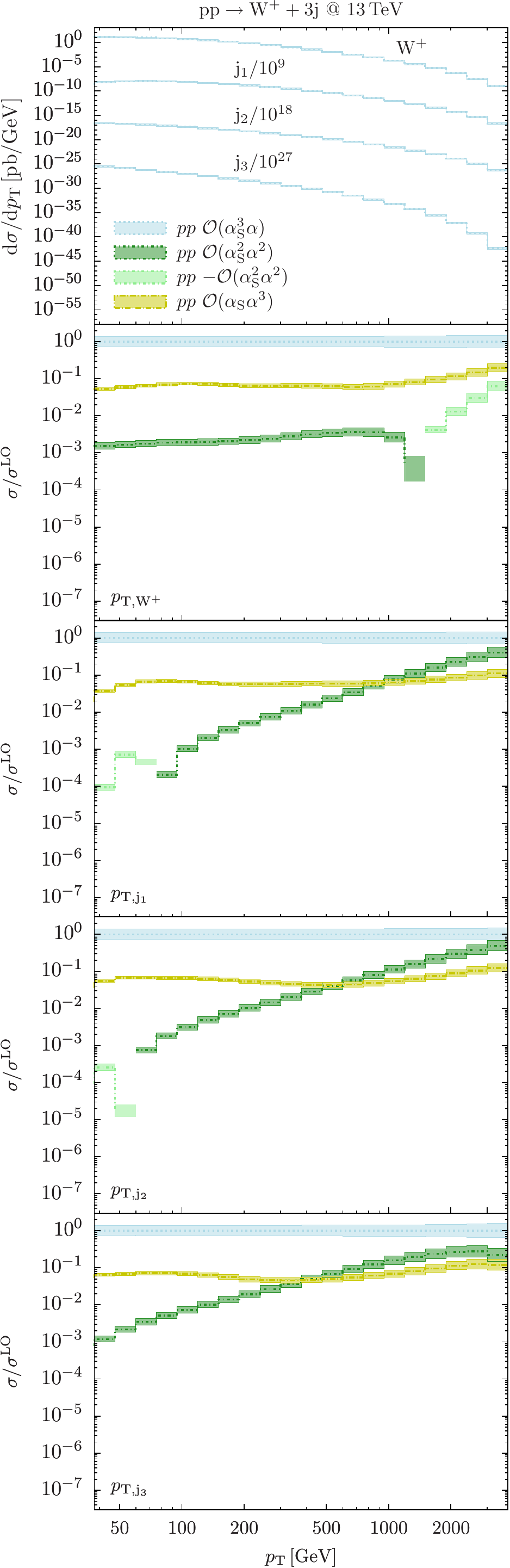}}
\qquad
\imagetop{\includegraphics[width=\relplotwidthBorn\textwidth]{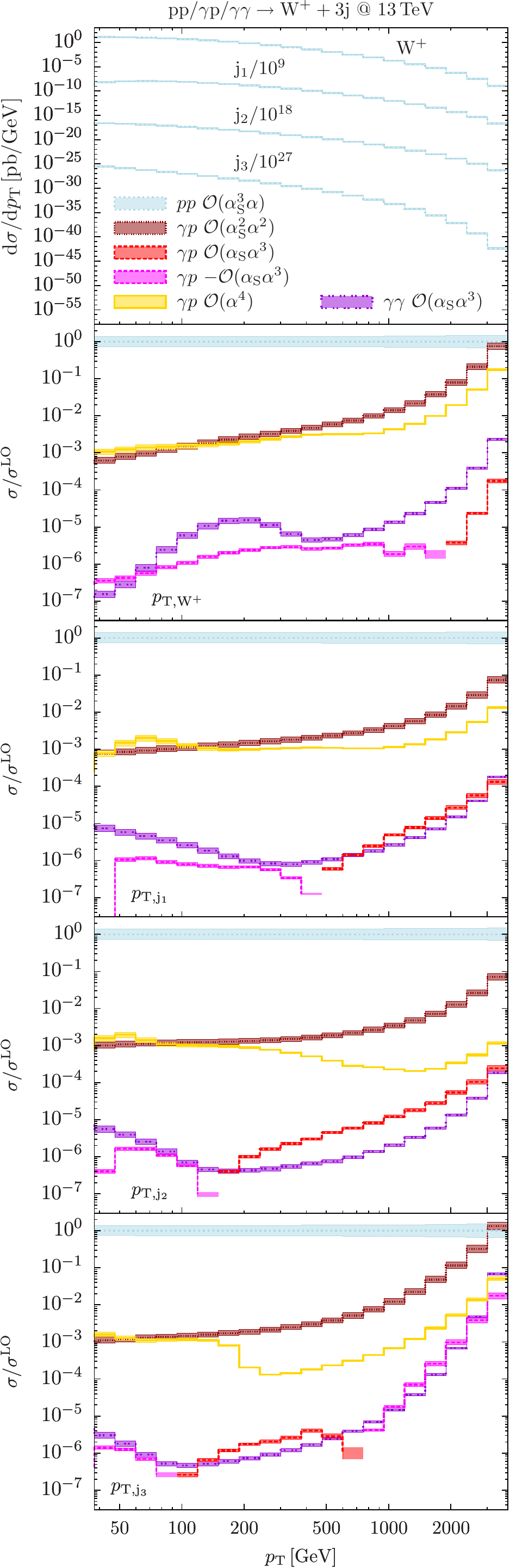}}
\caption{
Distributions in the transverse momenta of the $W$ boson and of the first $n$ jets
for inclusive \mbox{$W^++3\jet$} production. In the upper panels absolute predictions for the LO Born contribution at $\ord(\alphaS^3\alpha)$ (light blue) are shown.
The distributions in the $n$-th jet are
rescaled by a factor $10^{-9n}$.
In the lower panels, on the left, proton--proton induced mixed EW--QCD predictions at $\ord(\alphaS^2\alpha^2)$ (dark/light green depending on the sign) and resonant EW predictions at $\ord(\alphaS\alpha^3)$  (olive) relative to the LO Born contribution are shown. 
On the right, predictions for photon--proton induced production at $\ord(\alphaS^2\alpha^2)$ (dark red), $\ord(\alphaS\alpha^3)$ (red/magenta depending on the sign) and $\ord(\alpha^4)$ (yellow) are shown together with photon--photon induced production at $\ord(\alphaS\alpha^3)$ (violet) relative to the LO Born contribution.
Bands as in \reffi{fig:bornwj_pTall}.
}
\label{fig:bornwjjj_pTall}
\end{figure*}
\clearpage

\section{Summary and conclusions}
\label{se:conclusions}

The calculation of electroweak corrections is a central prerequisite for
precision tests of the Standard Model and for new-physics searches at the
energy frontier.  In particular, the strong impact of EW corrections on a
wide range of processes and observables at the TeV scale motivates the
extension of automated NLO generators from the QCD to the EW sector of the
Standard Model.

In this context, a systematic bookkeeping of all possible EW--QCD
interference terms at NLO is needed.  Standard NLO EW corrections of a
certain order $\alphaS^{n}\alpha^{m+1}$ arise via insertion of virtual or
real {\it electroweakly} interacting particles in squared tree amplitudes 
of order $g_\rS^{n} e^m$.  
But NLO EW corrections at the same order $\alphaS^{n}\alpha^{m+1}$
can also arise via insertion of virtual or real {\it strongly} interacting 
partons in interference terms between tree amplitudes of order 
$g_\rS^{n}e^m$ and $g_\rS^{n-2}e^{m+2}$.
In general, in order to obtain infrared-finite cross sections, 
all possible EW--QCD interference terms that contribute at a given 
order $\alphaS^{n}\alpha^{m+1}$ need to be included.

The cancellation of infrared singularities at NLO EW
requires the real emission of QED and possibly
also QCD partons, while the factorisation of initial-state
collinear singularities requires QED effects in the PDFs, including a photon density. 
Moreover, due to
the interplay of QED and QCD IR singularities associated with collinear
photon--quark and photon--gluon pairs inside jets, at NLO EW the infrared-safe
definition of jet observables and the separation of hard jets from hard
photons is nontrivial.
In this respect we have discussed a theoretical definition of jets based on
democratic jet clustering in combination with a photon--jet separation 
formulated in terms of the photon-energy fraction inside jets. In particular, 
we have shown that the cancellation of QED and QCD infrared
singularities can be achieved by a simple recombination prescription for
photon--quark pairs in a way that provides an excellent approximation
to a rigorous jet definition based on quark fragmentation functions.

The first key result presented in this paper is the complete automation of
NLO QCD+EW calculations within the \OpenLoops one-loop generator in
combination with the Monte Carlo programs \Munich and \Sherpa.  The
\OpenLoops program generates all relevant matrix-element
ingredients, i.e.~one-loop amplitudes, tree amplitudes for Born and
bremsstrahlung contributions, as well as colour-, charge-,
gluon-helicity and photon-helicity correlations 
that are needed for IR subtractions.  Tree and one-loop matrix elements 
can be generated at any desired order $\alphaS^n\alpha^m$, including all
relevant EW--QCD interferences, and full NLO 
Standard Model calculations are also possible.  
To automate one-loop EW calculations, all EW Feynman rules
have been implemented in the framework of the numerical \OpenLoops recursion
and complemented by counterterms associated with $R_2$ rational parts and
with the on-shell renormalisation of UV singularities.

All complementary tasks, i.e.~the bookkeeping of partonic processes, the
subtraction of IR singularities, and phase space integration, have been
automated within \Munich and \Sherpa.  These two alternative Monte Carlo
frameworks are based on the dipole-subtraction formalism, whose
implementation had to be extended from NLO QCD to NLO QED.  In combination
with \OpenLoops, these tools automate the full chain of operations---from process 
definition to 
collider observables---that enter NLO QCD+EW simulations 
at parton level.
As far as the efficiency of the simulations is concerned, 
the fact that \OpenLoops can evaluate one-loop EW matrix elements 
at a similarly high speed as in the QCD case opens the route
to NLO QCD+EW studies for a very wide range of processes, up to high particle
multiplicity.

As a first nontrivial application, we have presented NLO QCD+EW predictions
for $W$-boson production in association with one, two, and three jets at the
13\;TeV LHC.  This represents the first NLO EW calculation for an LHC
process with more than two jets and for $W+n$-jet production with 
$n=2$ and $n=3$. 
Since the EW corrections to $W+$jets production are
expected to be almost independent of the $W$-boson charge~\cite{Kuhn:2007cv},
we have restricted ourselves to the case of positively charged
$W$ bosons.
Our predictions include all $\ord(\alphaS^{n+1}\alpha)$ and $\ord(\alphaS^n\alpha^2)$ contributions to
$pp\to W^++n$\,jets with stable on-shell $W$ bosons. At this order, 
reconciling the on-shell approximation with the
presence of resonant $W$ propagators at amplitude level
is nontrivial. However, the fact that 
resonant amplitudes contribute only through interference with non-resonant
ones allowed us to regularise the poles of the relevant propagators 
with a technical width parameter, in a way 
that corresponds to a smooth and numerically negligible 
deformation with respect to the gauge-invariant on-shell limit.
Using this approach we are going to implement $W$ boson decays
in the narrow-width approximation in the near future.

We have presented various predictions for $W+$\,multijet cross
sections and distributions.  For $pp\to W+1\jet$, our NLO EW results
confirm the well known Sudakov behaviour.  The $W$-boson $p_\rT$
distribution receives large negative EW corrections, which reach $-40\%$ at
2\;TeV and are accompanied by NLO QCD corrections of similar size and
opposite sign.  Here, and in various other observables, the simultaneous
presence of large EW and QCD corrections implies 
a sizable uncertainty related to the unknown
EW$\times$QCD corrections of NNLO type.  For the 
distributions in the $p_\rT$ of the first jet and in $\HTtot$ this problem becomes
dramatic: in the TeV region NLO QCD corrections reach a factor ten, and the mere 
inclusion of NLO EW corrections at $\ord(\alphaS\alpha^2)$ 
is completely insufficient.
Actually, in the multi-TeV region we observe that NLO EW effects lead to a sizable positive correction,
which arises from mixed EW--QCD real-emission contributions, while the expected 
Sudakov correction is completely suppressed.

As is well known, the explosion of NLO QCD corrections at high jet-$p_\rT$
is due to the fact that $W+$\,jets production with a very hard jet is
dominated by $W+$\,multijet configurations where the $W$ boson 
tends to be produced at moderate transverse momentum, while the transverse energy of the
event is predominantly carried by two (or more) hard jets that recoil
against each other.  It is thus clear that, for a meaningful description of
the hard-jet regime, NLO EW corrections must be extended to
$W+n$-jet production with $n\ge 2$.

For $pp\to W^++2$\,jets, although $\HTtot$ remains quite sensitive to extra QCD
radiation, the distributions in the $W$-boson and in the jet 
transverse momenta feature a good stability with respect to NLO QCD effects.  
Thus NLO QCD+EW predictions start providing a reliable theoretical 
description for these observables. At the TeV scale, the $p_{\rT,W}$ 
distribution receives similar NLO EW corrections as 
in $W+1\jet$ production, and also the jet-$p_\rT$ distributions feature 
the expected Sudakov behaviour.
The high relevance of the $\HTtot$ variable for new-physics searches 
and its strong sensitivity to QCD radiation motivate the extension of 
NLO QCD+EW calculations up to $pp\to W+3\jet$, where $\HTtot$ 
starts to be stable with respect to NLO QCD corrections, thereby
rendering NLO QCD+EW predictions more reliable.  Similarly as for
$W+2\jet$, NLO EW corrections to $W+3\jet$ are characterised by
the expected Sudakov suppression in
all $p_\rT$ distributions. However the actual size of the corrections 
varies significantly, depending on the jet multiplicity of the considered process  
and on the individual $\pT$-distribution.
The magnitude of EW corrections at high energy 
can strongly depend on the type of observable as well.
For instance, dijet invariant-mass
distributions turn out to be completely insensitive to EW corrections,
all the way up to the multi-TeV region.
Finally, we pointed out that 
also photon-induced processes and
subleading Born terms
of $\ord(\alpha^{n-1}\alpha^2)$
and $\ord(\alpha^{n-2}\alpha^3)$, 
which result from EW contributions to the
matrix elements, can have a sizable impact in the TeV region.

In summary, EW correction effects in $W+$multijet production feature a
nontrivial dependence on the jet multiplicity, as well as on various
kinematical parameters. Their sizable impact at high energies will
play a key role for tests of the Standard Model and for many BSM
searches based on signatures with jets, leptons and missing energy at the TeV scale.
In a forthcoming publication we plan to present more detailed phenomenological 
results for vector-boson plus multi-jet production, including 
the case of $W^-$ and $Z$ bosons as well as leptonic vector-boson decays.
Our results motivate also further important developments of 
NLO QCD+EW simulations of vector-boson
production in association with multiple jets,
including matching to the parton shower and, ultimately, the 
extension of multi-jet merging techniques
to NLO QCD+EW  simulations.

\acknowledgments
We thank A.~Denner, S.~Dittmaier and L.~Hofer for providing us
with the one-loop tensor-integral library \Collier. 
We are grateful to F.~Cascioli for collaboration
at the initial stage of this project.
This research was supported in part by the Swiss
National Science Foundation (SNF) under contract PP00P2-128552
and by the Research Executive Agency (REA) of the European Union
under the Grant Agreements PITN--GA--2010--264564 ({\it
LHCPhenoNet}), PITN--GA--2012--315877 ({\it MCnet}) and 
PITN--GA--2012--316704 ({\it HiggsTools}). We
thank the Munich Institute for Astro- and Particle Physics
(MIAPP) of the DFG cluster of excellence ``Origin and Structure
of the Universe'' for the hospitality during the completion of
this work.

\bibliographystyle{JHEP}
\input{wjets_nloew_v2.bbl}

\end{document}

%% file: XStable.wj.tex
$\PWp+1\jet$
&  inclusive
&  $\tiny{\deltajj < 3\pi/4}$
&  $\tiny{\HTtot>2~\TeV}$
&  $\tiny{\pTjone>1~\TeV}$
&  $\tiny{p_{T,\PWp}>1~\TeV}$
\\[0.5mm] \hline
{$\sigma^{\NLO}_{\QCD}$~[pb]}
&  $\xs{15664}{(2.3)}^{+5\%}_{-5\%}$              
&  $\xs{13429}{(2.3926)}^{+2\%}_{-3\%}$           
&  $\xs{0.231}{(0.0000)}^{+27\%}_{-20\%}$         
&  $\xs{0.181}{(0.0000)}^{+25\%}_{-19\%}$         
&  $\xs{0.050}{(0.0000)}^{+10\%}_{-10\%}$         
\\
{$\sigma^{\NLO}_{\QCDpEW}$~[pb]}
&  $\xs{15621}{(2.3)}^{+5\%}_{-5\%}$              
&  $\xs{13380}{(2.3927)}^{+2\%}_{-3\%}$       
&  $\xs{0.245}{(0.0000)}^{+26\%}_{-19\%}$         
&  $\xs{0.195}{(0.0000)}^{+25\%}_{-18\%}$         
&  $\xs{0.040}{(0.0000)}^{+6\%}_{-8\%}$           
\\
{$\sigma^{\NLO}_{\QCDpEW}/\sigma^{\NLO}_{\QCD}$}
&  $\xs{1.00}{(0.000)}^{+5\%}_{-5\%}$            
&  $\xs{1.00}{(0.000)}^{+2\%}_{-3\%}$            
&  $\xs{1.06}{(0.000)}^{+26\%}_{-19\%}$           
&  $\xs{1.07}{(0.000)}^{+25\%}_{-18\%}$           
&  $\xs{0.80}{(0.000)}^{+6\%}_{-8\%}$             
\\
{$\sigma^{\NLO}_{\QCDtEW}/\sigma^{\NLO}_{\QCD}$}
&  $\xs{1.00}{(0.000)}^{+5\%}_{-5\%}$            
&  $\xs{1.00}{(0.000)}^{+2\%}_{-3\%}$            
&  $\xs{1.45}{(0.000)}^{+24\%}_{-18\%}$           
&  $\xs{1.41}{(0.000)}^{+23\%}_{-17\%}$           
&  $\xs{0.70}{(0.000)}^{+9\%}_{-10\%}$            
\\
{$\sigma^{\LO}/\sigma^{\NLO}_{\QCD}$}
&  $\xs{0.73}{(0.000)}^{+12\%}_{-10\%}$           
&  $\xs{0.86}{(0.000)}^{+12\%}_{-10\%}$           
&  $\xs{0.14}{(0.000)}^{+23\%}_{-18\%}$           
&  $\xs{0.18}{(0.000)}^{+23\%}_{-18\%}$           
&  $\xs{0.65}{(0.000)}^{+23\%}_{-18\%}$           
\\
\hline

%% file: XStable.wjj.tex
$\PWp+2\jet$
&  inclusive
&  $\tiny{\HTtot>2~\TeV}$
&  $\tiny{\pTjone>1~\TeV}$
&  $\tiny{p_{T,\PWp}>1~\TeV}$
\\[0.5mm] \hline
{$\sigma^{\NLO}_{\QCD}$~[pb]}
&  $\xs{4349}{(1.8810)}^{+0\%}_{-4\%}$            
&  $\xs{0.364}{(0.0001)}^{+12\%}_{-13\%}$         
&  $\xs{0.202}{(0.0001)}^{+2\%}_{-6\%}$           
&  $\xs{0.048}{(0.0000)}^{+0\%}_{-7\%}$           
\\
{$\sigma^{\NLO}_{\QCDpEW}$~[pb]}
&  $\xs{4316}{(1.8811)}^{+1\%}_{-4\%}$            
&  $\xs{0.329}{(0.0001)}^{+8\%}_{-10\%}$          
&  $\xs{0.173}{(0.0001)}^{+1\%}_{-9\%}$           
&  $\xs{0.033}{(0.0000)}^{+6\%}_{-25\%}$          
\\
{$\sigma^{\NLO}_{\QCDpEW}/\sigma^{\NLO}_{\QCD}$}
&  $\xs{0.99}{(0.001)}^{+1\%}_{-4\%}$             
&  $\xs{0.90}{(0.000)}^{+8\%}_{-10\%}$            
&  $\xs{0.86}{(0.001)}^{+1\%}_{-9\%}$             
&  $\xs{0.69}{(0.001)}^{+6\%}_{-25\%}$            
\\
{$\sigma^{\NLO}_{\QCDtEW}/\sigma^{\NLO}_{\QCD}$}
&  $\xs{0.99}{(0.001)}^{+1\%}_{-4\%}$             
&  $\xs{0.85}{(0.000)}^{+10\%}_{-11\%}$           
&  $\xs{0.84}{(0.001)}^{+0\%}_{-4\%}$             
&  $\xs{0.71}{(0.001)}^{+0\%}_{-8\%}$             
\\
{$\sigma^{\LO}/\sigma^{\NLO}_{\QCD}$}
&  $\xs{0.92}{(0.000)}^{+24\%}_{-17\%}$           
&  $\xs{0.64}{(0.000)}^{+31\%}_{-23\%}$           
&  $\xs{0.90}{(0.001)}^{+32\%}_{-23\%}$           
&  $\xs{1.05}{(0.001)}^{+33\%}_{-23\%}$           
\\
\hline

%% file: XStable.wjjj.tex
$\PWp+3\jet$
&  inclusive
&  $\tiny{\HTtot>2~\TeV}$
&  $\tiny{\pTjone>1~\TeV}$
&  $\tiny{p_{T,\PWp}>1~\TeV}$
\\[0.5mm] \hline
{$\sigma^{\NLO}_{\QCD}$~[pb]}
&  $\xs{1135}{(1.9331)}^{+1\%}_{-11\%}$           
&  $\xs{0.377}{(0.0004)}^{+0\%}_{-12\%}$          
&  $\xs{0.161}{(0.0004)}^{+6\%}_{-35\%}$          
&  $\xs{0.038}{(0.0001)}^{+0\%}_{-14\%}$          
\\
{$\sigma^{\NLO}_{\QCDpEW}$~[pb]}
&  $\xs{1120}{(1.9471)}^{+1\%}_{-12\%}$           
&  $\xs{0.313}{(0.0004)}^{+3\%}_{-26\%}$          
&  $\xs{0.123}{(0.0004)}^{+20\%}_{-65\%}$         
&  $\xs{0.026}{(0.0001)}^{+8\%}_{-40\%}$          
\\
{$\sigma^{\NLO}_{\QCDpEW}/\sigma^{\NLO}_{\QCD}$}
&  $\xs{0.99}{(0.003)}^{+1\%}_{-12\%}$            
&  $\xs{0.83}{(0.002)}^{+3\%}_{-26\%}$            
&  $\xs{0.76}{(0.004)}^{+20\%}_{-65\%}$           
&  $\xs{0.69}{(0.003)}^{+8\%}_{-40\%}$            
\\
{$\sigma^{\NLO}_{\QCDtEW}/\sigma^{\NLO}_{\QCD}$}
&  $\xs{0.99}{(0.003)}^{+1\%}_{-11\%}$            
&  $\xs{0.84}{(0.002)}^{+1\%}_{-14\%}$            
&  $\xs{0.83}{(0.004)}^{+9\%}_{-37\%}$            
&  $\xs{0.72}{(0.003)}^{+1\%}_{-14\%}$            
\\
{$\sigma^{\LO}/\sigma^{\NLO}_{\QCD}$}
&  $\xs{1.02}{(0.002)}^{+40\%}_{-26\%}$           
&  $\xs{1.05}{(0.001)}^{+42\%}_{-28\%}$           
&  $\xs{1.43}{(0.004)}^{+42\%}_{-28\%}$           
&  $\xs{1.09}{(0.002)}^{+43\%}_{-28\%}$           
\\
\hline

%% file: XStable.born_new.tex
Born order
& {$\PWp+1\jet$}
&  inclusive
&  $\tiny{\HTtot>2~\TeV}$
&  $\tiny{\pTjone>1~\TeV}$
&  $\tiny{p_{T,\PWp}>1~\TeV}$
\\[0.5mm] \hline
$\ord(\alpha^2)$
& $\sigma^{\mathrm{Born}}_{\gamma p}/\sigma^{\NLO}_{\QCD}$
&  $\xs{0.0031}{(1.e-06)}^{+14\%}_{-14\%}$       
&  $\xs{0.0173}{(8.e-06)}^{+1\%}_{-1\%}$         
&  $\xs{0.0221}{(1.0e-05)}^{+1\%}_{-1\%}$        
&  $\xs{0.0805}{(3.7e-05)}^{+1\%}_{-1\%}$        
\\[2mm]
\hline
Born order
& {$\PWp+2\jet$}
&  inclusive
&  $\tiny{\HTtot>2~\TeV}$
&  $\tiny{\pTjone>1~\TeV}$
&  $\tiny{p_{T,\PWp}>1~\TeV}$
\\[0.5mm] \hline
$\ord(\alphaS\alpha^2)$
& $\sigma^{\mathrm{Born}}_{pp}/\sigma^{\NLO}_{\QCD}$
&  $\xs{0.0008}{(2.e-06)}^{+14\%}_{-11\%}$       
&  $\xs{0.0659}{(4.7e-05)}^{+19\%}_{-15\%}$      
&  $\xs{0.1085}{(9.9e-05)}^{+19\%}_{-15\%}$      
&  \llap{$-$}$\xs{0.0006}{(2.0e-05)}^{-30\%}_{+21\%}$     
\\
$\ord(\alpha^3)$
& $\sigma^{\mathrm{Born}}_{pp}/\sigma^{\NLO}_{\QCD}$
&  $\xs{0.0345}{(1.6e-05)}^{+6\%}_{-8\%}$        
&  $\xs{0.0562}{(1.5e-05)}^{+11\%}_{-10\%}$      
&  $\xs{0.0792}{(4.0e-05)}^{+11\%}_{-10\%}$      
&  $\xs{0.0728}{(6.4e-05)}^{+12\%}_{-10\%}$      
\\
$\ord(\alphaS\alpha^2)$
& $\sigma^{\mathrm{Born}}_{\gamma p}/\sigma^{\NLO}_{\QCD}$
&  $\xs{0.0014}{(1.e-06)}^{+11\%}_{-10\%}$       
&  $\xs{0.0083}{(9.e-06)}^{+9\%}_{-8\%}$         
&  $\xs{0.0103}{(1.5e-05)}^{+9\%}_{-8\%}$        
&  $\xs{0.0426}{(7.6e-05)}^{+9\%}_{-8\%}$        
\\
$\ord(\alpha^3)$
& $\sigma^{\mathrm{Born}}_{\gamma\gamma}/\sigma^{\NLO}_{\QCD}$
&  $\xs{< 0.0001}{(0.000000)}$     
&  $\xs{< 0.0001}{(0.000000)}$       
&  $\xs{< 0.0001}{(0.000000)}$       
&  $\xs{0.0001}{(0.000000)}^{+9\%}_{-8\%}$       
\\[2mm]
\hline
Born order
& {$\PWp+3\jet$}
&  inclusive
&  $\tiny{\HTtot>2~\TeV}$
&  $\tiny{\pTjone>1~\TeV}$
&  $\tiny{p_{T,\PWp}>1~\TeV}$
\\[0.5mm] \hline
$\ord(\alphaS^2\alpha^2)$
& $\sigma^{\mathrm{Born}}_{pp}/\sigma^{\NLO}_{\QCD}$
&  $\xs{0.0018}{(6.e-06)}^{+23\%}_{-17\%}$       
&  $\xs{0.0796}{(9.4e-05)}^{+28\%}_{-21\%}$      
&  $\xs{0.1351}{(0.000352)}^{+28\%}_{-21\%}$     
&  $\xs{0.0016}{(1.2e-05)}^{+43\%}_{-30\%}$      
\\
$\ord(\alphaS\alpha^3)$
& $\sigma^{\mathrm{Born}}_{pp}/\sigma^{\NLO}_{\QCD}$
&  $\xs{0.0619}{(0.000115)}^{+10\%}_{-8\%}$      
&  $\xs{0.0670}{(8.0e-05)}^{+21\%}_{-16\%}$      
&  $\xs{0.0947}{(0.000248)}^{+21\%}_{-16\%}$     
&  $\xs{0.0831}{(0.000172)}^{+22\%}_{-17\%}$     
\\
$\ord(\alphaS^2\alpha^2)$
& $\sigma^{\mathrm{Born}}_{\gamma p}/\sigma^{\NLO}_{\QCD}$
&  $\xs{0.0011}{(3.e-06)}^{+21\%}_{-15\%}$       
&  $\xs{0.0057}{(1.1e-05)}^{+18\%}_{-14\%}$      
&  $\xs{0.0073}{(2.5e-05)}^{+18\%}_{-15\%}$      
&  $\xs{0.0197}{(4.5e-05)}^{+18\%}_{-14\%}$      
\\
$\ord(\alphaS\alpha^3)$
& $\sigma^{\mathrm{Born}}_{\gamma p}/\sigma^{\NLO}_{\QCD}\;$
&  $\xs{\lessgtr \pm 0.0001}{(0.000000)}$  
&  $\xs{\lessgtr \pm 0.0001}{(0.000000)}$       
&  $\xs{\lessgtr \pm 0.0001}{(0.000000)}$      
&  $\xs{\lessgtr \pm 0.0001}{(0.000000)}$ 
\\
$\ord(\alpha^4)$
& $\sigma^{\mathrm{Born}}_{\gamma p}/\sigma^{\NLO}_{\QCD}$
&  $\xs{0.0014}{(3.e-06)}^{+15\%}_{-15\%}$       
&  $\xs{0.0013}{(2.e-06)}^{+2\%}_{-2\%}$         
&  $\xs{0.0018}{(5.e-06)}^{+2\%}_{-2\%}$         
&  $\xs{0.0057}{(1.3e-05)}^{+2\%}_{-2\%}$        
\\
$\ord(\alphaS\alpha^3)$
& $\sigma^{\mathrm{Born}}_{\gamma\gamma}/\sigma^{\NLO}_{\QCD}\;$
&  $\xs{< 0.0001}{(0.000000)}$ 
&  $\xs{< 0.0001}{(0.000000)}$ 
&  $\xs{< 0.0001}{(0.000000)}$        
&  $\xs{< 0.0001}{(0.000000)}$       
\\[2mm]
\hline

%% file: wjets_nloew_v2.bbl
\providecommand{\href}[2]{#2}\begingroup\raggedright\endgroup

%% file: wjets_nloew_v2.bbl
\begin{thebibliography}{100}
\bibitem{munich}
{\sc Munich} is the abbreviation of ``MUlti-chaNnel Integrator at Swiss (CH)
  precision''---an automated parton level NLO generator by S.~Kallweit. In
  preparation.
\end{thebibliography}
